\documentclass[conference]{IEEEtran}
\IEEEoverridecommandlockouts
\usepackage[space]{cite}
\usepackage{amsmath,amssymb,amsfonts}
\usepackage{algorithmic}
\usepackage{graphicx}
\usepackage{epstopdf}
\usepackage{textcomp}
\usepackage[table]{xcolor}
\usepackage{xcolor}
\usepackage{subcaption}
\usepackage{setspace}
\usepackage{url}
\usepackage{hyperref}
\usepackage{breakurl}
\newcommand{\mycomment}[1]{}
\newcommand{\ignore}[1]{}

\usepackage{color}
\usepackage{soul}
\usepackage{xspace}

\def\BibTeX{{\rm B\kern-.05em{\sc i\kern-.025em b}\kern-.08em
    T\kern-.1667em\lower.7ex\hbox{E}\kern-.125emX}}
\begin{document}

\title{Faster than Flash: An In-Depth Study of System Challenges for Emerging Ultra-Low Latency SSDs\\
}

\vspace{-10pt}

\author{
{\normalfont Sungjoon Koh, Junhyeok Jang, Changrim Lee, Miryeong Kwon, Jie Zhang and Myoungsoo Jung } \\
       {\normalsize{Computer Architecture and Memory Systems Laboratory, KAIST}}\\
       {\normalsize \{skoh, jhjang, crlee, mkwon, jie\}@camelab.org, and mj@camelab.org}
       \vspace{-15pt}
}

\maketitle

\begin{abstract}
Emerging storage systems with new flash exhibit ultra-low latency (ULL) that can address performance disparities between DRAM and conventional solid state drives (SSDs) in the memory hierarchy. 
Considering the advanced low-latency characteristics, different types of I/O completion methods (polling/hybrid) and storage stack architecture (SPDK) are proposed. While these new techniques are expected to take costly software interventions off the critical path in ULL-applied systems, unfortunately no study exists to quantitatively analyze system-level characteristics and challenges of combining such newly-introduced techniques with real ULL SSDs. 

In this work, we comprehensively perform empirical evaluations with 800GB ULL SSD prototypes and characterize ULL behaviors by considering a wide range of I/O path parameters, such as different queues and access patterns. We then analyze the efficiencies and challenges of the polled-mode and hybrid polling I/O completion methods (added into Linux kernels 4.4 and 4.10, respectively) and compare them with the efficiencies of a conventional interrupt-based I/O path. In addition, we revisit the common expectations of SPDK by examining all the system resources and parameters. Finally, we demonstrate the challenges of ULL SSDs in a real SPDK-enabled server-client system. Based on the performance behaviors that this study uncovers, we also discuss several system implications, which are required to take a full advantage of ULL SSD in the future.

\end{abstract}


\setstretch{0.932}

\section{Introduction}
\label{sec:intro}
The state-of-the-art solid state drives (SSDs) begun to offer extremely high bandwidth by connecting with an on-chip processor controller, such as a memory controller hub or processor controller hub, via PCI Express (PCIe) buses \cite{jung2014exploring, jung2013challenges}. For example, NVMe SSDs (e.g., Intel 750 \cite{intel750series}) provide read and write bandwidths as high as 2.4 GB/s and 1.2 GB/s, respectively, which are approximately 4.3$\times$ and 2.4$\times$ higher than the bandwidth that a conventional SSD offers \cite{intelsata}. There also exist several industry prototypes that promise to deliver even much higher performance, ranging from 840 KIOPS to one million IOPS \cite{fadu, pm1725}. Thanks to such performance superiority, high-end NVMe SSDs are widely used for diverse computing domains such as data-intensive applications \cite{liu2017graphene, caulfield2009gordon, caulfield2010understanding, li2016hippogriffdb} and server systems as a disk drive cache \cite{hicken2000disk, oh2015enabling}, burst buffer \cite{liu2012role, kougkas2018hermes}, and in-memory computing storage \cite{kumar2019graphone, tseng2018morpheus, li2016hippogriffdb}.

Even though the bandwidth of such modern high-end SSDs almost reaches the maximum performance capability that PCIe buses can deliver, their system-level turn-around delays are still far different from those of other memory technologies or fast peripheral devices, such as DRAM and GPGPU. To bridge the latency disparity between the on-chip processor controller and underlying SSD, new flash based archives, called \emph{Z-SSD}, have come to the attention of both industry and academia \cite{koh2018exploring, zhang2018flashshare}. This new type of SSDs can provide \emph{ultra-low latency} (ULL) behaviors, which exhibit great potential to get block devices close to computational components \cite{jung2014exploring, seshadri2014willow, kang2013enabling}. Specifically, the new flash medium that ULL SSDs employ is a revised version of vertically-stacked 3D NAND flash (48 layers) whose memory read latency for a page is 3 $\mu$s \cite{samsung2017znand}, which is 8$\times$ faster than the fastest page access latency of modern multi-level cell (MLC) flash memory \cite{cheong2018512}. ULL SSDs are expected to satisfy a high level of service-level agreement (SLA) and quality of service (QoS) while providing a much higher storage capacity, compared with other types of new memory, such as resistive memory (ReRAM) \cite{hsu2003electrically}.

To take the advantage of such fast storage, advanced types of I/O completion methods are proposed and implemented in modern Linux systems \cite{yang2012poll, yang2017spdk, zhang2018flashshare}. For example, \cite{yang2012poll} reveals that a \emph{polled-mode} I/O completion method can be better than an interrupt-driven I/O completion. This claim is reasonably accepted in cases where  the device-level latency of high-performance NVMe SSDs is shorter than the latency delays of context switches and interrupt service routine (ISR) management. To make polling more efficient, a hybrid polling is also proposed, which sleeps the polling processes at a certain period and then performs it after the period \cite{polling17, eisenman2018reducing}. These polled mode and hybrid I/O completion methods are newly implemented and published in Linux kernel 4.4 \cite{pollingrelease} and 4.10 \cite{polling17}, respectively. While Linux has simply applied aforementioned new types of I/O completion methods, the benefits of all those studies are evaluated on a DRAM emulation platform. Thus, the system performance of real ULL SSDs may be different than expected by the previous studies.

On the other hand, kernel-bypass storage architecture, such as storage performance development kit (SPDK) \cite{yang2017spdk}, can be employed to directly expose the latency superiority of ULL SSDs to user-level applications. Specifically, SPDK moves most drivers into the userspace, which aims to avoid multiple system calls and to eliminate redundant memory copies from the application accesses. However, because of this user-level implementation, SPDK cannot efficiently manage system interrupt handlers, which in turn makes SPDK employ the polled-mode I/O completion method instead of the interrupts. The real system therefore may not fully enjoy the benefits of SPDK, but, the system-level challenges imposed by such practical challenges of SPDK are not quantitatively analyzed and studied in the literature by far.



An industry article and gray literature uncover the low device-level latency characteristics  of ULL SSDs \cite{sz985}. However, it is unfortunately difficult to estimate their actual performance behaviors by considering diverse system execution parameters, such as different block sizes and queue depths. In addition, from the viewpoint of the advanced I/O completions and storage architecture, it is non-trivial to analyze the system-level challenges that should be addressed to take full advantages of ULL behaviors since these new types of advanced SSD technologies are unavailable in a public market yet. In this work, we characterize the performance behaviors of a real 800GB Z-SSD prototype and describe a wide spectrum of challenges in integrating ULL SSDs into the current software storage stack. 

The main questions that we want to address and observations can be categorized into three groups.  

\subsection{System-Level Performance and ULL-Specific Behaviors}
Regarding new flash, there are a very few studies to demonstrate low-level circuitries and peripherals \cite{cheong2018512} or high-level marketing data \cite{znand17}. However, designing an I/O path with new technologies requires considering many combinations of system and storage parameters, such as diverse queue and block device configuration values. In addition, many system designers need to understand ULL-specific performance characteristics, such as garbage collection and read-and-write interference.

\begin{itemize}
\item How fast are ULL SSDs compared to NVMe SSDs?
\item How much can ULL SSDs cut the long-tail latency?
\item Is the traditional NVMe multi-queue mechanism also affordable for ULL SSDs?
\item Do ULL SSDs also have a critical path that most flash suffers from?
\end{itemize}

\ignore{
\subsubsection{System-Level Latency and Queue Design}
We observe that the ULL SSD shortens the system-level read and write latencies of a high-end NVMe SSD \cite{intel750series} by 36\% and 57\%, on average, respectively. Specifically, even though systems increase I/O queue depths, the ULL SSD provides a sustainable ultra-low latency (for both average and long tail) while the average and 99.999\% latencies of the high-end NVMe SSD seriously degrade. 
Our analysis also shows that the current rich queue mechanism of NVMe may be ``over-kill'' for ULL SSDs; 6$\sim$8 queue entries are good enough to maximize the bandwidth of ULL SSDs, on average, and even in the worst case, it consumes only 16 queue entries, which is well aligned with light queue mechanisms, such as native command queue (NCQ). 

\subsubsection{SSD-Specific Vulnerability}
We observe that the ULL SSD minimizes many I/O interferences between reads and writes, which are considered as a great performance bottleneck of conventional SSDs. Since flush operations or metadata writes that modern file systems must handle are usually intermixed with user requests, the performance of many SSDs are unfortunately not as much promising as SSD vendors promote, in real-life storage stack \cite{won2018barrier, jung2013revisiting, jung2014sprinkler}. 
}

\subsection{System Impacts with Polled-Mode Completion}
Modern polled-mode I/O completion and hybrid polling mechanism are designed to a new storage stack for high-performance SSDs. Since high-end NVMe SSDs have potential to serve most read and write requests in several $\mu$s, \cite{yang2012poll} studies that the polled-mode I/O completion method can reduce the latency further in addressing the software overheads related to interrupt handling. Motivated by this, Linux already implemented those new I/O completion methods in its kernel and published them \cite{pollingrelease, polling17}. 

\begin{itemize}
\item Is the polled-mode I/O completion faster than interrupts?
\item What are the practical overheads when systems employ the polled-mode? 
\item Will be a hybrid polling better than the polled-mode?
\end{itemize}

\ignore{
Our study reveals that, while still high-performance NVMe SSDs may not require a polling-based I/O completion routine (supported from Linux 4.4), ULL SSDs can take an advantage of polling (instead of using an interrupt). However, it raises several system-level challenges as the polling routine fully occupies one or more heavy CPU cores. Specifically, it consumes CPU cycles more than 99\% of the total I/O execution (based on 1 core). In addition, the load/store instructions that access system memory consume more than 95\% of the total pipeline slots for just polling the I/O completion. Even though the polling-based NVMe I/O service can shorten the latency for the most case, we also observe that the current polling technique should be optimized further; it can potentially hurt overall system performance as its overheads have a great impact on making five nine latency (99.999\%) much longer than that of the interrupt-based I/O service (as high as 115\%).
}

\subsection{Advanced Storage Stack Analysis} 
Linux storage stack is considered as the performance bottleneck when ones use high-end NVMe SSDs \cite{yang2017spdk, kim2016nvmedirect}. Intel SPDK reorganizes the storage stack such that users can bypass the kernel and directly access the underlying storage. A common expectation behind this kind of kernel-bypass schemes is to reduce software intervention and shorten the storage latency.

\begin{itemize}
\item Can SPDK really eliminate the latency overheads, imposed by the storage stack? What about ULL SSDs?
\item Will SPDK be the best option for future low-latency storage? Would it have any side effects?
\item How much can SPDK reduce the system-level latency under a real server execution environment?
\end{itemize}


\section{Background}
\label{sec:background}

\begin{table}[]
\small
\centering
\begin{tabular}{|c|c|c|c|}
\hline
\rowcolor[HTML]{EFEFEF} 
\footnotesize{\textbf{3D NAND}}      & \footnotesize{\textbf{BiCS}}      & \footnotesize{\textbf{V-NAND}}    & \cellcolor[HTML]{FFCE93}\footnotesize{\textbf{Z-NAND}}   \\ \hline
\begin{tabular}[c]{@{}c@{}} \footnotesize{\# layer}\end{tabular} & \footnotesize{48}        & \footnotesize{64}        & \footnotesize{48}       \\ \hline
\footnotesize{t\textsubscript{R}}                 & \footnotesize{45$\mu$s}      & \footnotesize{60$\mu$s}      & \footnotesize{3$\mu$s}      \\ \hline
\footnotesize{t\textsubscript{PROG}}              & \footnotesize{660$\mu$s}     & \footnotesize{700$\mu$s}     & \footnotesize{100$\mu$s}    \\ \hline
\footnotesize{Capacity}           & \footnotesize{256Gb}     & \footnotesize{512Gb}     & \footnotesize{64Gb}     \\ \hline
\footnotesize{Page Size}          & \footnotesize{16KB/Page} & \footnotesize{16KB/Page} & \footnotesize{2KB/Page} \\ \hline
\end{tabular}
\caption{Analysis of 3D flash characteristics \cite{cheong2018512}.}
\label{tbl:nand_table}
\vspace{-15pt}
\end{table}

\subsection{Ulta-Low Latency SSD}
\subsubsection{New Flash}
\begin{figure*}
	\centering
	\begin{minipage}[b]{.36\linewidth}
		\centering
		{\includegraphics[width=1\linewidth]{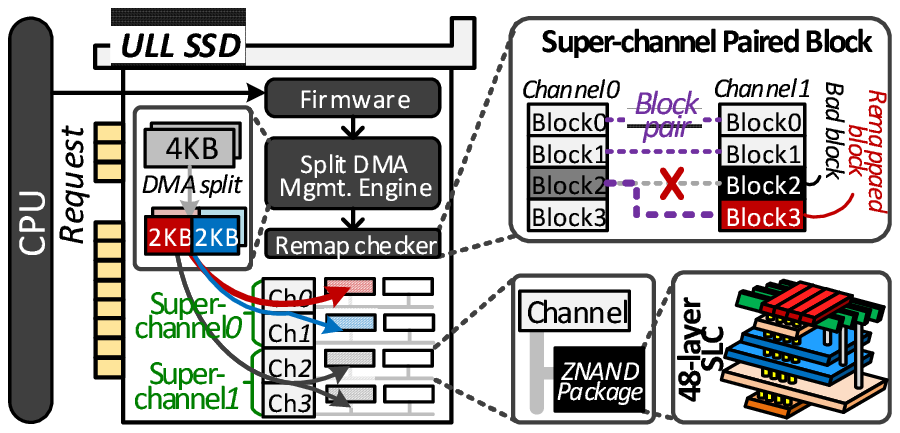}}
		\caption{ULL SSD internals and Split DMA.}\label{fig:bck_dm}\vspace{-15pt}
	\end{minipage}
	\begin{minipage}[b]{.26\linewidth}
		\centering
		{\includegraphics[width=\linewidth]{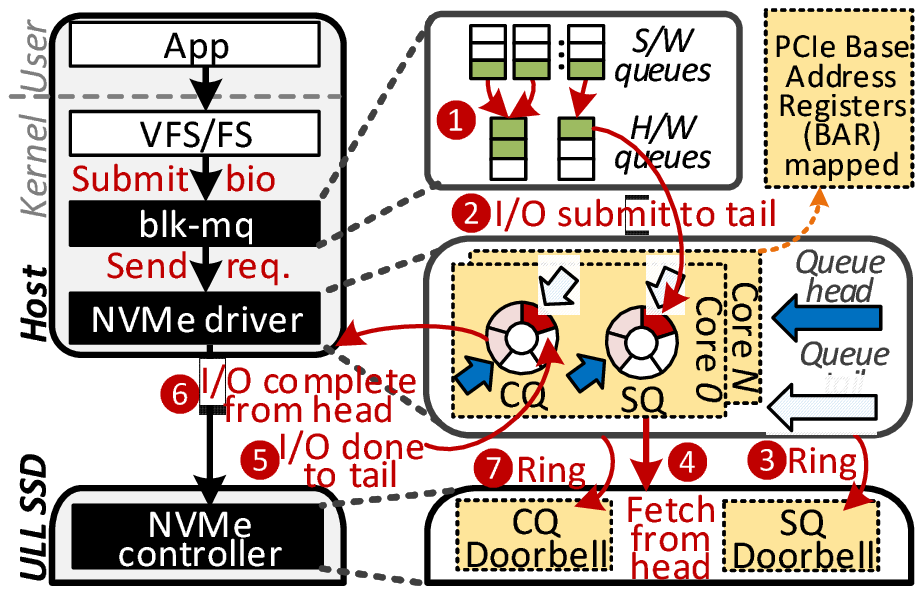}}
		\caption{NVMe storage stack.}\label{fig:bck_storage_stack}\vspace{-15pt}
	\end{minipage}
	\begin{minipage}[b]{.35\linewidth}
		\centering
		{\includegraphics[width=\linewidth]{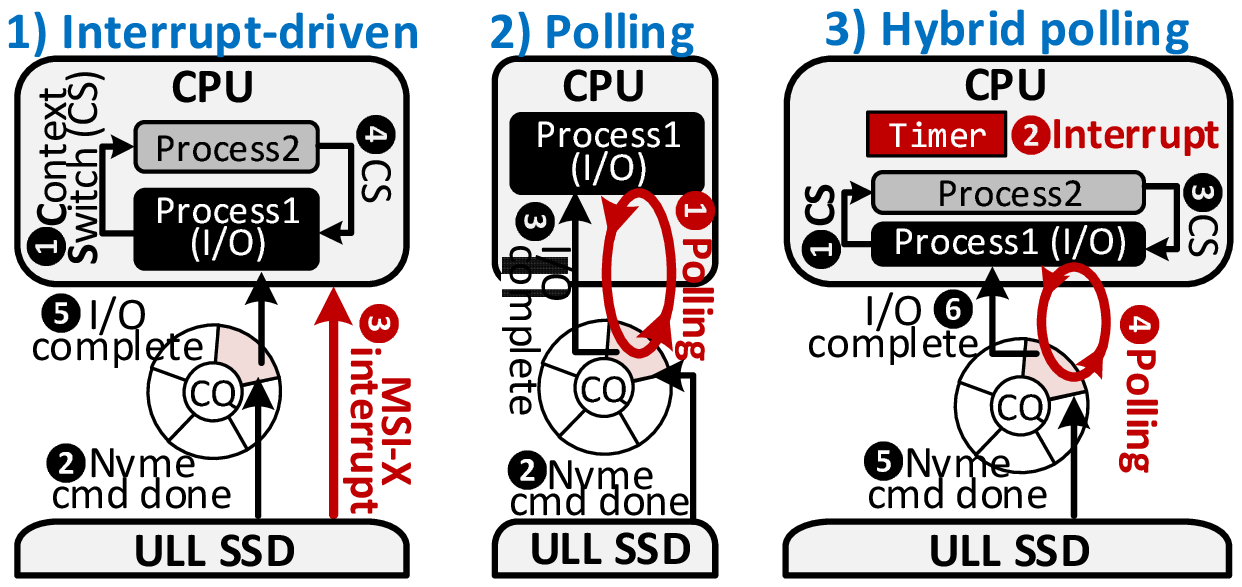}}
		\caption{NVMe I/O completion methods.}\label{fig:bck_completion_timeline}\vspace{-15pt}	
	\end{minipage}
\end{figure*}

Modern SSDs can satisfy a high bandwidth requirement of users by exploiting various architectural supports, such as internal parallelism, I/O queuing/scheduling and DRAM buffers. However, shortening the latency for a basic unit of I/O operation requires low-level memory design changes and device updates. As shown in the right most of Figure \ref{fig:bck_dm}, new flash memory, called \emph{Z-NAND}, leverages single-level cell (SLC) based 3D flash design, but optimizes several I/O circuitries, such as a default I/O page size and DDR interfaces, to offer the low flash latency and shorter data transfer delays, respectively \cite{samsung2017znand}. Table \ref{tbl:nand_table} summarizes the device-level characteristics of three different state-of-the-art 3D flash technologies: i) Bit Cost Scaling (\emph{BiCS}) 3D flash \cite{yamashita201711} ii) Vertically stacked (\emph{V-NAND}) 3D flash \cite{kim2018512} and iii) ULL-tailored flash (\emph{Z-NAND}) \cite{cheong2018512}. Z-NAND uses 48 stacked word-line layer, which exhibits 3$\mu$s and 100$\mu$s for a read operation and a write operation, respectively. The write latency of Z-NAND is shorter than that of BiCS and V-NAND by 6.6$\times$ and 7$\times$, respectively, while its read latency is 15$\sim$20 $\times$ shorter than those of such two modern 3D flash technologies. Even though the storage capacity and page size of Z-NAND are smaller than those of BiCS/V-NAND, ULL SSDs can offer a bigger storage volume with shorter latency by putting more Z-NAND packages into their device platform as a scale-out solution. 

\subsubsection{Multiple Super-Channel Architecture}
The storage architecture of high-end SSDs consists of multiple system buses, referred to as \emph{channels}, each employing many flash packages via multiple datapaths, called \emph{ways} \cite{eshghi2013ssd, jung2017exploring, jung2012physically}. Similarly, ULL SSDs adopt this multi-channel and multi-way architecture, but further optimize the datapaths and its channel-level striping method. As described in Table \ref{tbl:nand_table}, the basic I/O unit of Z-NAND (i.e., page) is much smaller than that of other flash technologies, which in turn can introduce a higher level of device-level and flash-level parallelism by serving a host request with finer granular operations \cite{jung2017exploring, jung2012evaluation, jung2012physically}. Specifically, ULL SSDs split a 4KB-sized host request into two operations and issue them to two different channels simultaneously, as shown in Figure \ref{fig:bck_dm} (left). These two channels always handle flash transactions together as a pair of system buses, which are referred to as a \emph{super-channel}. To efficiently manage data transfers and flash transactions upon the super-channel architecture, ULL SSDs exploit an optimized circuit that automatically adjusts data-flow and manages a timing skew on individual channels \cite{cheong2018512}. This circuit, called \emph{split-DMA} management engine, can reduce the read access time thereby tailoring ULL further. One of the concerns behind the super-channel architecture is to manage wear-out blocks (i.e., bad blocks) each potentially existing in different channels. Since two operations spread across a pair of system buses based on their offsets (within a super-channel), flash firmware can waste the storage space if the bad blocks appear in one of the channel pairs. To address these challenges, the split-DMA management engine also employs a remap checker to automatically remap the physical block addresses between a bad block and a clean block. The remap checker exposes this semi-virtual address space to the flash firmware, such that the storage space can be fully utilized with the super-channel technique.

\subsubsection{Suspend and Resume Support}
\label{section:suspend}
To reduce the read latency more, Z-SSDs also apply a suspend/resume DMA technique \cite{cheong2018512}. While the split-DMA management engine can shorten the latency with a finer granular data access, it may not be able to immediately serve a read request if the target super-channel, associated with the data, is busy to transfer a write request, which was issued at an earlier time. This resource conflict can increase the latency of such reads, waiting for a service in a device-level pending queue. The suspend/resume DMA method of ULL SSDs pauses the write service in progress for the super-channel that the read targets by storing an exact point of the flash write operation into a controller-side small buffer. 
The suspend/resume DMA engine then issues the read command to its target flash package without a wait for resolving the resource conflict. Once the read has been successfully committed, which takes only a few cycles, the engine resumes the write by restoring the context (e.g., the stored write point). Thus, this suspend/resume mechanism can maximize the resource utilization of multi-way of super-channels and reduce the overall latency of ULL SSDs, thereby satisfying diverse levels of QoS and SLA. 

\subsection{Advanced Storage Stack}
\subsubsection{NVMe Storage Stack}
In conventional systems, most block I/O services should be requested to kernel-level I/O modules through system call(s). The requests are delivered at first to virtual file system (VFS) and native file system (e.g., ext4). They not only provide compatible user interfaces and file management, but also perform page caching and guarantee ACID (atomicity, consistency, isolation, durability) of the underlying storage. Underneath the file systems, a multi-queue block layer, \texttt{blk-mq} \cite{bjorling2013linux}, schedules I/O requests while an NVMe driver manages block storage protocols. The \texttt{blk-mq} employs two queues, each corresponding to a software queue and hardware queue. The software queue exists per CPU core and is responsible for file system's block request (i.e., \texttt{bio} structure) handling. On the other hand, the number of existing hardware queues equals the number of queues provided by the underlying NVMe driver provides and is used for delivering \texttt{bio} requests to the NVMe driver. 


\subsubsection{NVMe Multi-Queue Mechanism}
All NVMe SSDs, including ULL SSDs, expose a certain memory region, which is used for multi-queue communications to the host-side NVMe driver through \emph{PCIe base address registers} (BARs) \cite{nvme121}. 
When a host issues an NVMe command (as a form of NVMe queue entry) to the underlying NVMe SSD, it is enqueued into the target \emph{submission queue} (SQ), and SQs are managed in a first-in first-out manner. To complete the corresponding I/O service, the NVMe SSD requires issuing a completion command to the corresponding \emph{completion queue} (CQ).
Each queue can accommodate $2^{16}$ NVMe command entries, and all the queues that the target SSD can support are mapped to the BARs. To synchronize the SQ/CQ states between the host and SSD, NVMe also specifies a pair of \emph{doorbell} (DB) registers, each being allocated to an SQ and CQ, respectively. If a host module or SSD firmware module updates any entry of the SQ or CQ, it can simply inform such update by writing to the corresponding DB; as all the DBs are mapped to the BARs, they are visible for both the host and SSD. While this multi-queue mechanism is affordable to fully utilize the PCIe bandwidth capability, the queue structure itself is very rich and complicated, which can increase queue waiting time at some extent. 


\subsubsection{I/O Completion Methods for NVMe}
When there is an update for a CQ, the SSD device informs its completion to the host through a \emph{message signal interrupt} (MSI). If an MSI is reported, the host's NVMe driver is required to complete the request by checking the target CQ and handling ISR. This interrupt-driven I/O path is sufficient to manage slow block devices, but as NVMe SSDs provide 3.2$\times$ higher bandwidth and 1.8$\times$ shorter latency than conventional SSDs (e.g., SATA SSD), a polled-mode I/O completion method has attracted the attention of both academia \cite{yang2012poll,zhang2018flashshare,eisenman2018reducing} and industry \cite{polling17, eisenman2018reducing} instead of interrupts. Typically, \texttt{blk-mq} polls hardware queues to check the target entry by retrieving the corresponding queue and request information from a bitmap mask, called \emph{cookie} (returned at the submission time). As the hardware queues are one-to-one matched with the NVMe queues (offered by the SSD device), the NVMe driver provides a polling interface, called \texttt{nvme\_poll()}, to \texttt{blk-mq}. In this function, the corresponding CQ entry is iteratively checked. If an update of the CQ entry is detected during polling process, it stops polling and completes the corresponding I/O process by returning the results to the users. 

This polling method is implemented in Linux 4.4 kernel, but from 4.10 kernel, Linux supports a hybrid polling \cite{polling17, eisenman2018reducing}. The hybrid polling sleeps for a while in an attempt to reduce the number of polling the queues. Specifically, the hybrid polling method checks the queue like the polled-mode I/O completion, but it adds up a certain time of sleep before jumping to the poll process. In the latest version of kernel that we tested (Linux 4.14), the hybrid polling calculates an average time for the previous I/O completions and then sleeps the poll process as long as the half of the average time, monitored. 


\subsubsection{Kernel-bypass (SPDK)}
To reduce the overheads brought by the storage stack, kernel-bypass schemes are applied in modern NVMe SSDs \cite{yang2017spdk, kim2016nvmedirect, openmpdk2018, unvme19, unvmemicron}. Intel SPDK is a practical approach that implements the NVMe driver at user-level \cite{yang2017spdk}. To this end, SPDK unbinds the underlying NVMe device and rebinds it to a \emph{user-space I/O driver} (\texttt{uio}) or \emph{virtual function driver} (\texttt{vfio}) driver. SPDK maps PCIe BARs to a huge page \cite{spdkhuge}, which is allocated and managed by a kernel and network bypass-framework, called \emph{DPDK} \cite{intel2014data}. Since the huge page is mostly not swapped out, \texttt{uio/vfio} of SPDK can directly check NVMe queues (SQs/CQs) and doorbell registers from the user-level, which makes it bypass the storage stack and allows user applications directly to access the underlying NVMe SSD. However, ISR cannot be handled from the user level driver, only polling can be adopted as an I/O completion routine.


\vspace{-5pt}
\section{Evaluation Setup}
\vspace{-5pt}

\subsection{Benchmark}
To characterize NVMe and ULL SSDs, we use FIO (v3.13) as our microbenchmark suite\cite{axboe2017flexible}. We set an \uppercase{O\_DIRECT} flag for all evaluations to bypass page caches and directly serve the I/O requests to/from the underlying SSDs. In this test, we also use Linux native AIO (\texttt{libaio}) \cite{bhattacharya2008linux} as an I/O engine to generate asynchronous block I/O requests. Even though we test the SSDs with different block I/O sizes, ranging from 4KB to 32KB, for specific evaluations, such as performance analysis and CPU utilization, we configure the default block size as 4KB. On the other hand, synchronous \texttt{preadv2}/\texttt{pwritev2} (\texttt{pvsync2}) is enabled by the  I/O engine to analyze the system impacts brought by different types of I/O completion methods.

\begin{figure}
	\centering
	\begin{subfigure}{0.49\linewidth}
		\includegraphics[width=\linewidth]{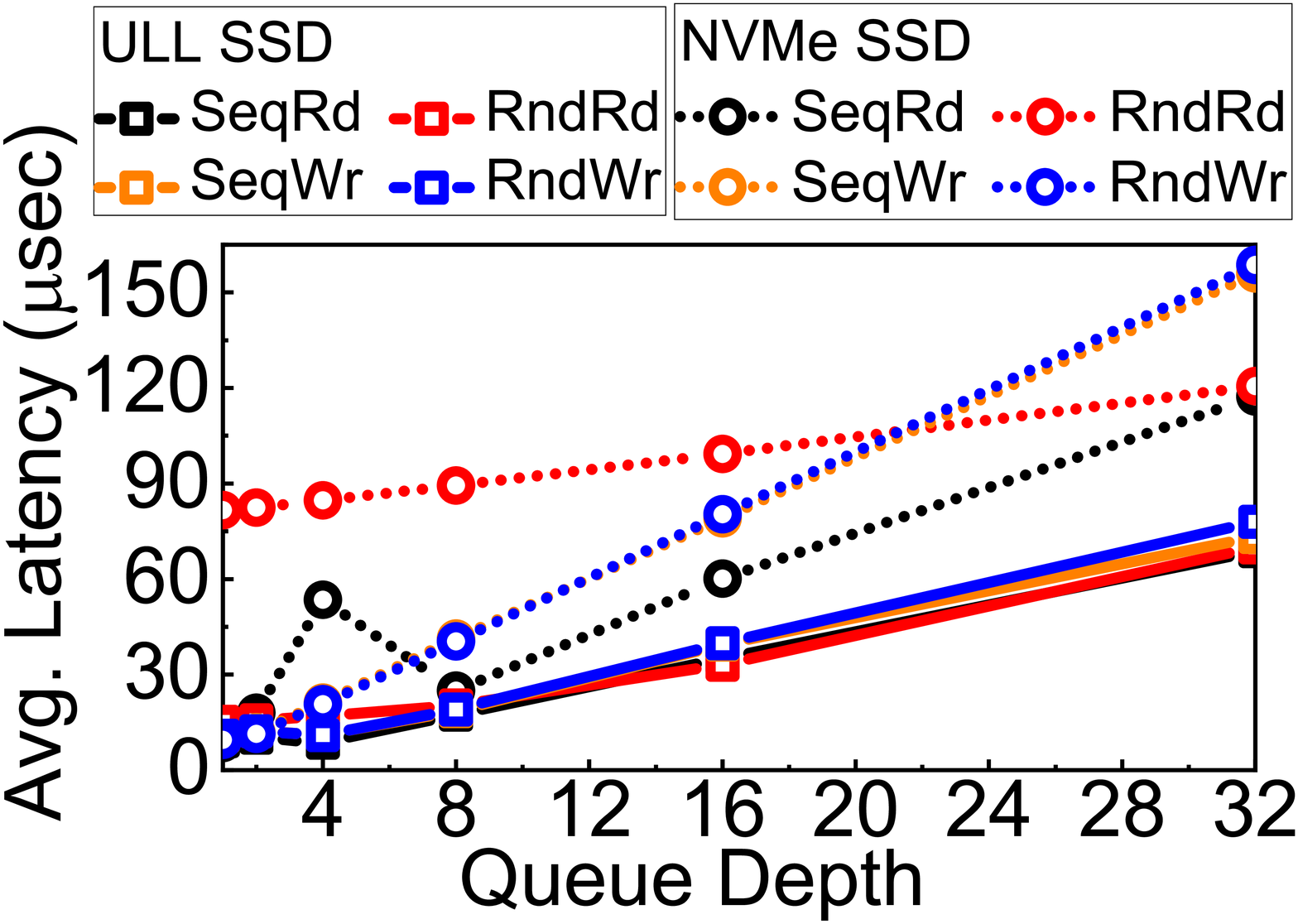}
		\caption{Average latency.}
		\label{fig:qd_avg_lat}
	\end{subfigure}
	\begin{subfigure}{0.49\linewidth}
		\includegraphics[width=\linewidth]{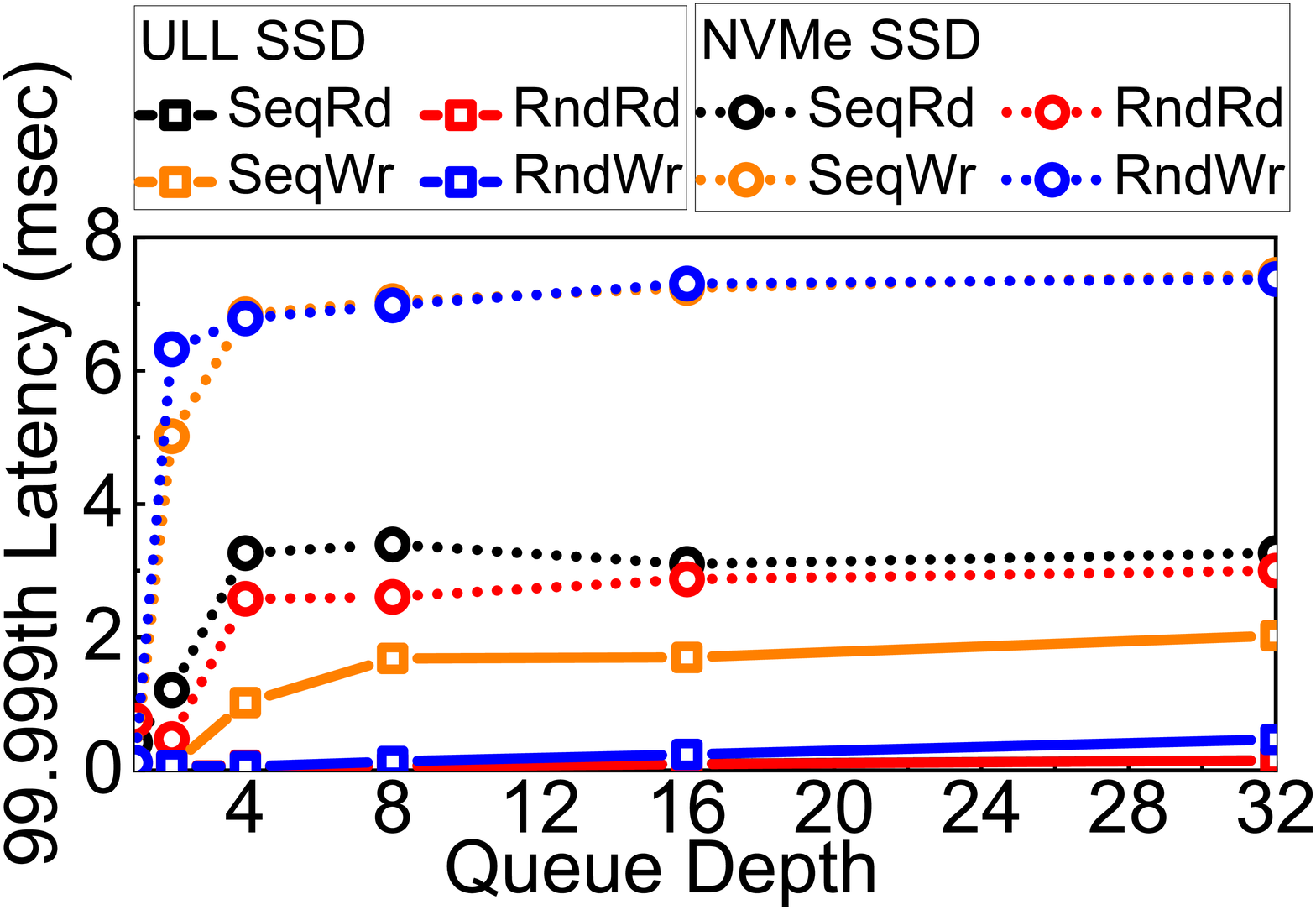}
		\caption{99.999th latency.}
		\label{fig:99999_lat}
	\end{subfigure}	
	\vspace{-3pt}
	\caption{Latency analysis of NVMe SSD and ULL SSD.}
	\label{fig:lat_cmp}
	\vspace{-20pt}
\end{figure}

\subsection{Device Configuration and Profilers}
We use a testbed that employs a 4.6 GHz, 6-core Intel i7 processor (i7-8700) and 16GB DDR4 DRAM. All OS modules and executables are stored and executed from an external 400GB SSD \cite{intel750series}. We evaluate an NVMe SSD by using Intel 750 SSD \cite{intel750series} as it is the only commercially available device that uses the PCIe interface (without a M.2 bridge) over a standard NVMe protocol in the market. To make the underlying NVMe SSD up-to-date, we use the latest firmware (8EV101H0). 
We evaluate a 800GB Z-SSD prototype as an ULL SSD. In this test, all the SSDs are connected to the host of the testbed via PCIe 4x 3.0 lanes. For this study, we use CentOS 7.6 and Linux kernel 4.14, which contains the most recent and stable version of NVMe storage stack. To make sure that CPU cores do not reside on the critical path, we configure the policy of \texttt{CPUfreq} governors (in the Linux kernel) with ``performance'', which statically sets CPU to the highest frequency in the range of target CPU scaling. 
Lastly, we use the FIO report and Intel Vtune Amplifier 2019 \cite{zoneintel} to characterize memory access patterns that the NVMe storage stack exhibits. 


\subsection{Storage Stack}
To evaluate the polled-mode I/O completion, we set the polling flag (\texttt{queue\_io\_poll}) of Linux pseudo file system (\texttt{sysfs}) as '1'; 
otherwise, it is configured with '0' for all the tests that we performed. In addition, the priority of I/O requests (i.e., \texttt{hipri} flag) is set with a value, higher than that of other tasks; we make sure that only one core, running at the maximum frequency, is utilized to manage incoming I/O requests. For kernel-bypass, we use SPDK 19.07 \cite{intelspdk} and configure the size of a huge page with 2MB. The huge page is directly mapped to PCIe BARs by DPDK 19.02 \cite{intel2014data} in our evaluation. For the user-level NVMe driver, we use \texttt{uio} 0.01.0 \cite{koch2011userspace} and utilize \texttt{fio\_plugin()} (provided by SPDK) for our SPDK-enabled benchmark executions.



\section{System-Level Analysis for ULL SSDs} 
\subsection{How fast are ULL SSDs compared to NVMe SSDs?}
Figure \ref{fig:qd_avg_lat} shows the overall latency characteristics of ULL SSD and NVMe SSD that we tested with varying I/O depths, ranging from 1 to 32. With a low I/O queue depth (1$\sim$4), the write latency of NVMe SSD is around 14.1$\mu$s, which is slightly worse than that of ULL SSD; ULL SSD offers 12.6$\mu$s and 11.3$\mu$s for reads and writes, on average, respectively. The reason why NVMe SSD can exhibit much shorter write latency than its actual flash execution time is that it caches and/or buffers the data using their large size of internal DRAM. Nevertheless, NVMe SSD cannot hide the long latency, imposed by the low-level flash, in the case of random reads (82.9$\mu$s). Specifically it is 5.2$\times$ slower than ULL SSD (15.9$\mu$s). This is because the low locality of random reads enforces the internal cache to frequently access the underlying flash media, which makes NVMe SSD expose the actual flash performance to the host. Similarly, as the queue depth increases, the execution time characteristics of NVMe SSD significantly get worse, and its latency increases as high as 121$\mu$s and 159$\mu$s for random writes and reads, respectively. In contrast, as shown in the figure, ULL SSD provides reasonably sustainable performance even in the test with the high I/O queue depth.

\subsection{How much can ULL SSDs cut the long-tail latency?}
The performance difference between NVMe SSD and ULL SSD becomes more notable when we examine their long tail latency behaviors. Figure \ref{fig:99999_lat} analyzes \emph{five-nines} (99.999\%) latency for both NVMe SSD and ULL SSD. The evaluation results show the worst case performance brought by each SSD's low-level flash due to internal tasks, such as garbage collection, and/or internal DRAM cache misses. 
For example, even though the average latency of random/sequential writes is better than that of reads, in this long tail latency evaluation, the writes are worse than the random reads by 2.1$\times$, on average. In overall, NVMe SSD increases the five-nines latency of reads and writes than the average latency of them by 17.9$\times$ and 108$\times$, respectively.  The reason behind this significant performance degradation is that the the five-nines latency cannot take an advantage of NVMe SSD's architectural supports due to many systemic problems (e.g., resource conflicts, insufficient internal buffer size), which are insufficient to accommodate all the incoming I/O requests, and heavy internal tasks \cite{jung2013revisiting, jung2012taking}. In contrast, ULL SSD exhibits much shorter latencies, ranging from a few $\mu$s to hundreds $\mu$s, for both the reads and writes. In contrast to NVMe SSD, the backend flash media of ULL SSD offers significantly low latency by new memory design and device-level optimizations. These low-level optimizations and Z-NAND characteristics not only can reduce memory-level latency further but also offer a better opportunity to fully reap the benefits of multi-channel and multi-way architectural support. 

\begin{figure}
	\centering
	\begin{subfigure}{0.36\linewidth}
		\includegraphics[width=\linewidth]{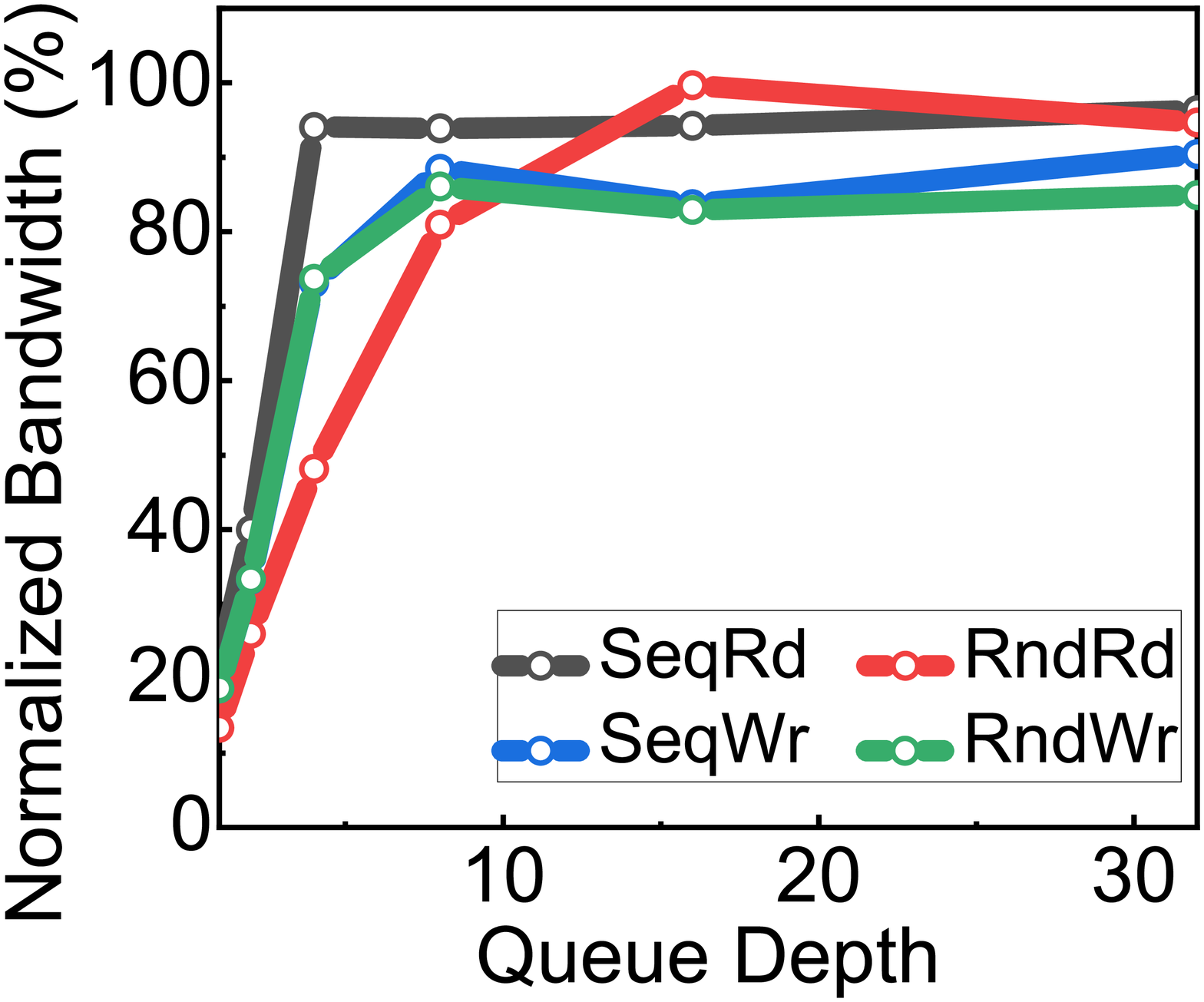}
		\caption{ULL SSD.}
		\label{fig:norm_bw_ull}
	\end{subfigure}
	\begin{subfigure}{0.62\linewidth}
		\includegraphics[width=\linewidth]{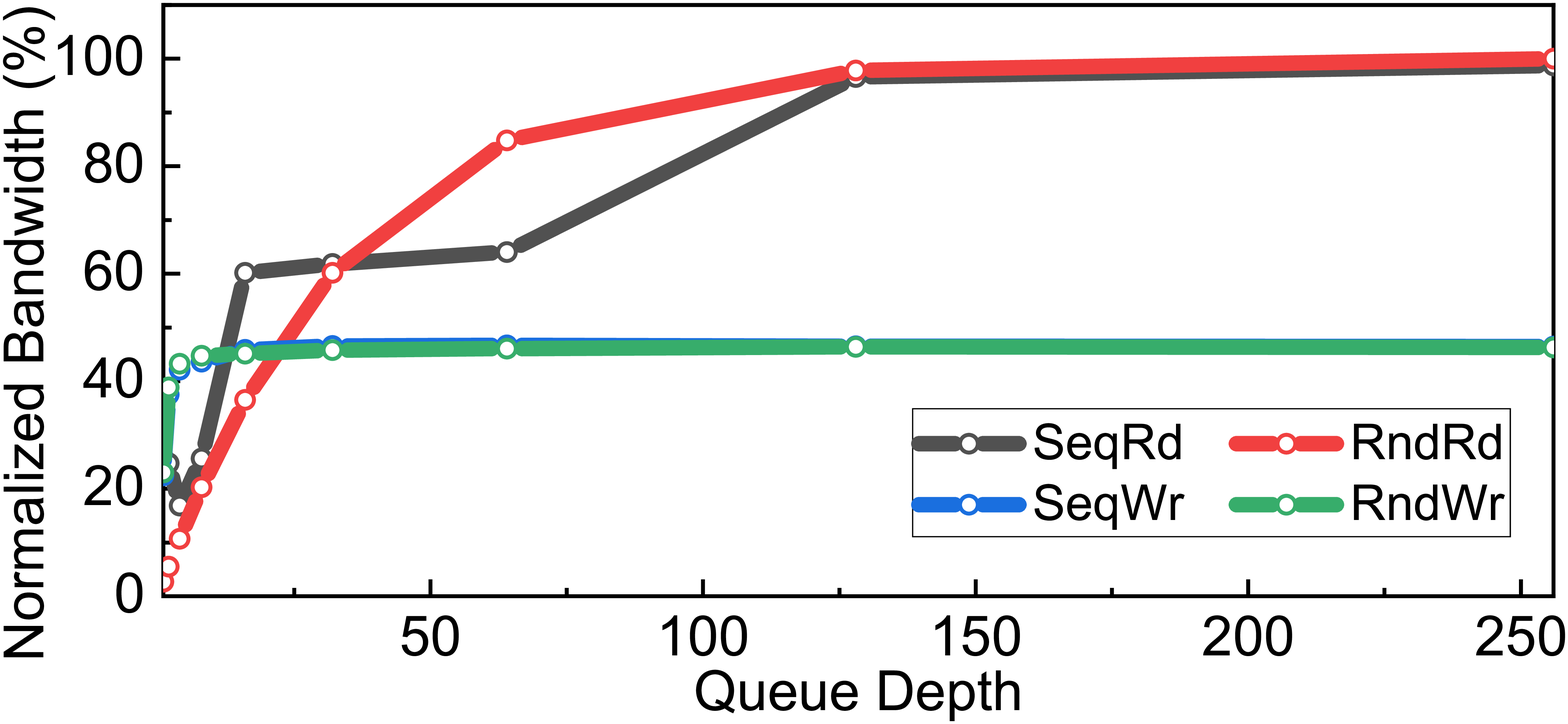}
		\caption{NVMe SSD.}
		\label{fig:norm_bw_nvme}
	\end{subfigure}
	\vspace{-5pt}
	\caption{Bandwidth analysis (normalized to max B/W).}
	\vspace{-10pt}
	\label{fig:bw_cmp}
\end{figure}
\begin{figure}
	\centering
	\begin{subfigure}{0.49\linewidth}
		\includegraphics[width=\linewidth]{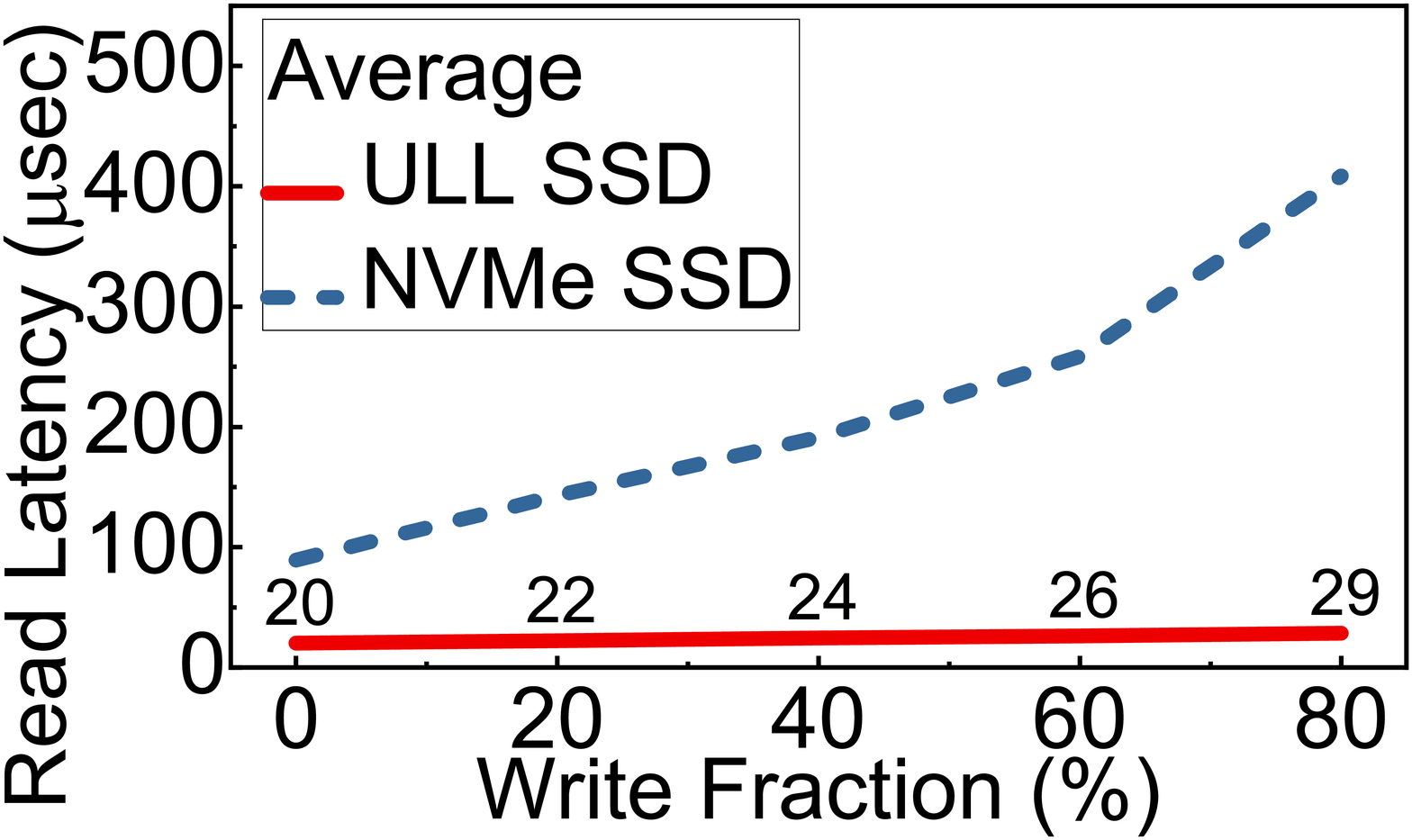}
		\caption{Average latency.}
		\label{fig:mixed_avg_lat}
	\end{subfigure}
	\begin{subfigure}{0.49\linewidth}
		\includegraphics[width=\linewidth]{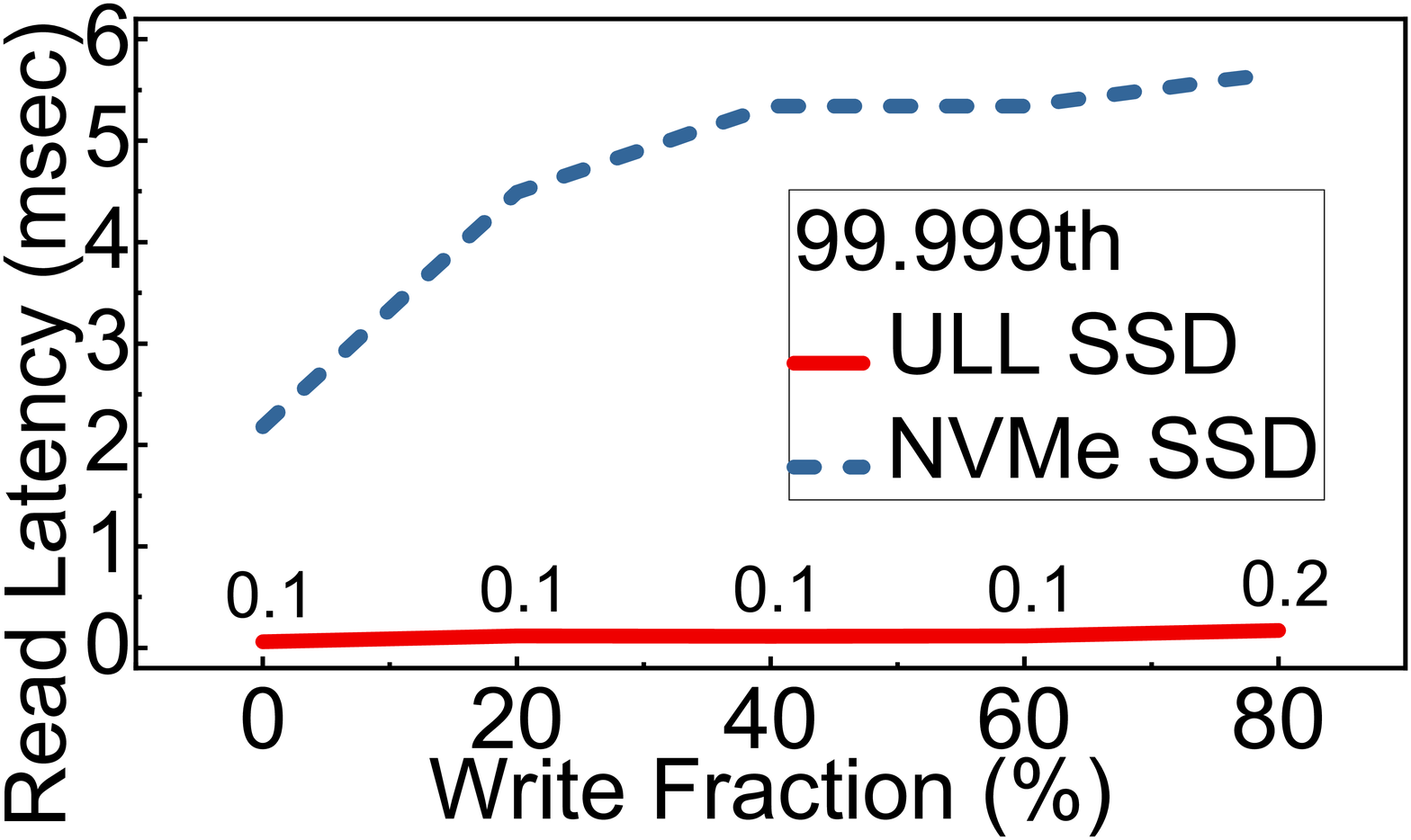}
		\caption{99.999th latency.}
		\label{fig:mixed_99999_lat}
	\end{subfigure}
	\vspace{-3pt}
	\caption{I/O interference analysis.}
	\label{fig:mixed_lat}
	\vspace{-20pt}
\end{figure}

\subsection{Is the traditional NVMe multi-queue mechanism also affordable for ULL SSDs?} 
Figure \ref{fig:bw_cmp} shows the bandwidth utilizations of NVMe SSD and ULL SSD with varying I/O depths (1$\sim$256). While ones can expect that the level of parallelism can increase as the queue depth increases (thereby higher bandwidth), unfortunately, NVMe SSD cannot reach the maximum bandwidth for the service of many 4KB-sized I/O requests. NVMe SSD only utilizes 40\% of the total performance capacity for the 4KB-sized writes. Interestingly, in contrast to the previous read latency evaluations, the bandwidth of NVMe SSDs for random reads outperforms that of all other I/O patterns (that we tested) with higher queue depths (more than 128). This is because, with more I/O requests, scheduled in the queue, SSDs can easily find out a set of flash media that can simultaneously serve multiple I/O requests (by spreading the requests across different flash dies in parallel). Thus, random and sequential reads on NVMe SSD can offer the maximum bandwidth (1.8GB/s). In contrast, the bandwidth utilization of ULL SSD bumps against the maximum bandwidth for all the read test scenarios, and even for the sequential and random writes, ULL SSD can utilize 90\% and 87\% of the total bandwidth, on average, respectively.

It is worthwhile to report that, ULL SSD needs only ``8 queue entries'' for the sequential accesses; even in the worst case, 16 queue entries are sufficient to achieve the maximum bandwidth that ULL SSD offers. We believe that the rich queue mechanism and software-based protocol management of NVMe (which are managed by an NVMe driver and/or Linux \texttt{blk-mq}) are very reasonable design strategies to maximize the benefits of modern high performance SSDs in the sense that such SSDs require securing many I/O resources for higher parallelism. However, once the latency becomes shorter by employing new flash or memory technologies (similar to ULL SSD), we believe that the rich queue and existing NVMe protocol specification are \emph{overkilled}; a future ULL-enabled system may require to have a lighter queue mechanism and simpler protocol, such as NCQ \cite{huffman2003serial} of SATA \cite{serial2007international}.

\begin{figure}
	\centering
	\begin{subfigure}{0.49\linewidth}
		\includegraphics[width=\linewidth]{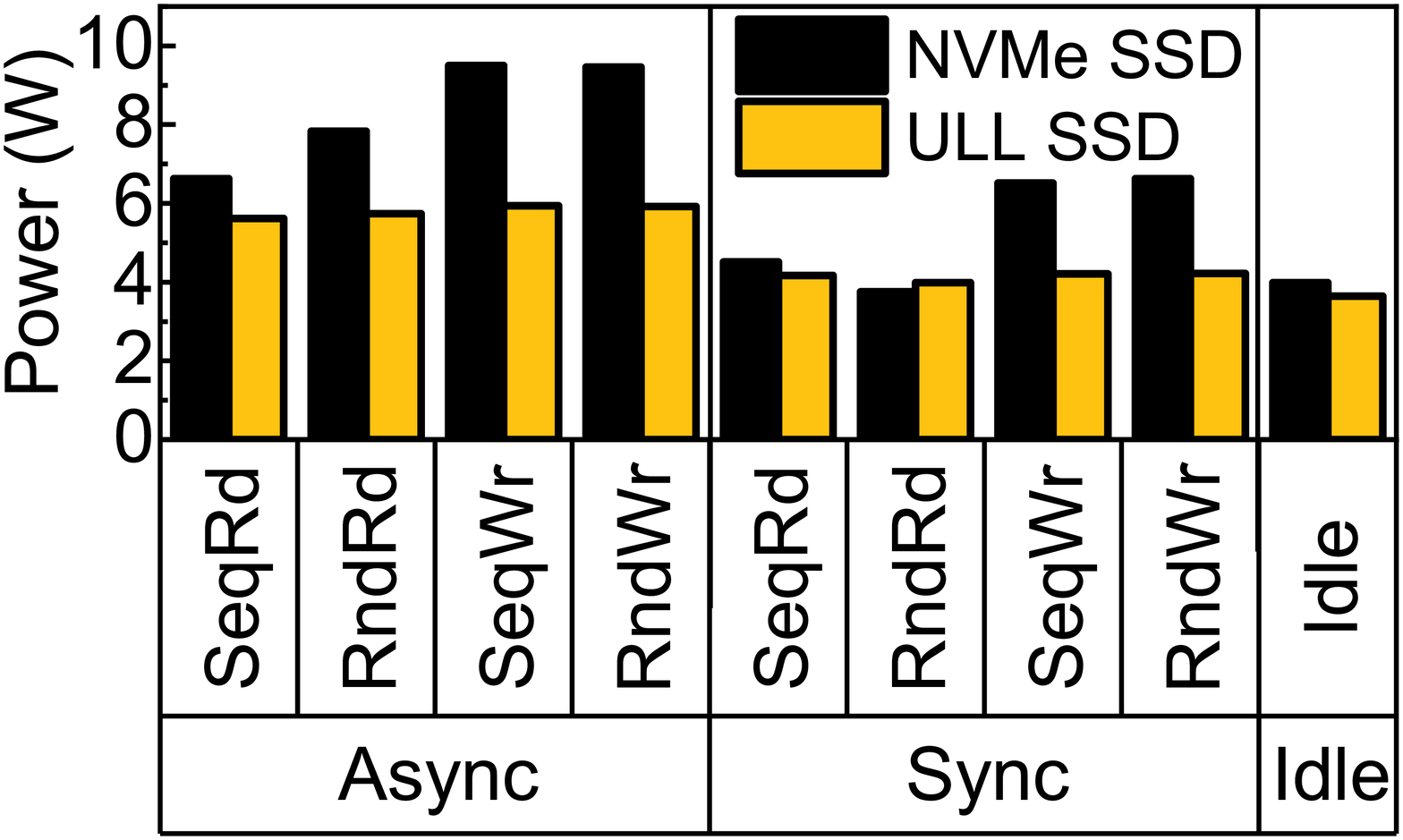}
		\caption{Average power consumption.}
		\label{fig:overall_avg_power}
	\end{subfigure}
	\begin{subfigure}{0.49\linewidth}
		\includegraphics[width=\linewidth]{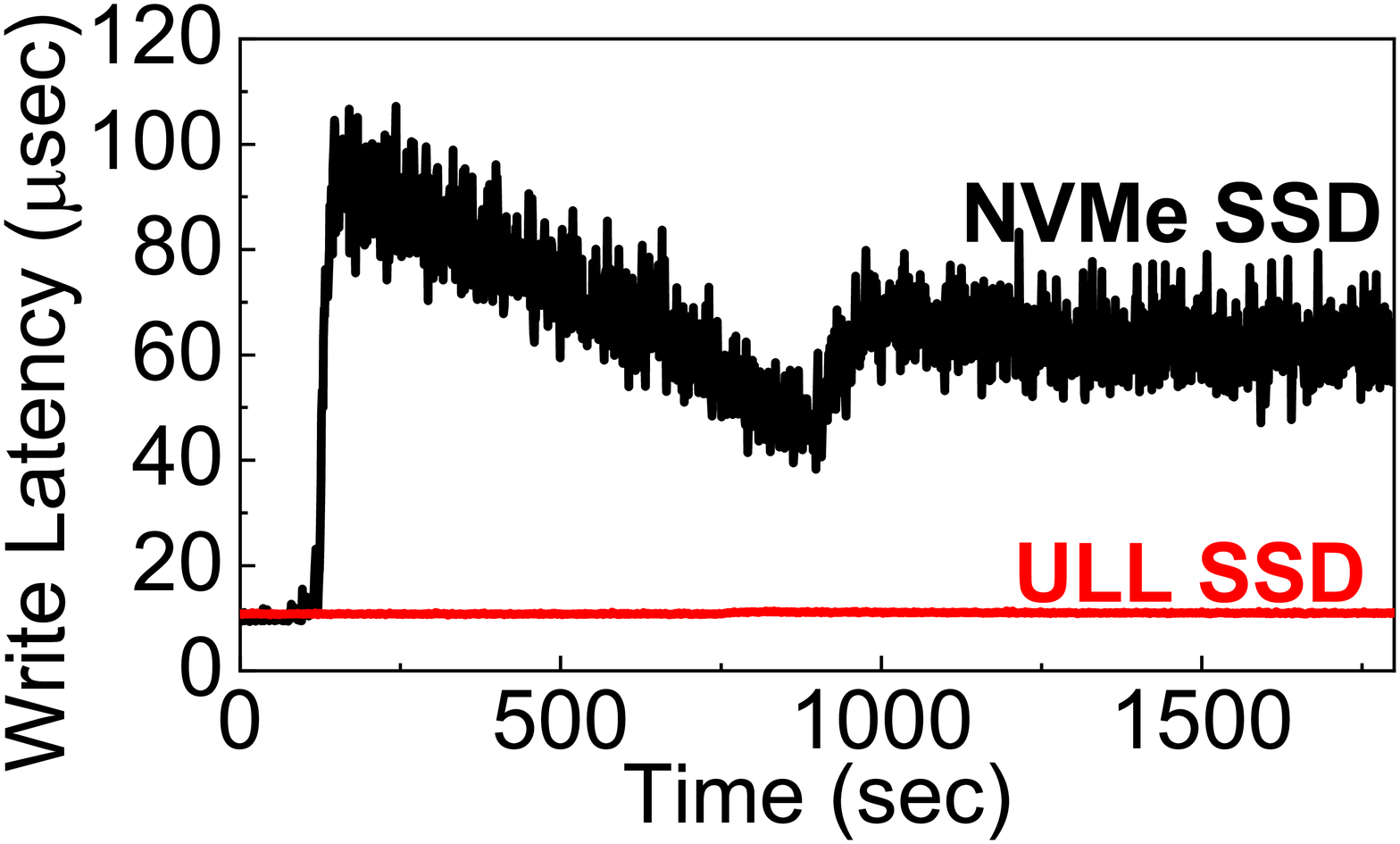}
		\caption{Latency degradation by GC.}
		\label{fig:qd1_gc}	
	\end{subfigure}
	\vspace{-5pt}
	\caption{Power analysis and garbage collection.}
	\label{fig:power_analysis}
	\vspace{-15pt}
\end{figure}

\vspace{-5pt}
\subsection{Do ULL SSDs also have a critical path that most flash suffers from?}
\subsubsection{I/O Interference Impact}
Figure \ref{fig:mixed_lat} analyzes the degree of I/O interference when reads and writes are intermixed. For this analysis, we randomly read data from NVMe SSD and ULL SSD by sporadically issuing the writes among many read requests. In addition, we increase the write fraction of the total I/O executions, ranging from 20\% to 80\%, in an attempt to analyze the different I/O interference behaviors. As shown in Figure \ref{fig:mixed_avg_lat}, the average read latency of NVMe SSD linearly increases as the write fraction (in the intermixed workloads) increases. Even when only 20\% writes are interleaved with the reads, the writes seriously interfere I/O services of such reads, which can make the read latency worse than that of read-only workloads by 1.6$\times$ (54$\mu$s, on average). There are two root causes. First, a write operation of a conventional flash (at memory-level) takes a significantly longer time than a read operation (19$\times$ at most), and the write blocks all other subsequent read services. Second, data transfers for the write (4KB) also occupy a specific channel for around 60$\mu$s, which prevents an incoming read command from being issued to a target memory in the channel. In contrast, we observe that ULL SSD exhibits sustainable latency behaviors irrespective of the amount of write operations (interleaved with the reads). These anti-interference characteristics of ULL SSD are captured by our five-nines latency analysis as well. As shown in Figure \ref{fig:mixed_99999_lat}, while the five-nines read latency of NVMe SSD increases as high as 4.5ms even with 20\% sporadic writes, ULL SSD maintains its latency under 118$\mu$s. Since modern file systems and OS kernels are required to periodically write metadata or to perform journaling \cite{shen2014journaling}, we believe that ULL SSD can be a more desirable solution even in cases where many data-intensive applications intensively reads multiple data. 

\begin{figure}
	\centering
	\begin{subfigure}{0.49\linewidth}
		\includegraphics[width=\linewidth]{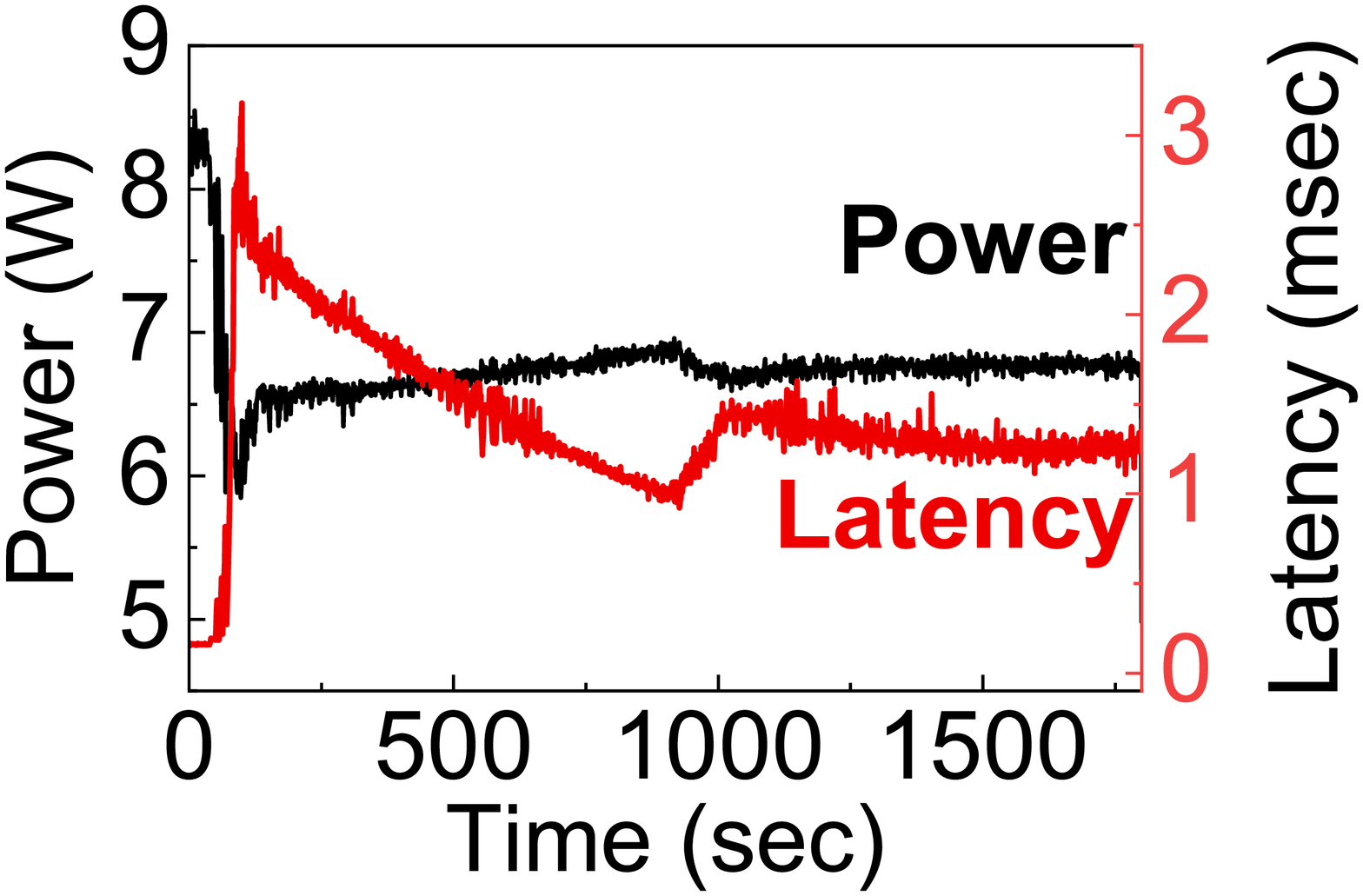}
		\caption{NVMe SSD.}
		\label{fig:nvme_gc_power}
	\end{subfigure}
	\begin{subfigure}{0.49\linewidth}
		\includegraphics[width=\linewidth]{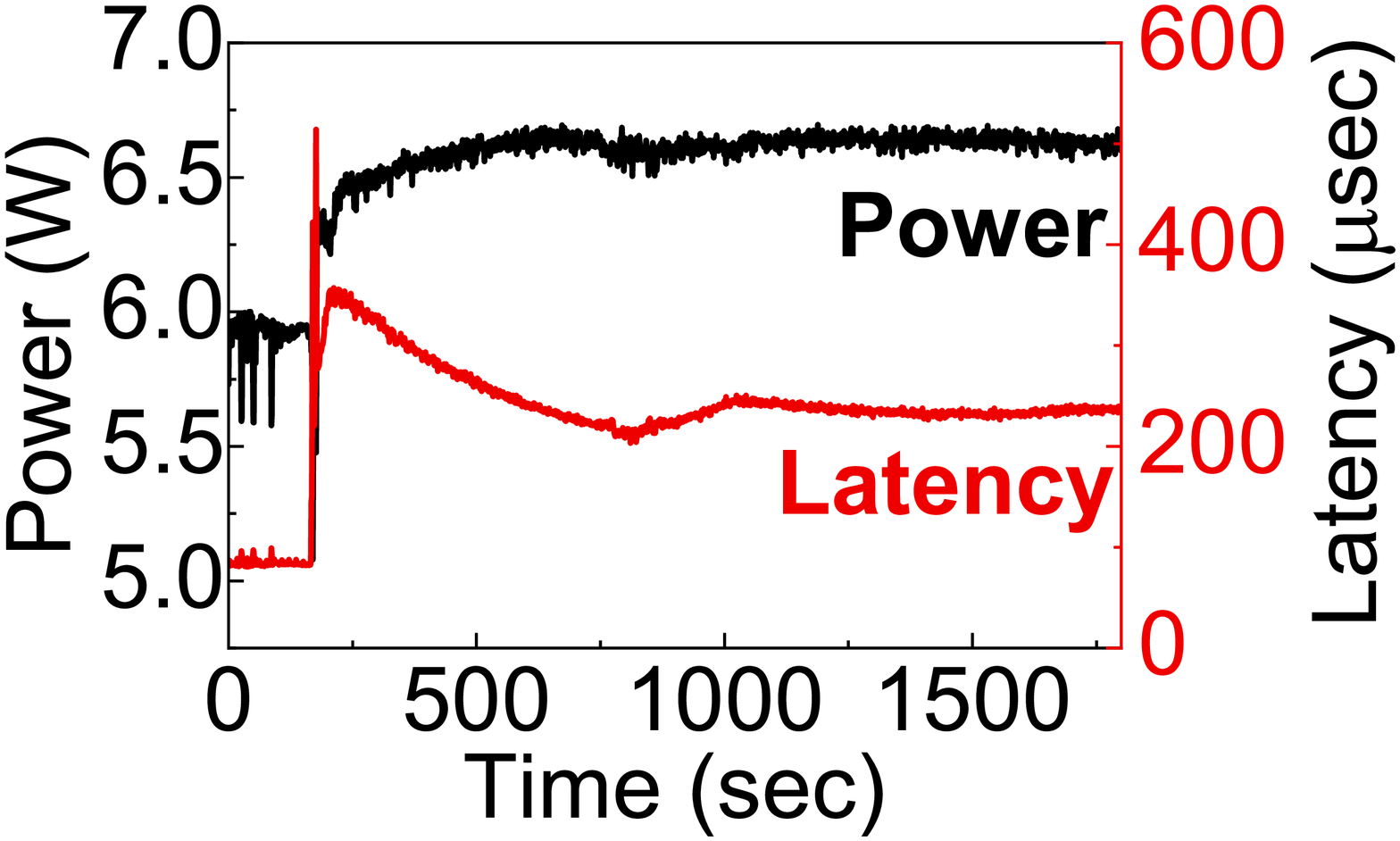}
		\caption{ULL SSD.}
		\label{fig:ull_gc_power}
	\end{subfigure}
	\vspace{-5pt}
	\caption{Power consumption during garbage collection.}
	\vspace{-10pt}
	\label{fig:gc_power}
\end{figure}

\subsubsection{Power and Garbage Collection}
Figure \ref{fig:overall_avg_power} analyzes power consumption characteristics of NVMe and ULL SSDs. ULL SSD consumes power less than NVMe SSD for asynchronous I/O operations by 30\%, on average. In particular, disparities of the power consumption trend between NVMe SSD and ULL SSD become more notable for the write services (compared with reads). We conjecture that this is because of SLC-like-Z-NAND; SLC in practice employs much fewer steps of writes (per request) on the target flash cell, compared to MLC \cite{choi2018invalid, suh19953, wu2012reducing, chang2015achieving}, which exhibits lower write latency. On the other hand, the power consumptions in idle states and sequential/random reads are similar to each other; 3.8W and 4.1W are consumed for the idles and read services, respectively. Flash reads require enabling only internal peripherals to sense out the target data (whose dynamic power is much lower than write power). This makes the power of internal DRAM buffers and controllers dominant for the I/O services rather than the power of backend flash.

One of the performance bottlenecks that most SSDs suffer from is garbage collections \cite{jung2013revisiting, jung2012taking, jung2019design, shahidi2016exploring, choi2018parallelizing}. Since flash does not allow to overwrite data without erasing a flash block, flash firmware \cite{lee2008last, lee2006fast, jung2012physically} forwards all incoming overwrite requests to a new space (i.e., block), which was erased in a previous stage, and remaps the target addresses, associated with the request, to the new block. If there is no available page in the remaining block to serve the overwrites, the firmware should reclaim pages by erasing a block. It then reads the corresponding data and writes them to the erased new block, which is called a \emph{garbage collection} (GC). Since it introduces multiple extra reads and writes, GCs exhibit long latency, which in turn can consume more power than a normal operation. 

Figure \ref{fig:qd1_gc} shows a time series analysis, which keep randomly writing 4KB data block after writing the entire address range of the underlying SSD. One can observe from this figure that the write latency of NVMe SSD sharply increases once GCs begin to reclaim flash blocks (107$\mu$s). The overall write latency with GCs is 6.3$\times$ higher than the average latency of writes (52$\mu$s). In contrast, ULL SSD exhibits a very sustained latency behavior with the same evaluation scenario that we tested for NVMe SSD (less than 1$\mu$s). We conjecture that there are three reasons behind ULL SSD does not suffer from GCs. First, as analyzed earlier, most of the flash reads and writes of ULL SSD are simply served faster than those of NVMe SSD. This flash-level performance superiority can shorten the long latency, imposed by GCs. Second, ULL SSD employs multiple super-channels, which can reclaim flash blocks with a higher parallelism by enabling more flash chips at a given time. Third, the suspend and resume operations (cf. Section \ref{section:suspend}) can make GCs interleaved with many incoming write requests, which makes the GC latency invisible to users at some extent. 
To be precise, we also analyze the power consumption behaviors of GCs. As shown in Figure \ref{fig:nvme_gc_power}, power consumption of NVMe SSD decrease when it starts GCs; we believe that this is because a few flash chips are involved in GCs per I/O request arrival (thereby low GC performance). In addition, as the data migration between the old flash block and new flash block makes multiple SSD resources busy (e.g., channels, die and planes), NVMe SSD cannot serve I/O requests while GC is in progress.
As shown in Figure \ref{fig:ull_gc_power} ULL SSD shows completely different story, compared with the power consumption trend of NVMe SSD. While the GC latency of NVMe SSD is as high as \emph{3ms}, that of ULL SSD is around 500$\mu$s, which is even faster than a write latency of conventional flash. Once a GC is invoked, ULL SSD consumes 12\% more power than a non-GC workload execution, on average. As described earlier, this is because ULL SSD enables many flash chips and perform GCs in parallel, so incoming requests targeting a specific flash chip can still be serviced by leveraging the suspend/resume operations.




\begin{figure}
	\centering
	\begin{minipage}{0.49\linewidth}
	\begin{subfigure}{0.49\linewidth}
		\includegraphics[width=\linewidth]{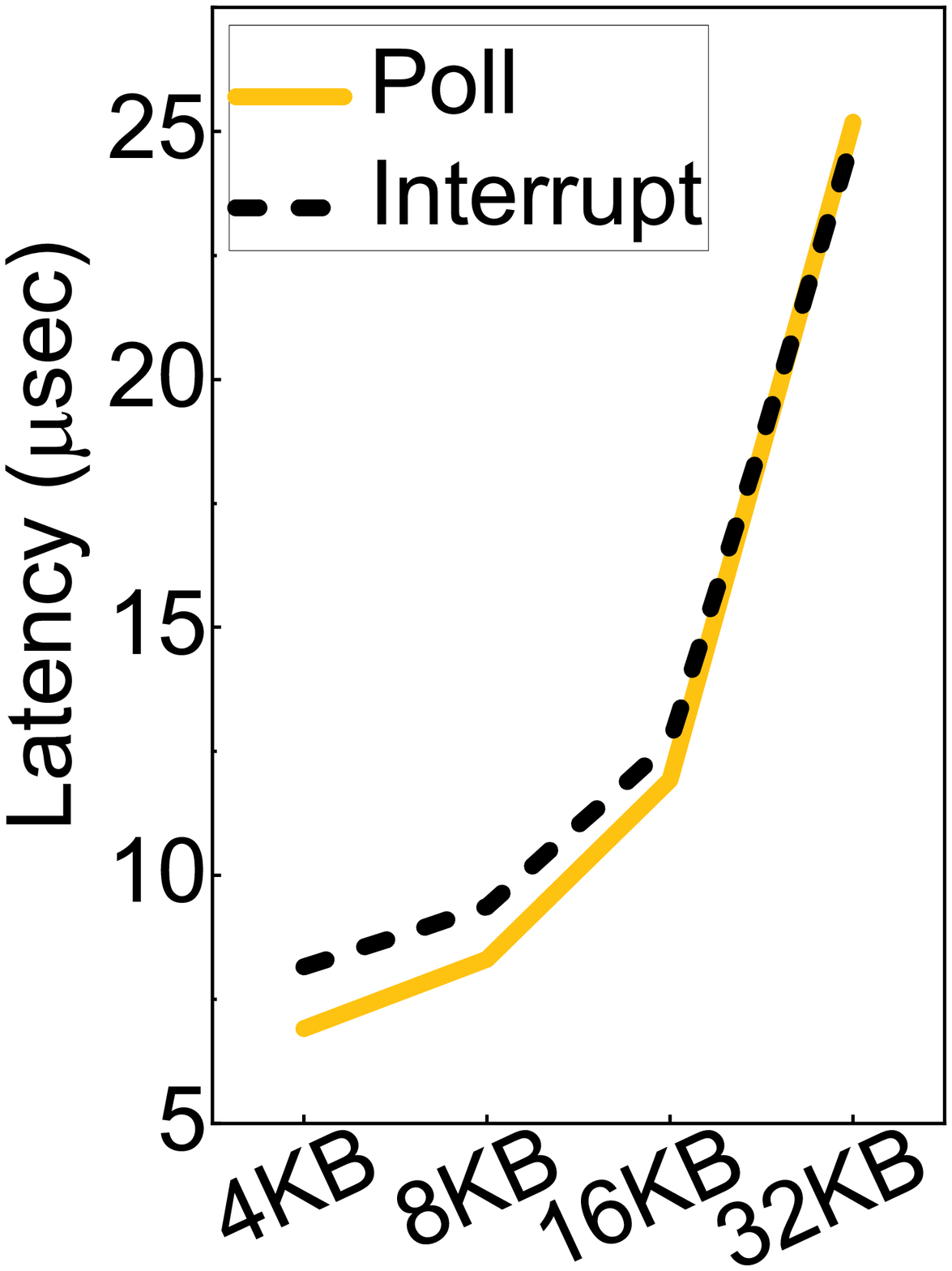}
		\caption{Seq. reads.\centering}
		\label{fig:seqrd_lat_nvme}
	\end{subfigure}
	\begin{subfigure}{0.49\linewidth}
		\includegraphics[width=\linewidth]{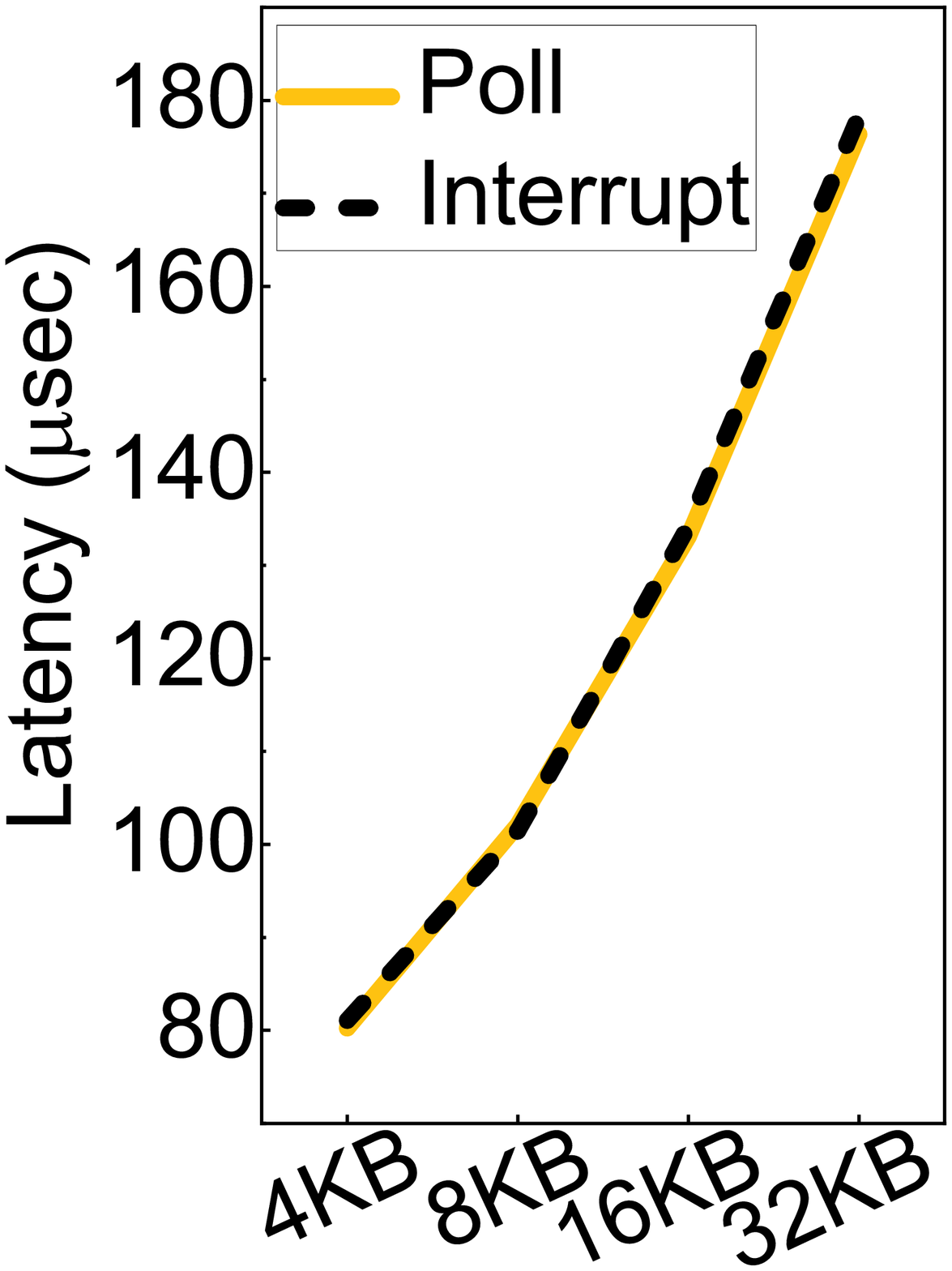}
		\caption{Rnd. reads.\centering}
		\label{fig:rndrd_lat_nvme}
	\end{subfigure}
	\end{minipage}
	\begin{minipage}{0.49\linewidth}
	\begin{subfigure}{0.49\linewidth}
		\includegraphics[width=\linewidth]{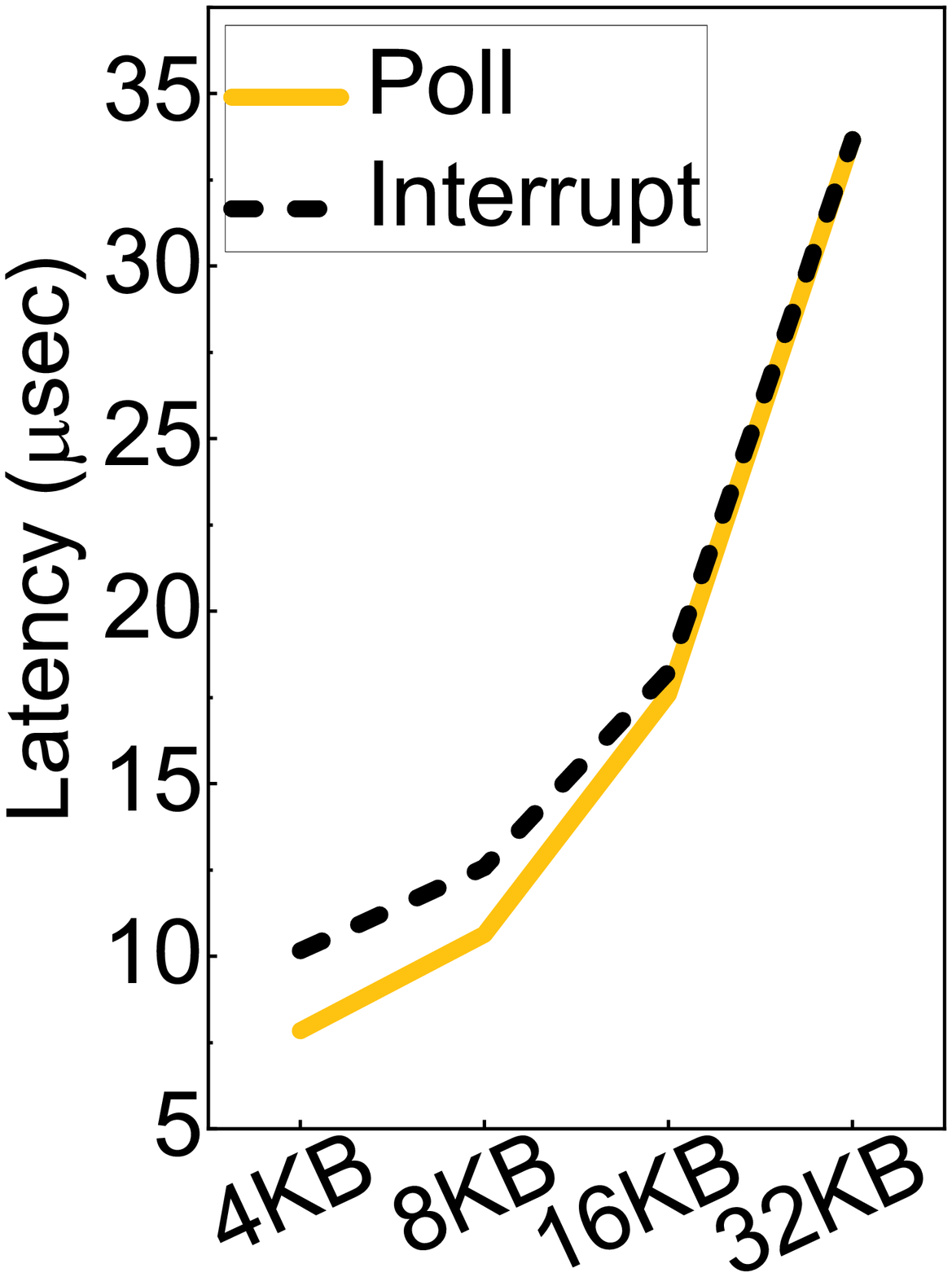}
		\caption{Seq. writes.\centering}
		\label{fig:seqwr_lat_nvme}
	\end{subfigure}
	\begin{subfigure}{0.49\linewidth}
		\includegraphics[width=\linewidth]{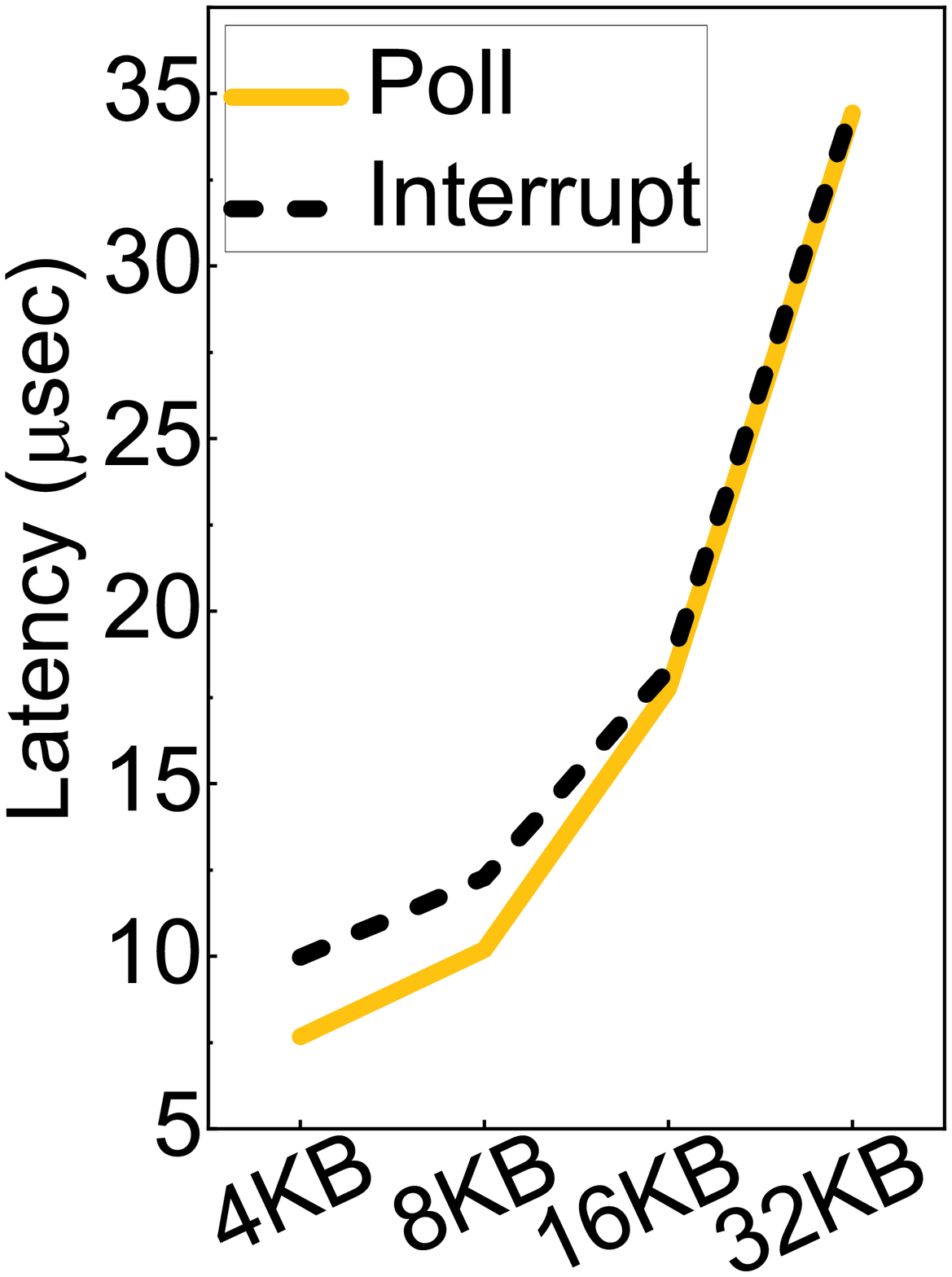}
		\caption{Rnd. writes.\centering}
		\label{fig:rndwr_lat_nvme}
	\end{subfigure}
	\end{minipage}
	\vspace{-5pt}
	\caption{Latency comparison (interrupt vs. poll) in NVMe SSD.}
	\vspace{-10pt}
	\label{fig:poll_avg_lat_nvme}
\end{figure}
\begin{figure}
	\centering
	\begin{minipage}{0.49\linewidth}
		\begin{subfigure}{0.49\linewidth}
		\includegraphics[width=\linewidth]{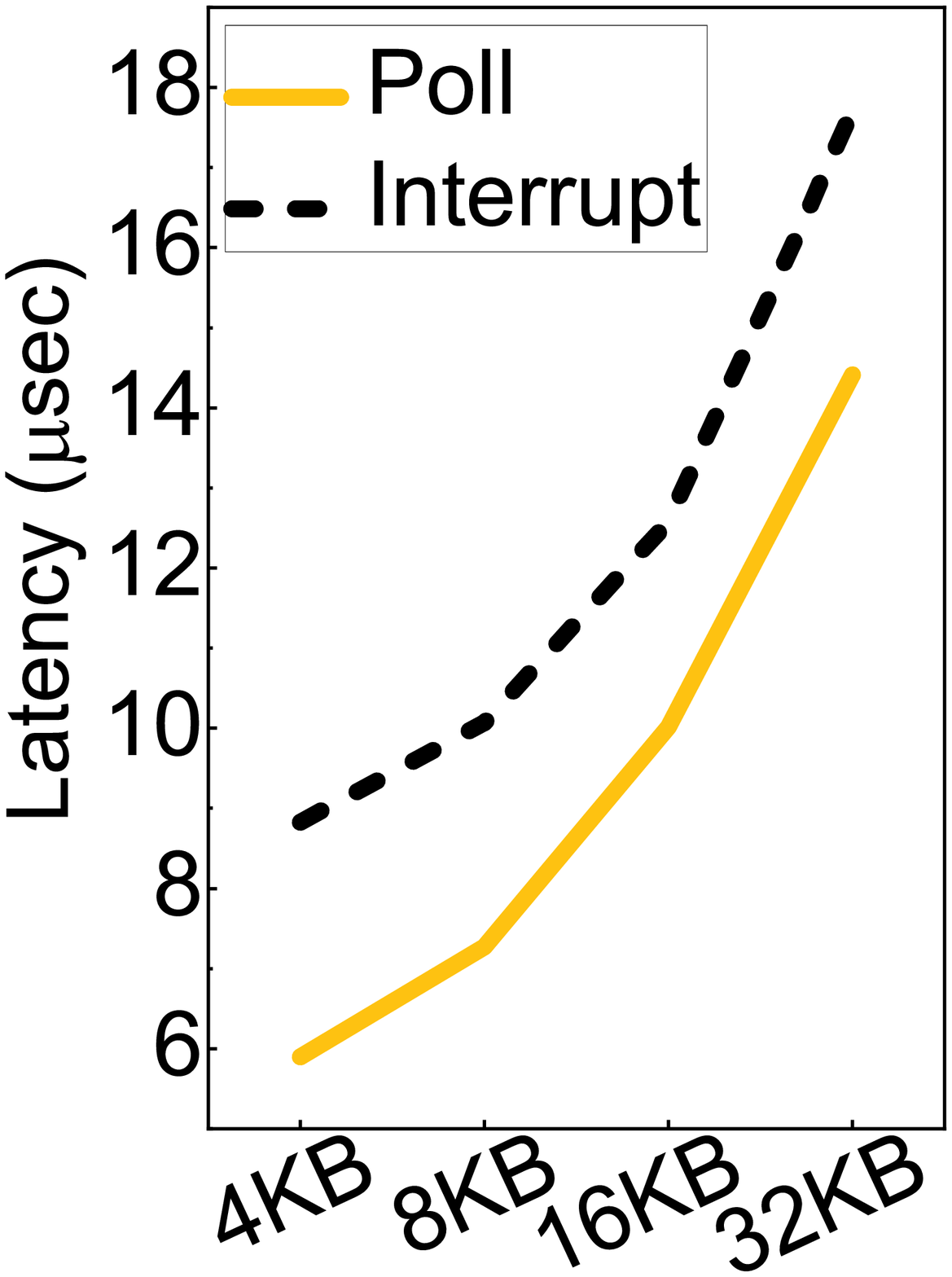}
		\caption{Seq. reads.\centering}
		\label{fig:seqrd_lat_ull}
	\end{subfigure}
	\begin{subfigure}{0.49\linewidth}
		\includegraphics[width=\linewidth]{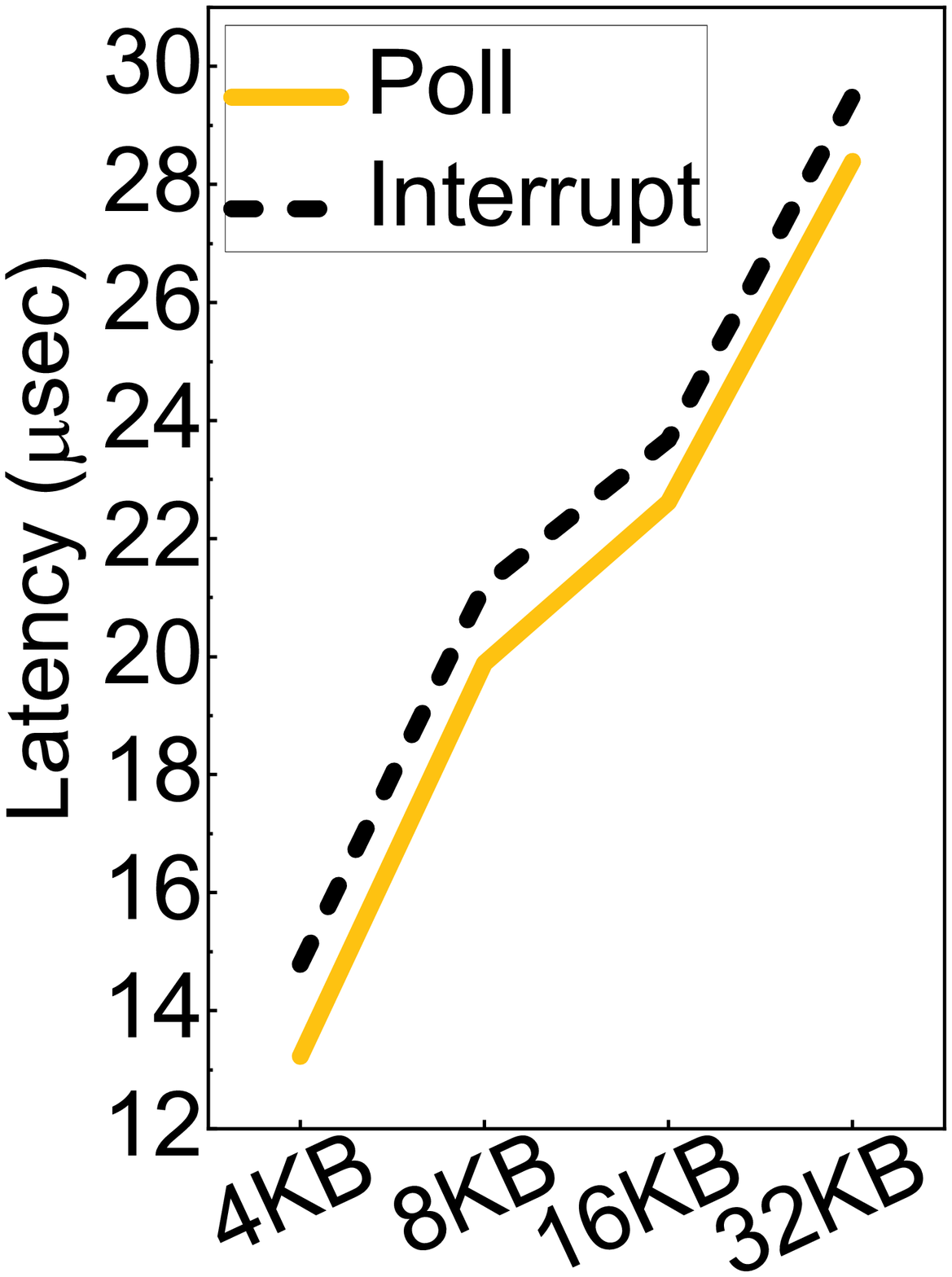}
		\caption{Rnd. reads.\centering}
		\label{fig:rndrd_lat_ull}
	\end{subfigure}
	\end{minipage}
	\begin{minipage}{0.49\linewidth}
	\begin{subfigure}{0.49\linewidth}
		\includegraphics[width=\linewidth]{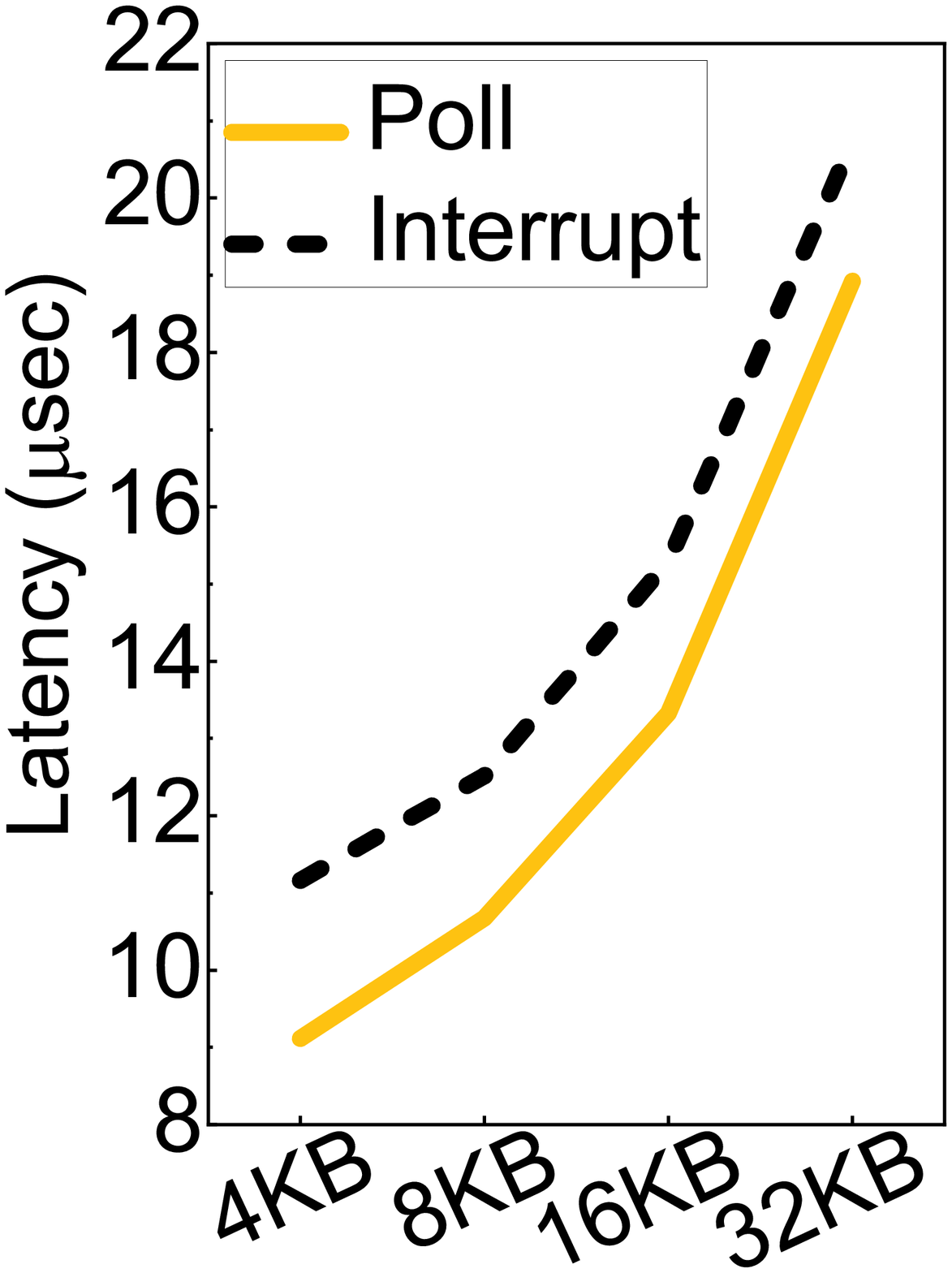}
		\caption{Seq. writes.\centering}
		\label{fig:seqwr_lat_ull}
	\end{subfigure}
	\begin{subfigure}{0.49\linewidth}
		\includegraphics[width=\linewidth]{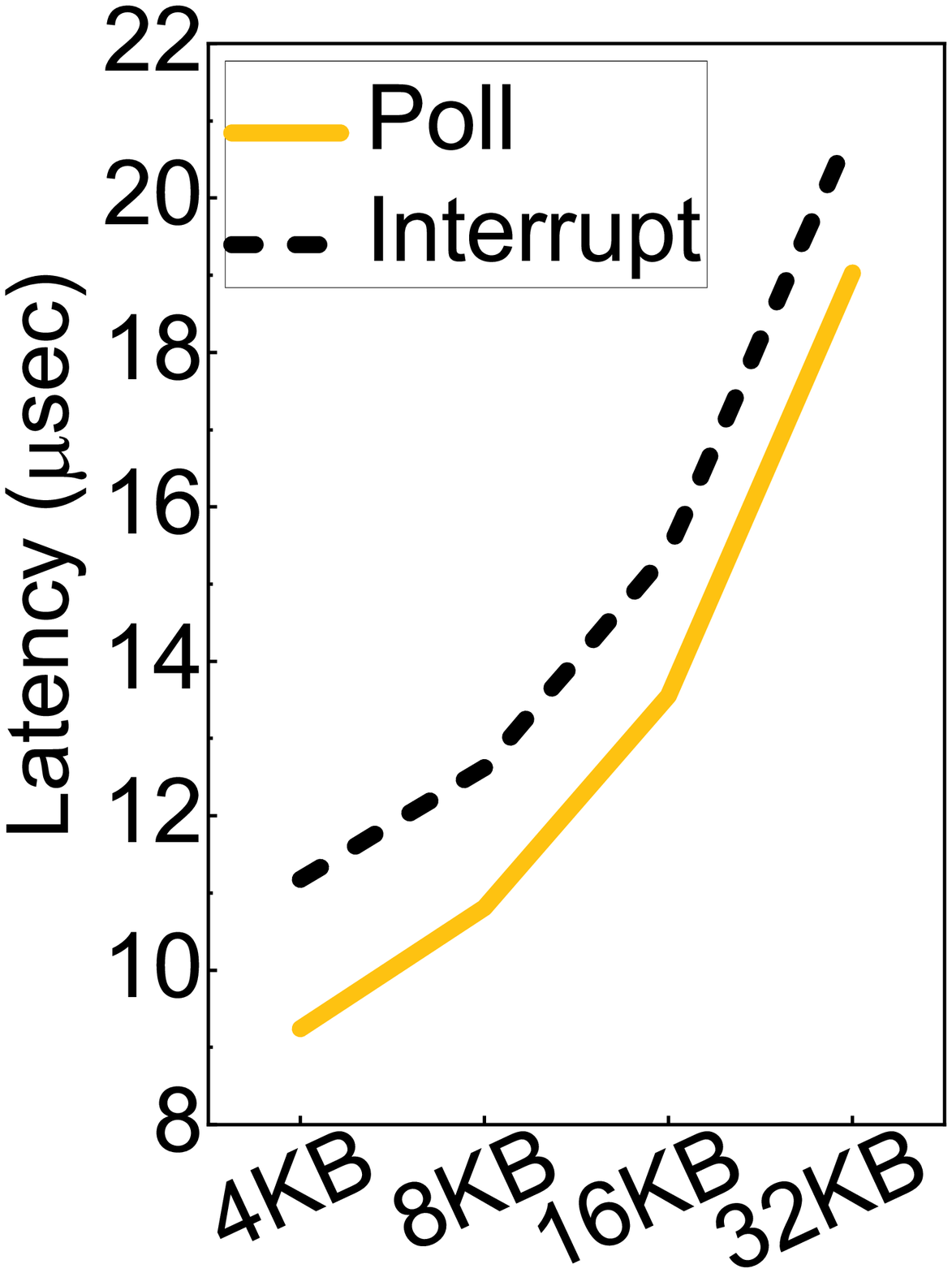}
		\caption{Rnd. writes.\centering}
		\label{fig:rndwr_lat_ull}
	\end{subfigure}
	\end{minipage}
	\vspace{-5pt}
	\caption{Average latency of interrupt and polling based I/O in ULL SSD.}
	\vspace{-10pt}
	\label{fig:poll_avg_lat_ull}
\end{figure}

\section{I/O Completion Methods and Challenges}
\label{sec:evaluation}

\subsection{Is the polled-mode I/O completion faster than interrupts?}
\subsubsection{Overall Latency Comparison}
Figure \ref{fig:poll_avg_lat_nvme} shows the latency difference between polling-based and interrupt-based I/O services with NVMe SSD. While polling gets attention from both industry and academia as a promising solution to expose the true latency of PCIe based SSDs to user-level applications, one can observe from the figure that polling unfortunately has no performance impact with the modern high-end SSD technology. Specifically, the latency difference of reads and writes brought by the interrupt-based and polling-based I/O services is less than 2.2\% and 11.2\%, on average, respectively. In contrast, as shown in Figure \ref{fig:poll_avg_lat_ull}, polling improves the performance compared to the conventional interrupt-based I/O completion when ULL SSD is applied to the current NVMe stack. The average read and write latency of ULL SSD with polling are respectively 9.6$\mu$s and 9.2$\mu$s (for 4KB-sized requests), while those with interrupts are 11.8$\mu$s and 11.2$\mu$s, respectively. Even though polling shortens the read and write latency by only 16.3\% and 13.5\% on average, respectively, we believe that the polling mechanism in the NVMe stack can bring an advantage for latency sensitive applications. The benefits can be more notable with future SSDs that employ faster NVM technologies such as resistive random access memory (ReRAM) \cite{zhang2017novel, zhang2016leader, chi2016prime, ji2019fpsa}. To appropriately exploit the polling method in the future systems, we believe that there exists several system-level challenges that the conventional NVMe storage stack should address; we will analyze the challenges shortly. 

\begin{figure}
	\centering
	\begin{minipage}{0.48\linewidth}
	\begin{subfigure}{0.49\linewidth}
		\includegraphics[width=\linewidth]{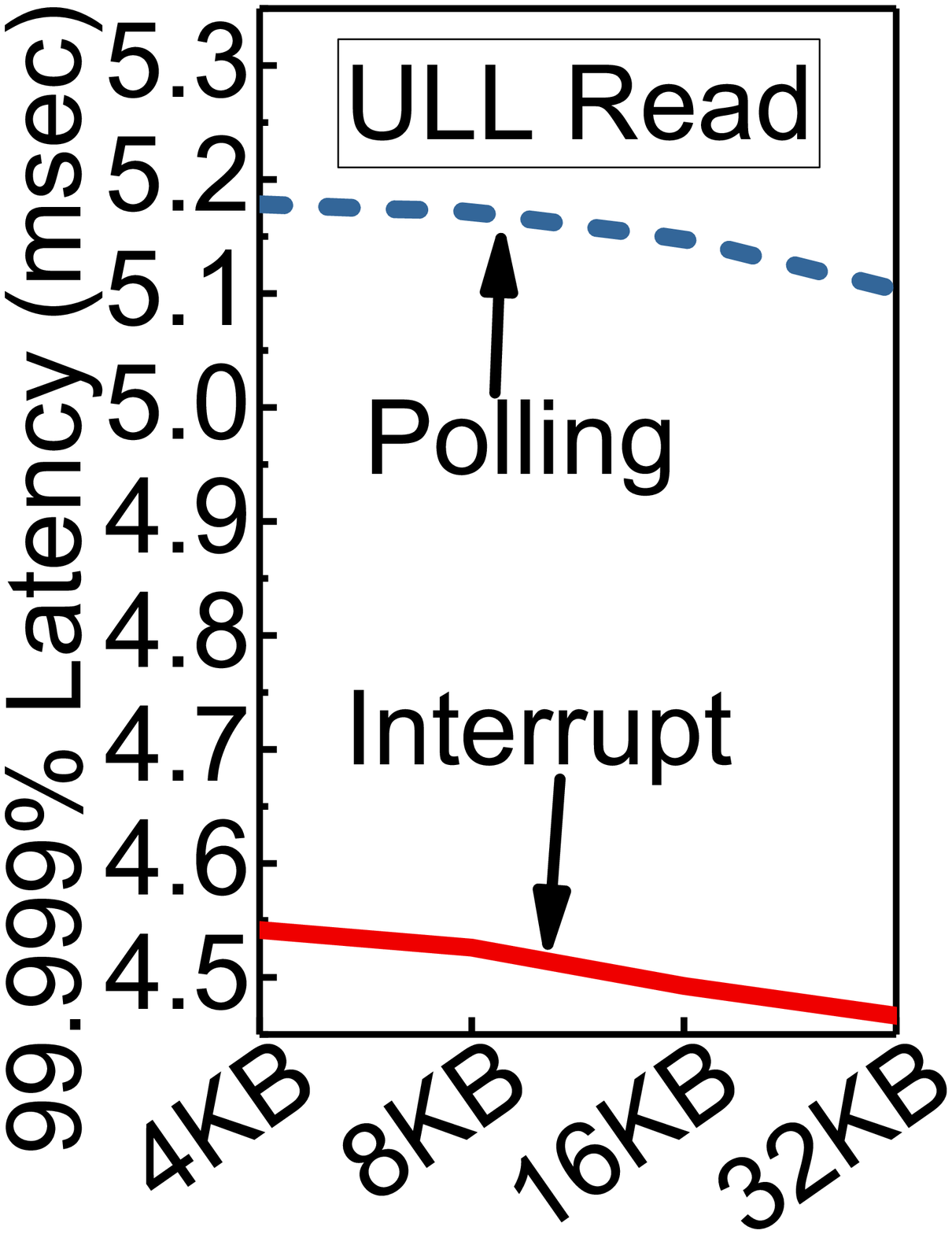}
		\caption{Reads.}
		\label{fig:poll_read_lat_99999}
	\end{subfigure}
	\begin{subfigure}{0.49\linewidth}
		\includegraphics[width=\linewidth]{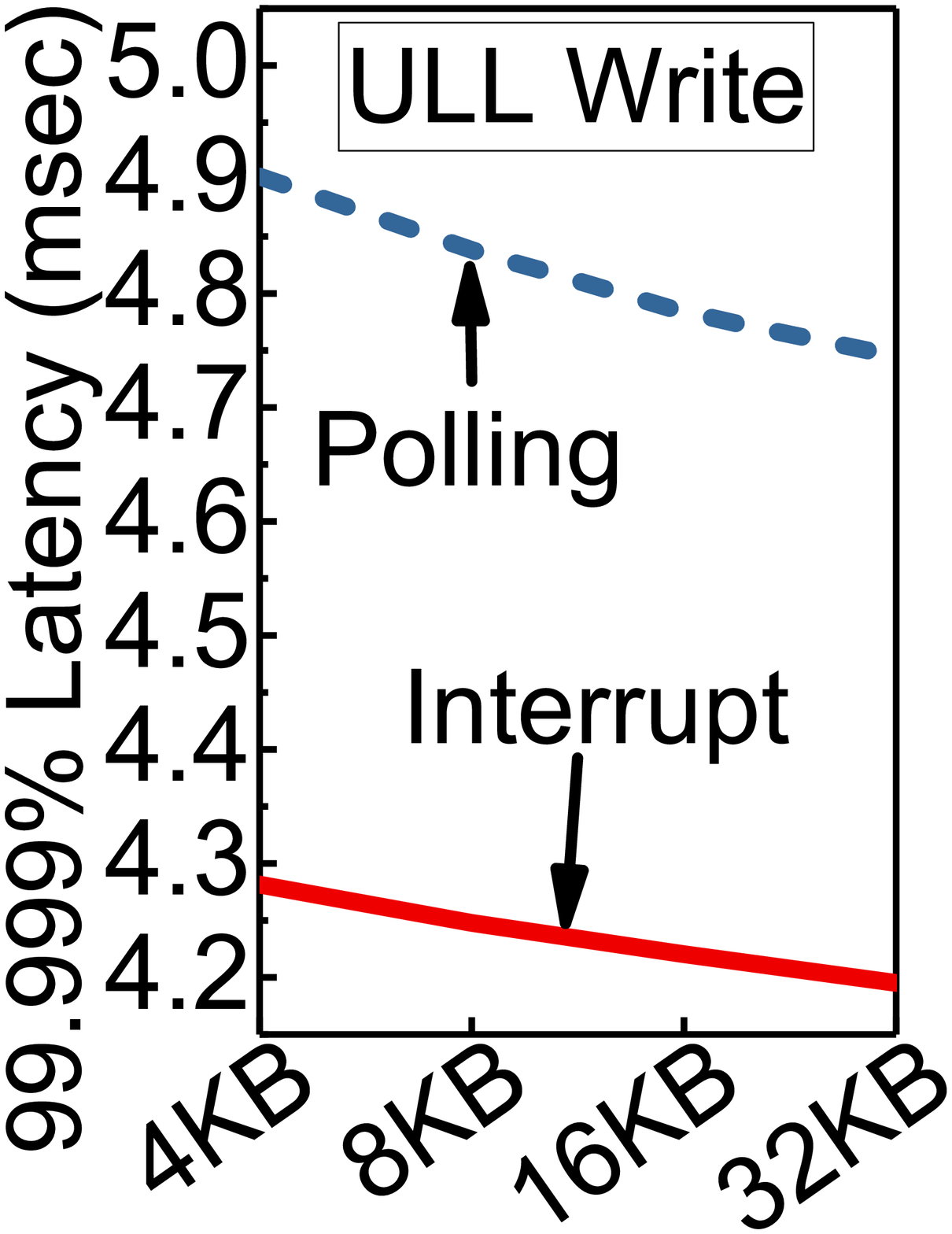}
		\caption{Writes.}
		\label{fig:poll_write_lat_99999}
	\end{subfigure}
	\vspace{-5pt}
	\caption{99.999th latency of ULL SSD.}
	\end{minipage}
	~
	\begin{minipage}{0.48\linewidth}
	\begin{subfigure}{\linewidth}
		\includegraphics[width=\linewidth]{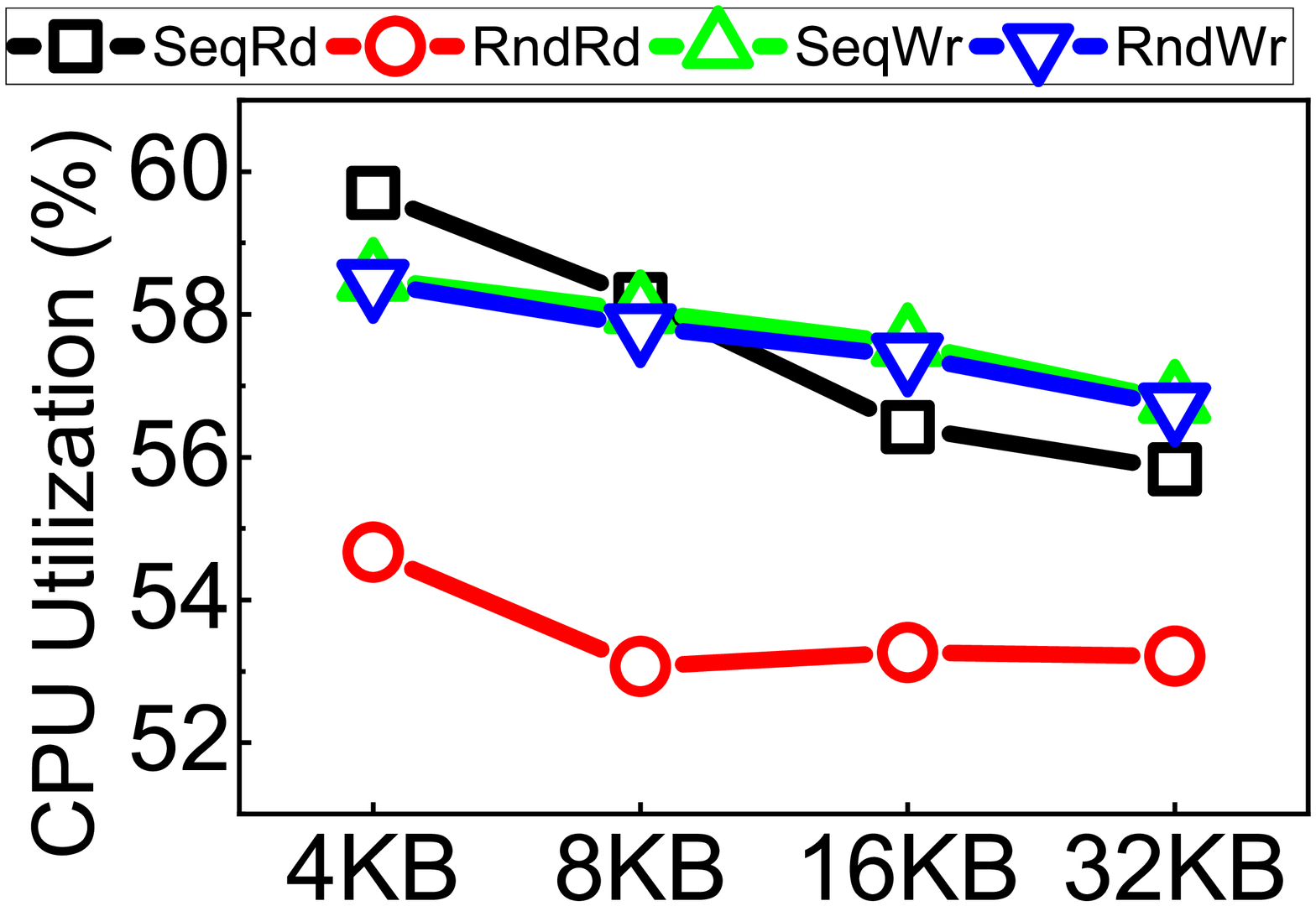}
		\label{fig:hybrid_cpu}
	\end{subfigure}
	\vspace{-5pt}
	\caption{CPU utilization of hybrid polling.}
	\label{fig:hybrid_cpu}
	\end{minipage}
	\vspace{-10pt}
\end{figure}
\begin{figure}
	\centering
	\begin{subfigure}{0.24\linewidth}
		\includegraphics[width=\linewidth]{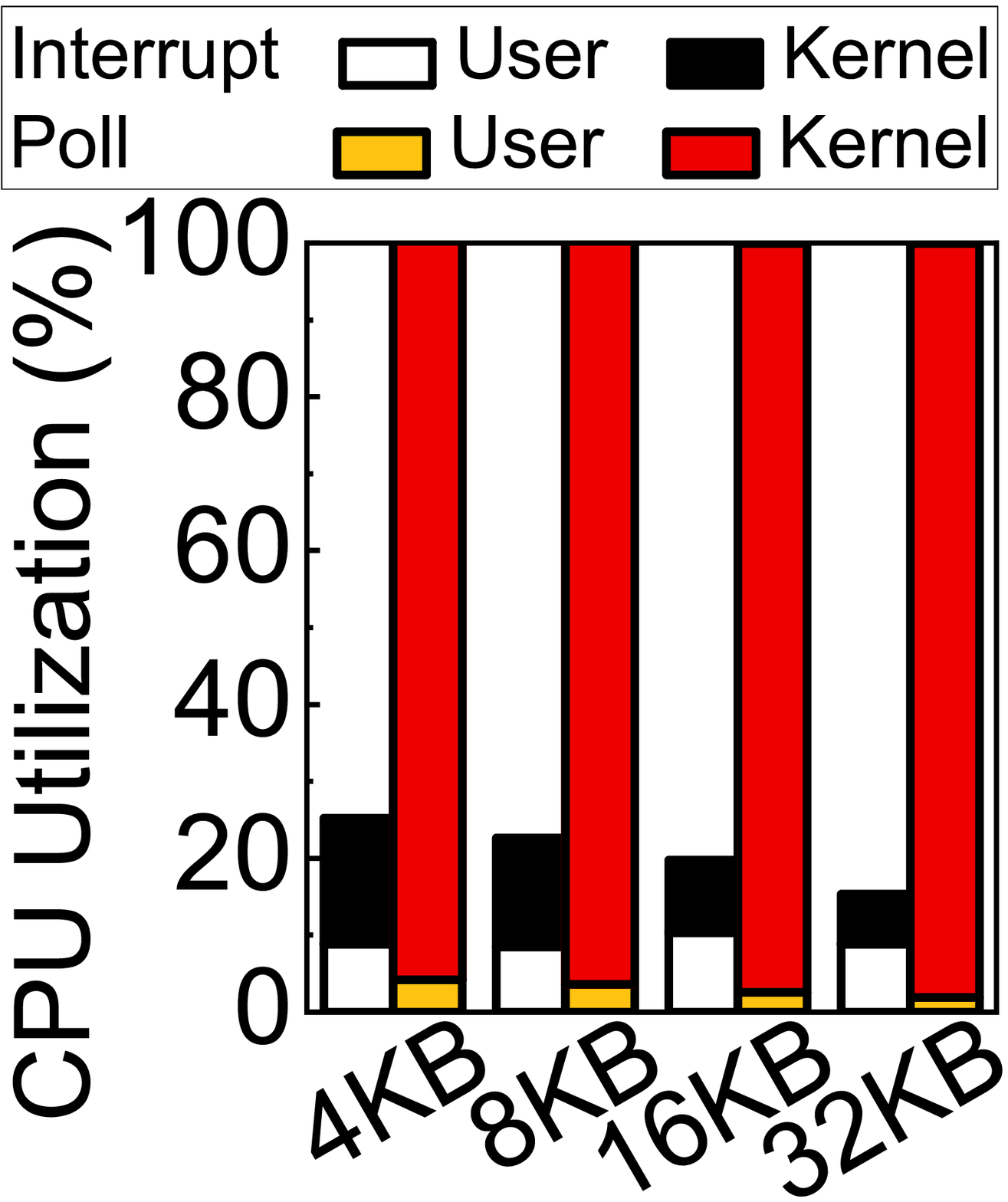}
		\caption{Seq. reads.\centering}
		\label{fig:seqrd_cpu}
	\end{subfigure}
	\begin{subfigure}{0.24\linewidth}
		\includegraphics[width=\linewidth]{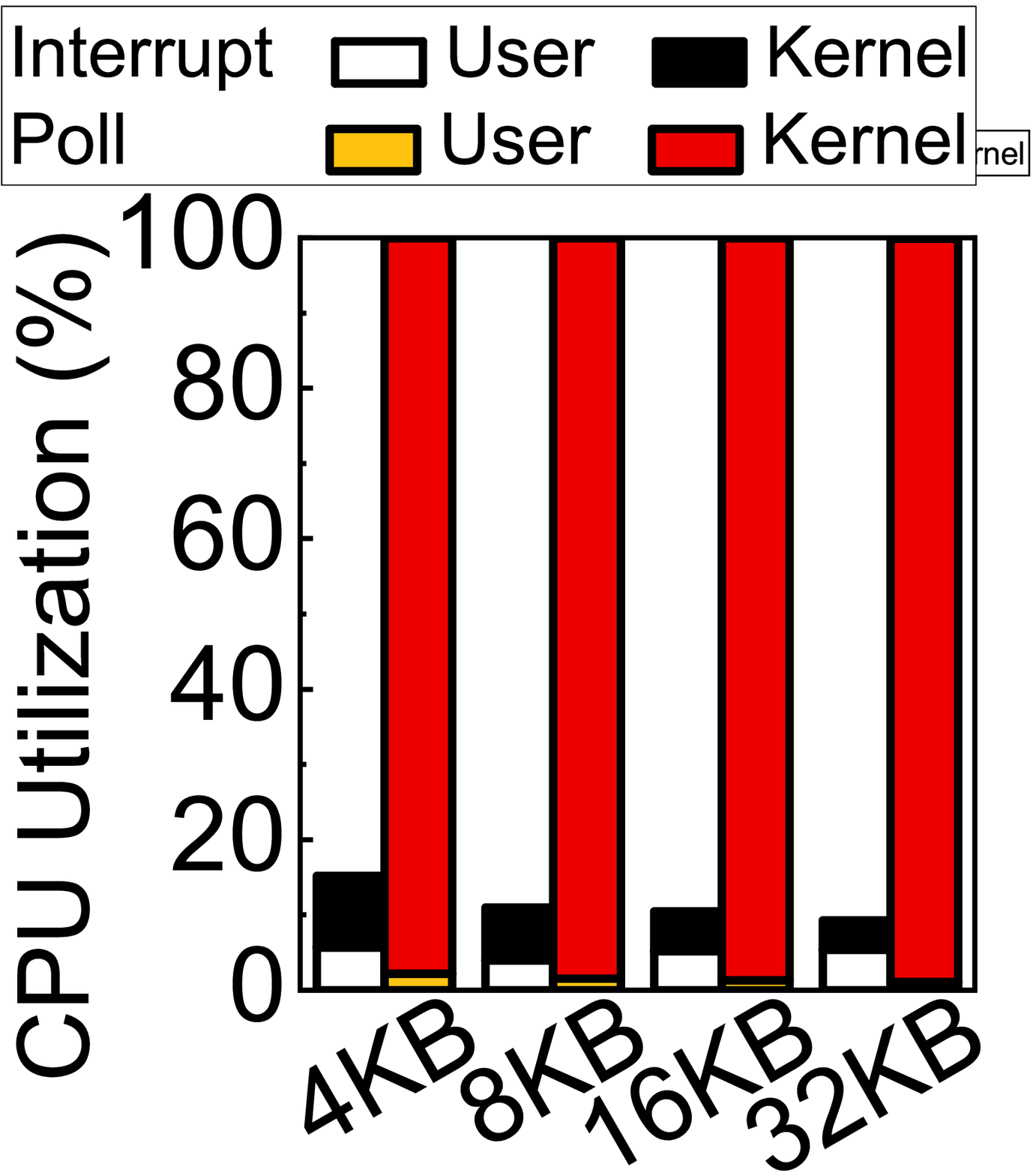}
		\caption{Rnd. reads.\centering}
		\label{fig:rndrd_cpu}
	\end{subfigure}
	\begin{subfigure}{0.24\linewidth}
		\includegraphics[width=\linewidth]{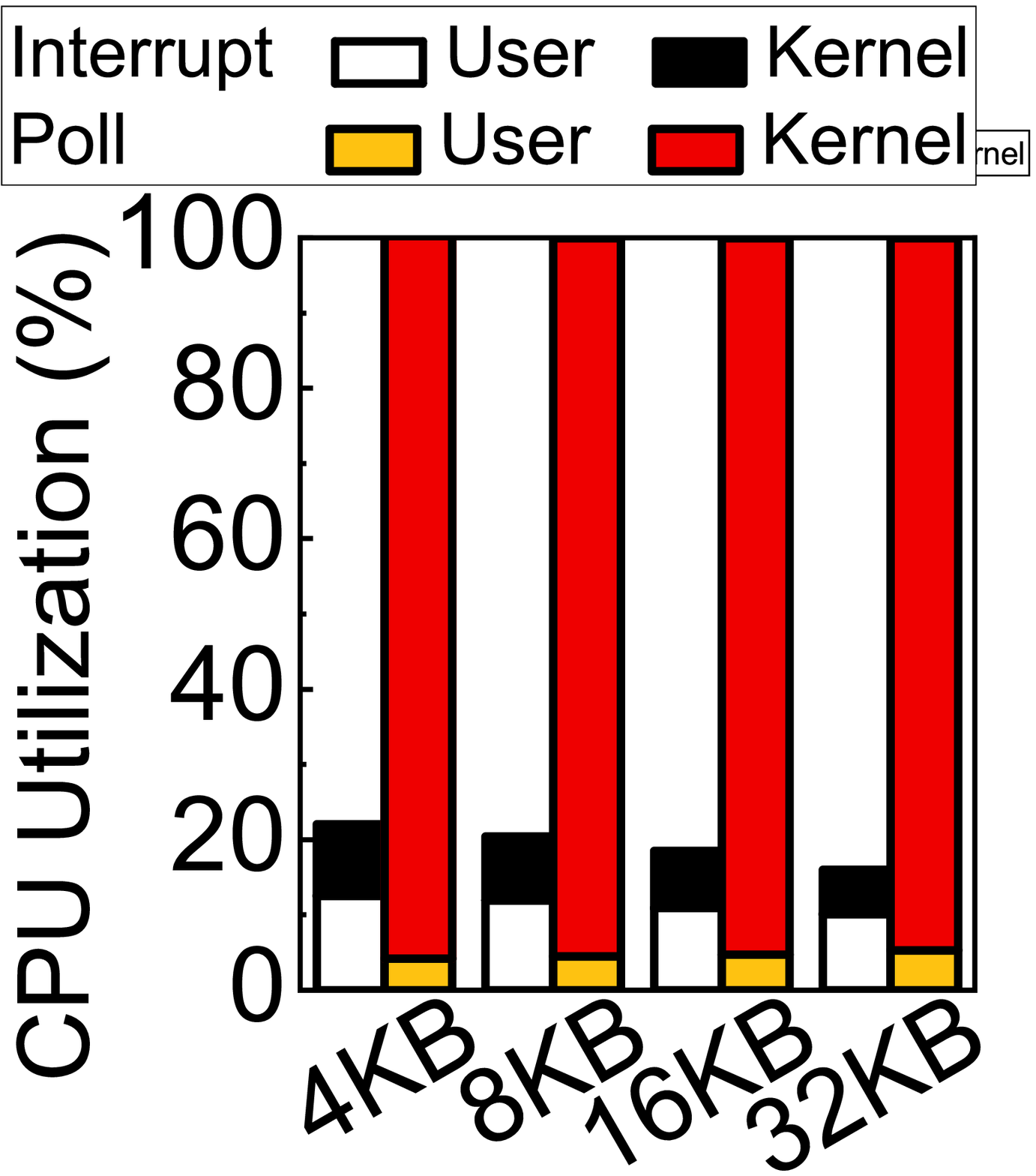}
		\caption{Seq. writes.\centering}
		\label{fig:seqwr_cpu}
	\end{subfigure}
	\begin{subfigure}{0.24\linewidth}
		\includegraphics[width=\linewidth]{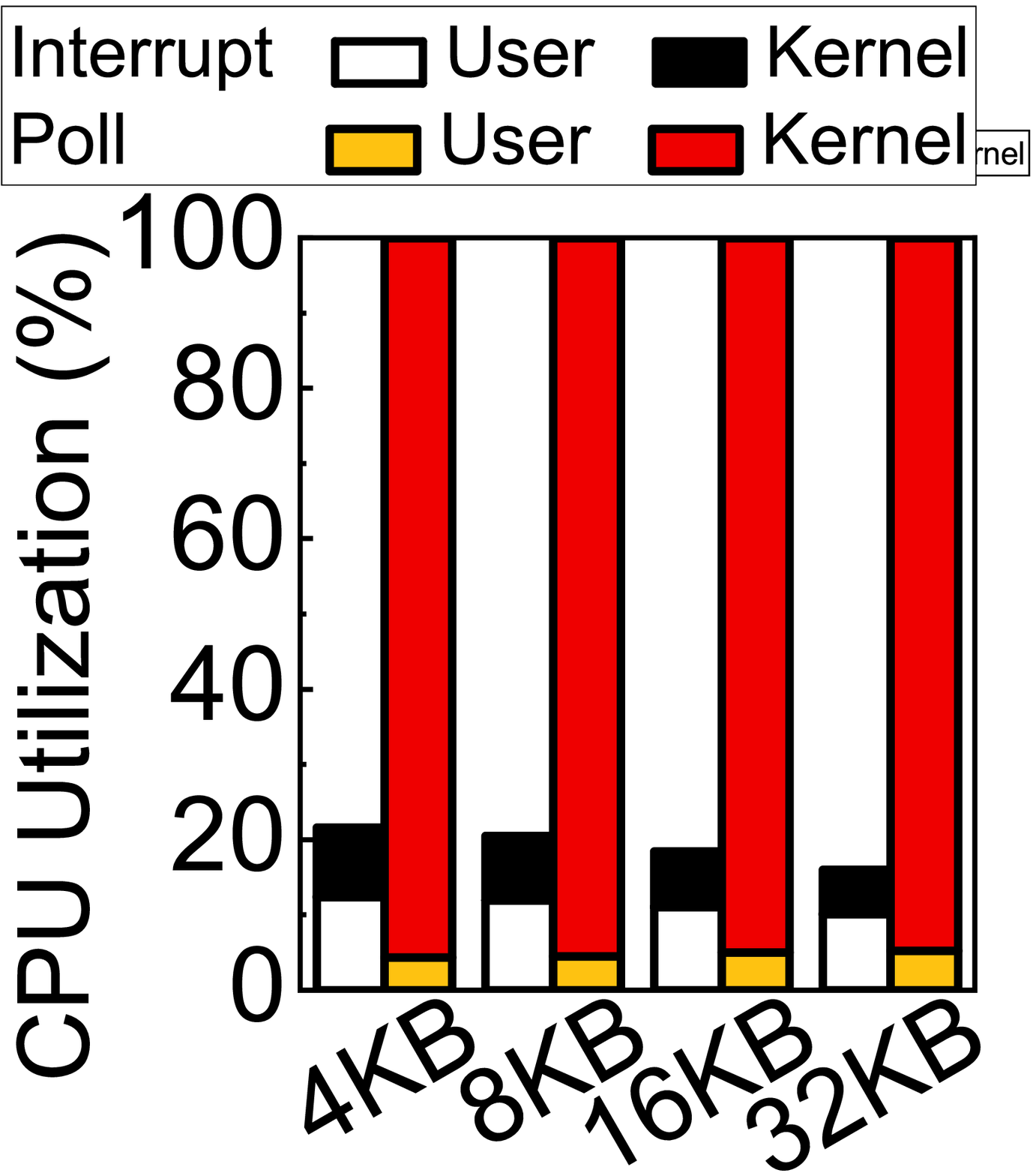}
		\caption{Rnd. writes.\centering}
		\label{fig:rndwr_cpu}
	\end{subfigure}
	\vspace{-5pt}
	\caption{CPU utilization of interrupt/polling based I/O.}
	\vspace{-10pt}
	\label{fig:poll_cpu}
\end{figure}


\subsubsection{Five-Nines Latency} Figures \ref{fig:poll_read_lat_99999} and \ref{fig:poll_write_lat_99999} analyze five-nines latency for reads and writes with two different I/O completion methods, respectively. In contrast to our previous observations, the long tail latency of polling for reads and writes is worse than those of interrupts by 12.5\% and 11.4\%, on average, respectively. This should be addressed in the future NVMe storage stack, as well. The key functions of polling (cf. \texttt{nvme\_poll} and \texttt{blk\_poll}) require acquiring spin locks when they process CQ(s), which should be iterated until the target request is completed. The polling may not release and relax CPU as spin locks are used for polling NVMe queues. Thus, other incoming requests will be pending as polling does not allow a context switch while its I/O request is outstanding. CPU scheduling should be revised to address the aforementioned shortcomings, exposed by the current polled-mode operation \cite{pollpatch}.

\subsection{What are the practical overheads when systems employ the polled-mode?}
\subsubsection{CPU Utilization Analysis}
\label{pollcpu}
Figure \ref{fig:poll_cpu} compares the CPU utilization of two systems, each employing the interrupt-based and polling-based I/O services. In this evaluation, we measure the CPU usages by reading/writing 4KB-sized data from/to ULL SSD with those two different I/O completion methods. As shown in the figure, the interrupt-based I/O services only take 8.4\% and 9.2\% of the total CPU cycles for kernel mode and user mode, on average, respectively. However, the polling shows a completely different trend on the CPU utilization analysis, compared to that on interrupts. While the CPU cycles consumed by polling at the user-level are similar to those of interrupts, such CPU cycles (of polling) at the kernel-level accounts for 96.4\% of the entire application execution, on average. This is because the NVMe driver does not release the CPU and keeps iterating the completion procedure. Specifically, it looks up NVMe's completion queue (CQ) head/tail pointers and checks all phase tags within the CQ's entries when the driver polls the I/O completion.

\begin{figure}
	\centering
	\begin{minipage}{0.47\linewidth}
	\begin{subfigure}{0.49\linewidth}
		\includegraphics[width=\linewidth]{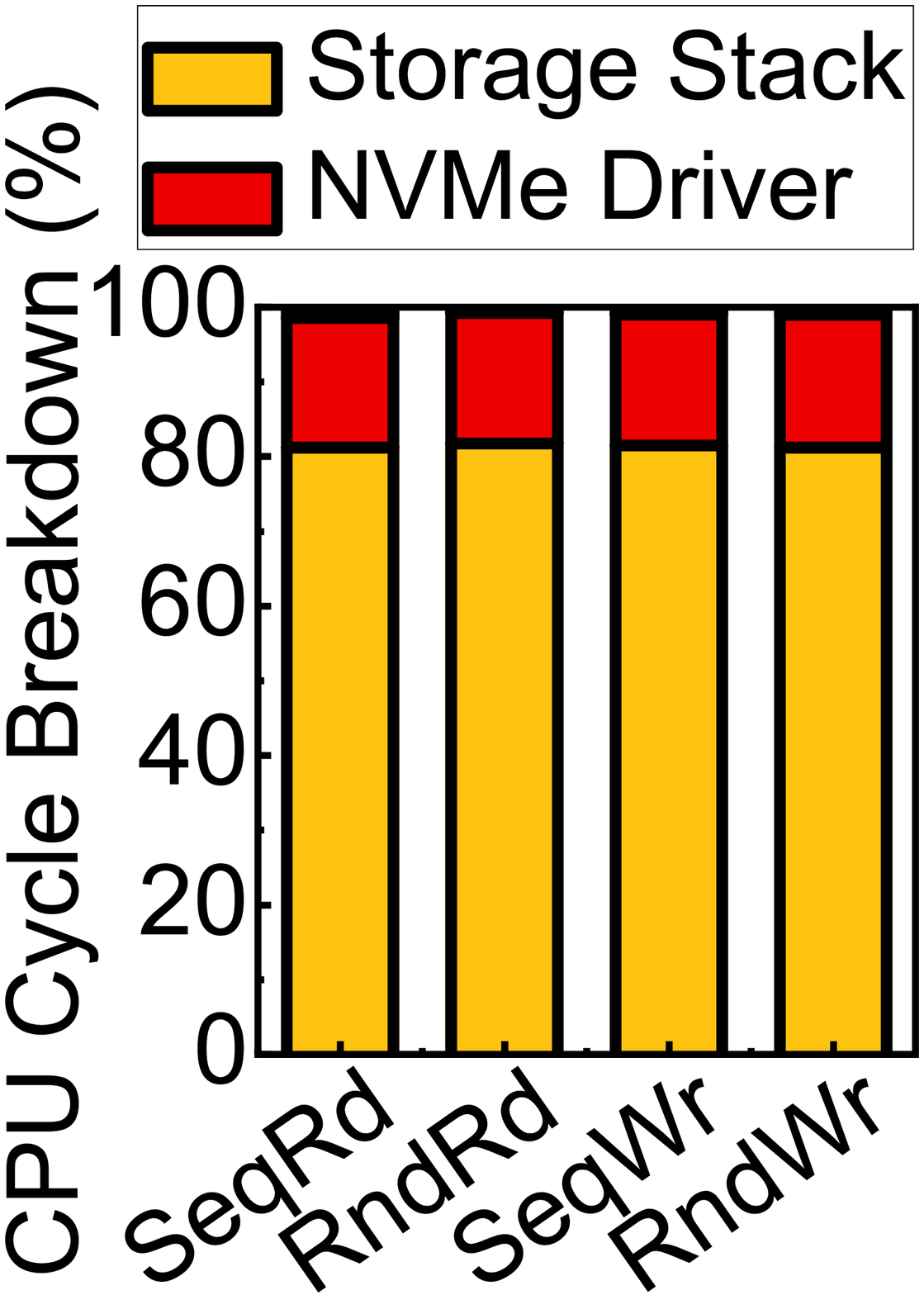}
		\caption{Module.}
		\label{fig:module_brkdown}
	\end{subfigure}
	\begin{subfigure}{0.49\linewidth}
		\includegraphics[width=\linewidth]{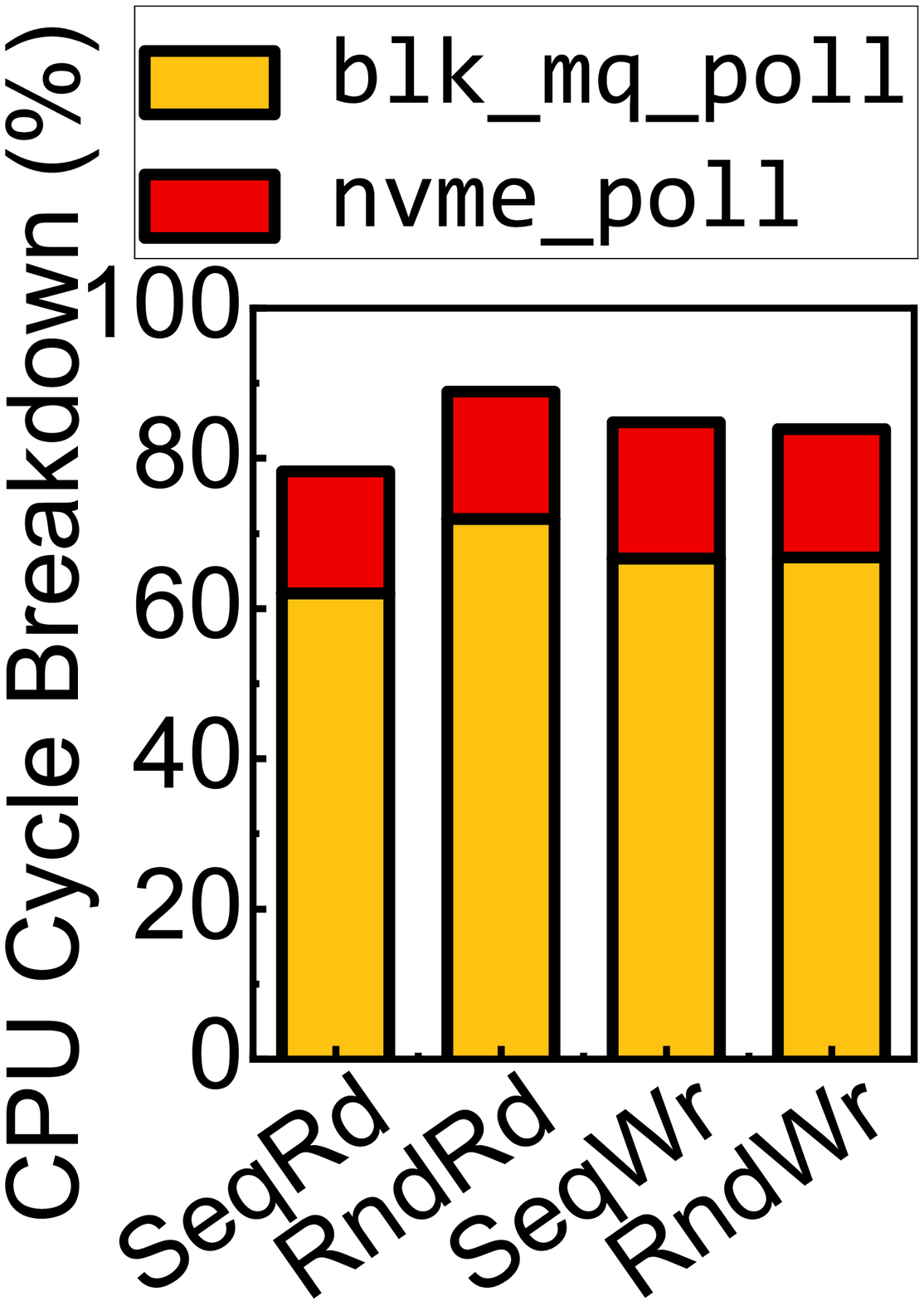}
		\caption{Function.}
		\label{fig:func_brkdown}
	\end{subfigure}
	\label{fig:cpu_brkdown}
	\vspace{-5pt}
	\caption{CPU utilization breakdown.}
	\end{minipage}
	~
	\begin{minipage}{0.47\linewidth}
	\begin{subfigure}{0.49\linewidth}
		\includegraphics[width=\linewidth]{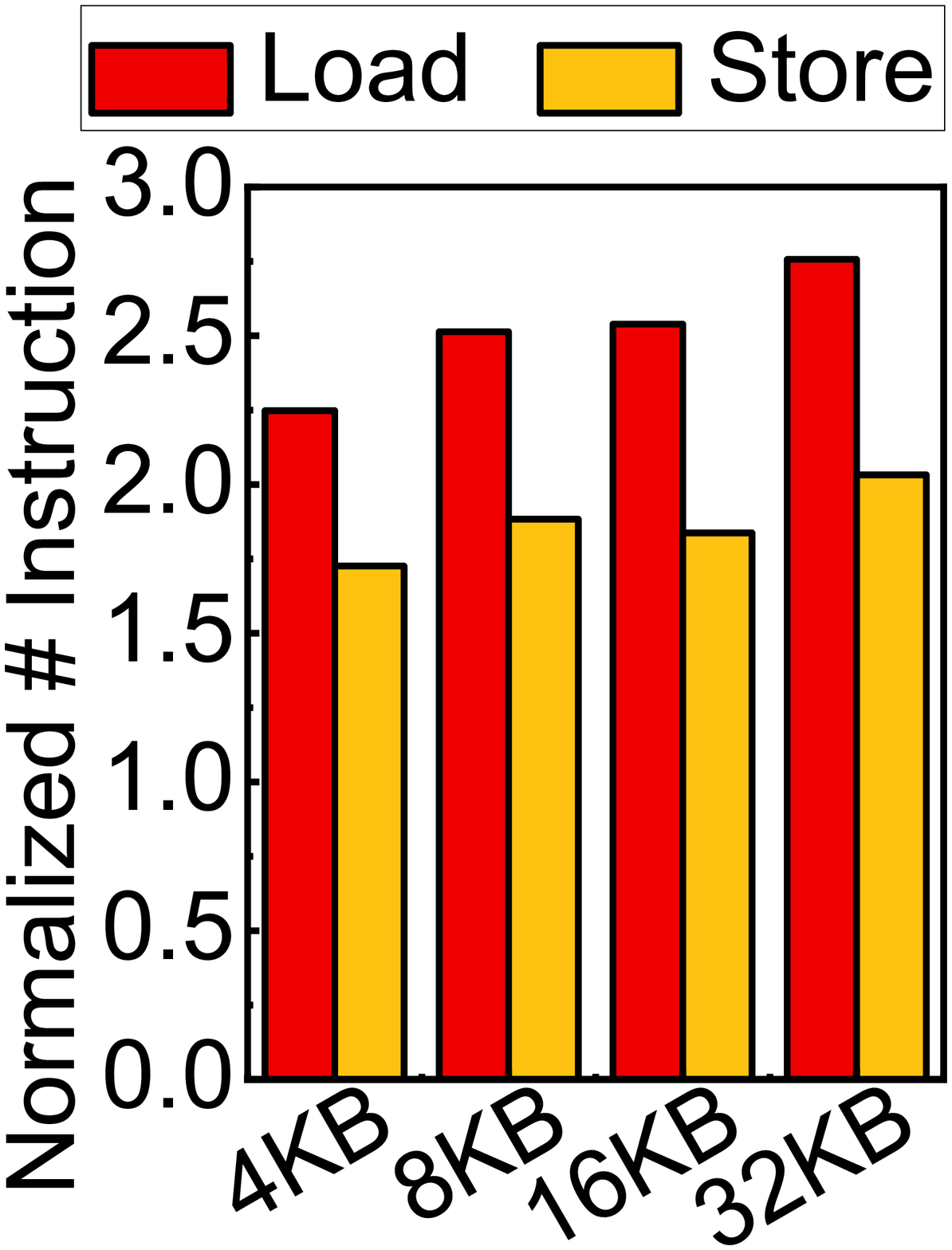}
		\caption{Reads.\centering}
		\label{fig:read_loadstore}
	\end{subfigure}
	\begin{subfigure}{0.49\linewidth}
		\includegraphics[width=\linewidth]{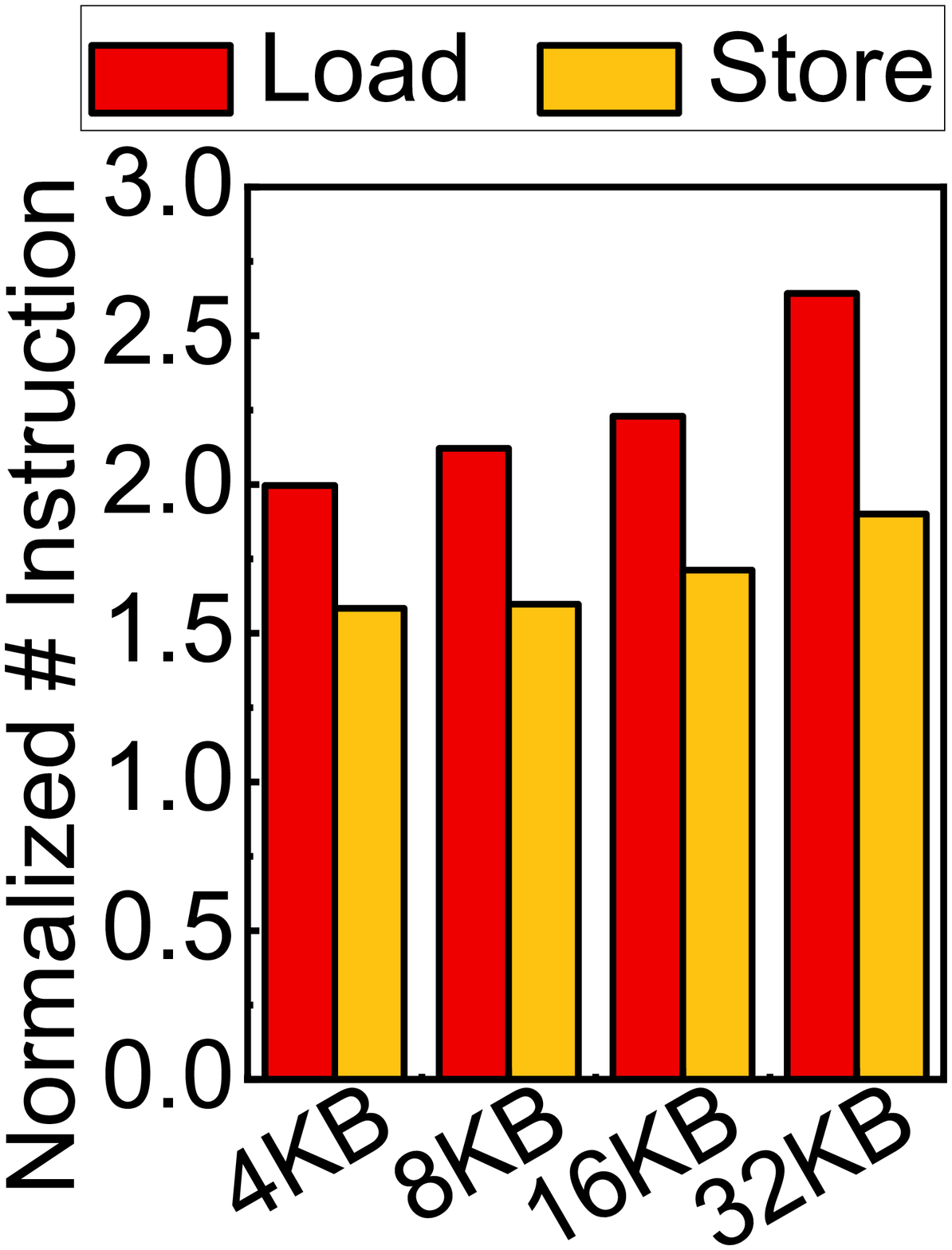}
		\caption{Writes.\centering}
		\label{fig:write_loadstore}
	\end{subfigure}
	\vspace{-5pt}
	\caption{Normalized memory instruction count of poll.}
	\label{fig:poll_loadstore}	
	\end{minipage}
	\vspace{-20pt}
\end{figure}

Figure \ref{fig:module_brkdown} decomposes the kernel-side CPU cycles consumed by a single NVMe driver and other modules (such as \texttt{blk-mq}). We found out that the NVMe driver uses only 17.5\% of the total CPU executions for all I/O access patterns. Then, we decomposes CPU utilization into function-level to study how the NVMe driver and I/O related modules spend the CPU cylces cycles. Figure \ref{fig:func_brkdown} shows CPU utilization breakdown results, obtained by VTune amplifier performance profiler; the CPU cycles consumed by different stack functions are normalized to the total CPU cycles used for the I/O services. It shows two major functions: i) \texttt{blk\_mq\_poll()} and ii) \texttt{nvme\_poll()}. These two functions are respectively associated to \texttt{blk\_mq} of the storage stack and the NVMe driver. Surprisingly, only these two functions consume 84\% of the total CPU cycles among all kernel modules in the storage stack. 
\texttt{blk\_mq\_poll()} takes around 67\% CPU cycles to check if the current thread needs to be rescheduled and whether a pending request exists, coming from other kernel modules or processes. 
\texttt{nvme\_poll()}, the callee of \texttt{blk\_mq\_poll()}, simply loads the target CQ entries and checks up if the corresponding I/O requests finish. Since the target entries should be written by the underlying SSD (over DMA), it is necessary to keep loading the CQ entries, repeatedly. This simple load and check up procedure takes 17\% of entire CPU cycles, on average, and this accounts for 97\% of cycles spent by NVMe module. We believe that, even though polling can shorten the device-level latency, allocating an entire core to refer the I/O completions can hurt the overall system performance as it prevents several computational tasks from running on the host system in a timely manner. 


\subsubsection{Memory Requirements} 
Figure \ref{fig:poll_loadstore} normalizes the number of memory instructions (load/store) for the polled-mode system executions to that of a conventional interrupt-driven system.
As polling requires checking up all target CQ entries, it exhibits 137\% more load instructions (to compare and check the I/O completion) than what the interrupt-driven system introduces. Note that the target CQ entries are managed by the underlying SSD devices, and thus, loads should go through all CPU caches and bring them back to the caches again; the caches always hold stale data for the CQ entries' information.
Thus, actual execution of the load instructions is significantly inefficient compared to the instructions which are usually performed to manage conventional memory subsystems. Similarly, polling exhibits 78\% more store instructions to serve I/O operations than interrupts, on average. Even though polling spends most times to examine the CQ entries' states, polling also keeps metadata, indicating how many loop iterations are performed for a poll invocation (at \texttt{blk-mq}). Note that these activities of polling is not observed in the interrupt-driven method. We believe that this kind of store instructions can be reduced by optimizing memory management routine of \texttt{blk-mq} and NVMe driver for the polled-mode I/O completion in the near future.



\begin{figure}
	\centering
	\begin{subfigure}{0.24\linewidth}
		\includegraphics[width=\linewidth]{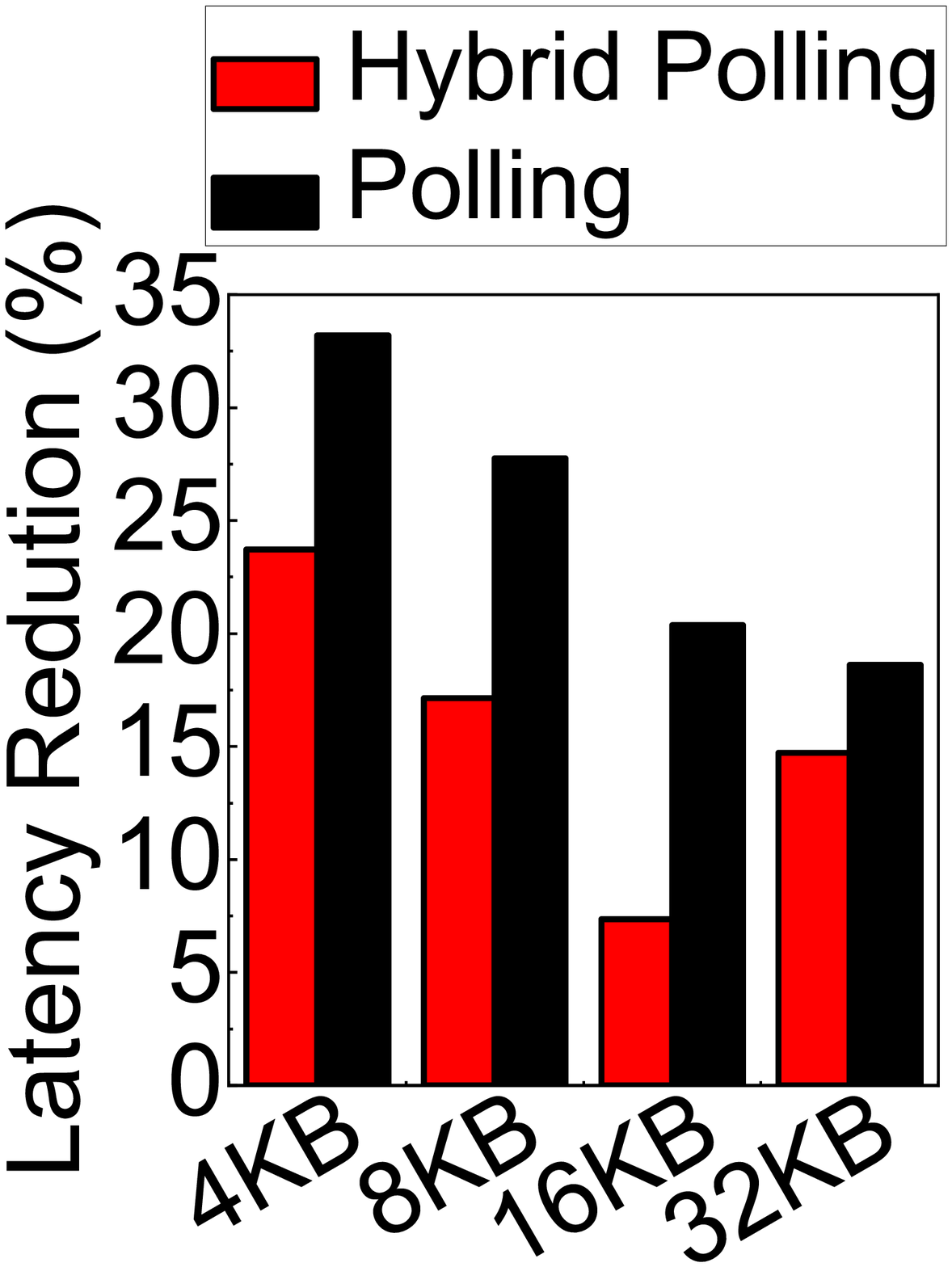}
		\caption{Seq. reads.\centering}
		\label{fig:seqrd_gain_hybrid}
	\end{subfigure}
	\begin{subfigure}{0.24\linewidth}
		\includegraphics[width=\linewidth]{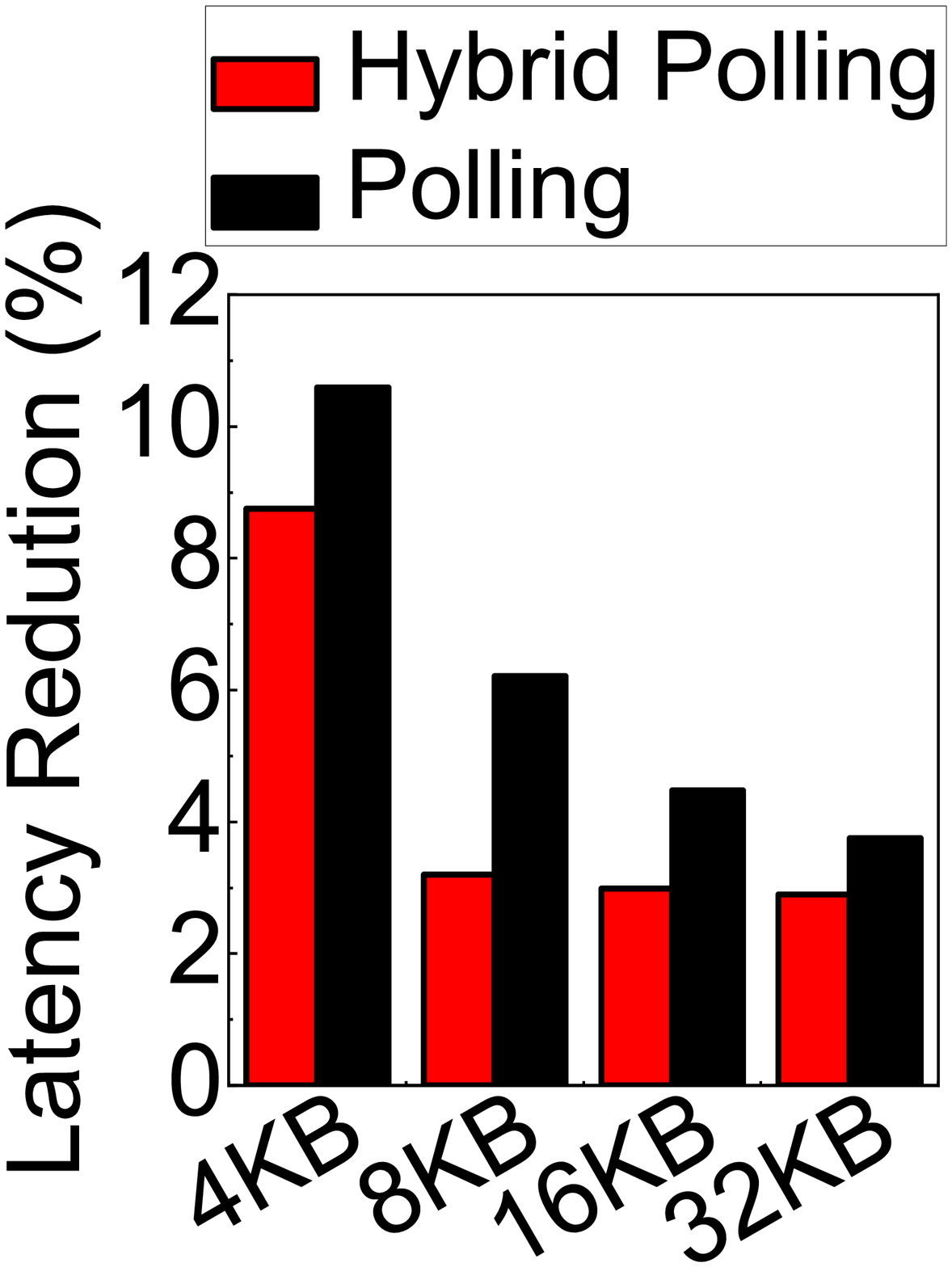}
		\caption{Rnd. reads.\centering}
		\label{fig:rndrd_gain_hybrid}
	\end{subfigure}
	\begin{subfigure}{0.24\linewidth}
		\includegraphics[width=\linewidth]{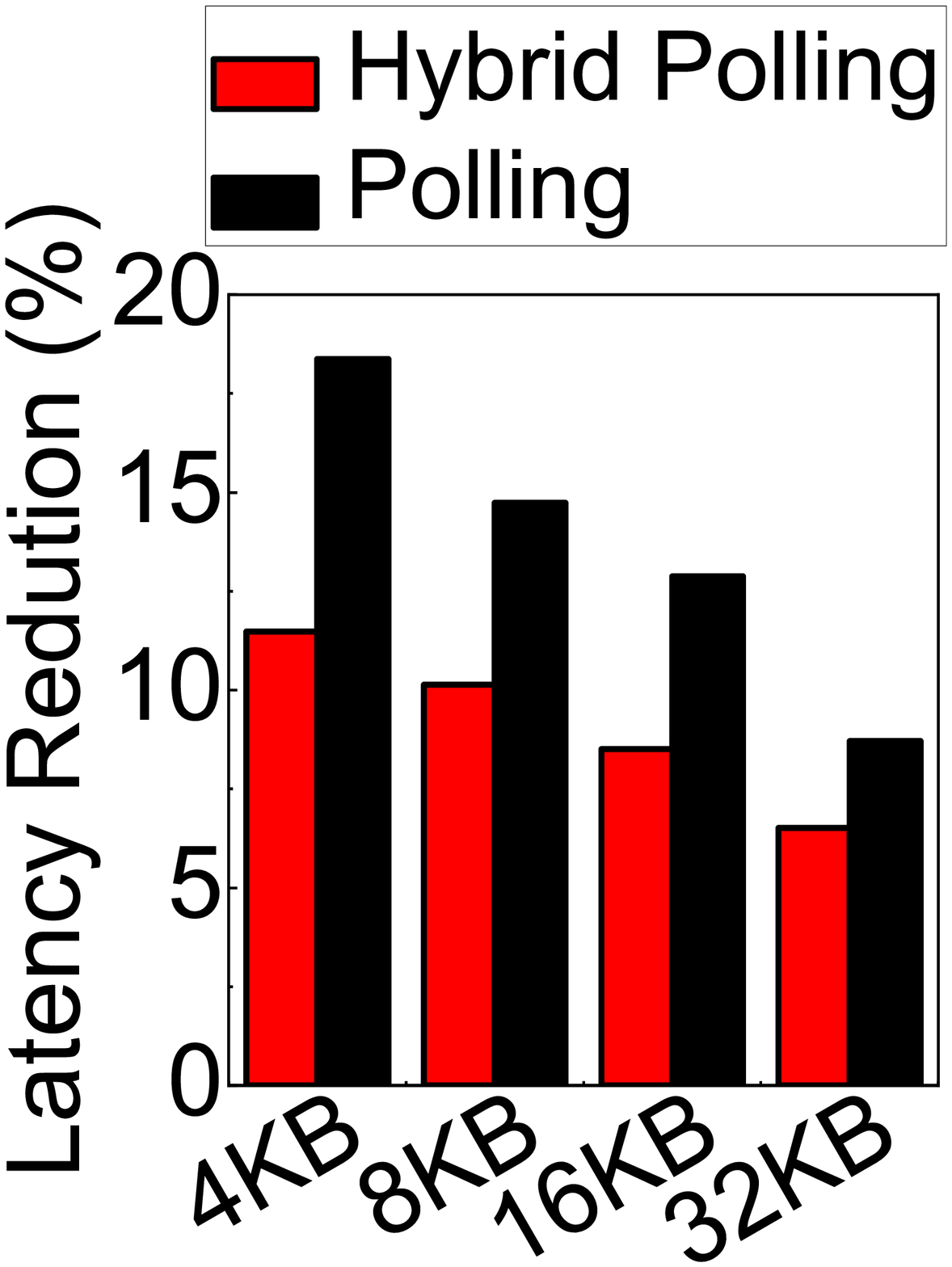}
		\caption{Seq. writes.\centering}
		\label{fig:seqwr_gain_hybrid}
	\end{subfigure}
	\begin{subfigure}{0.24\linewidth}
		\includegraphics[width=\linewidth]{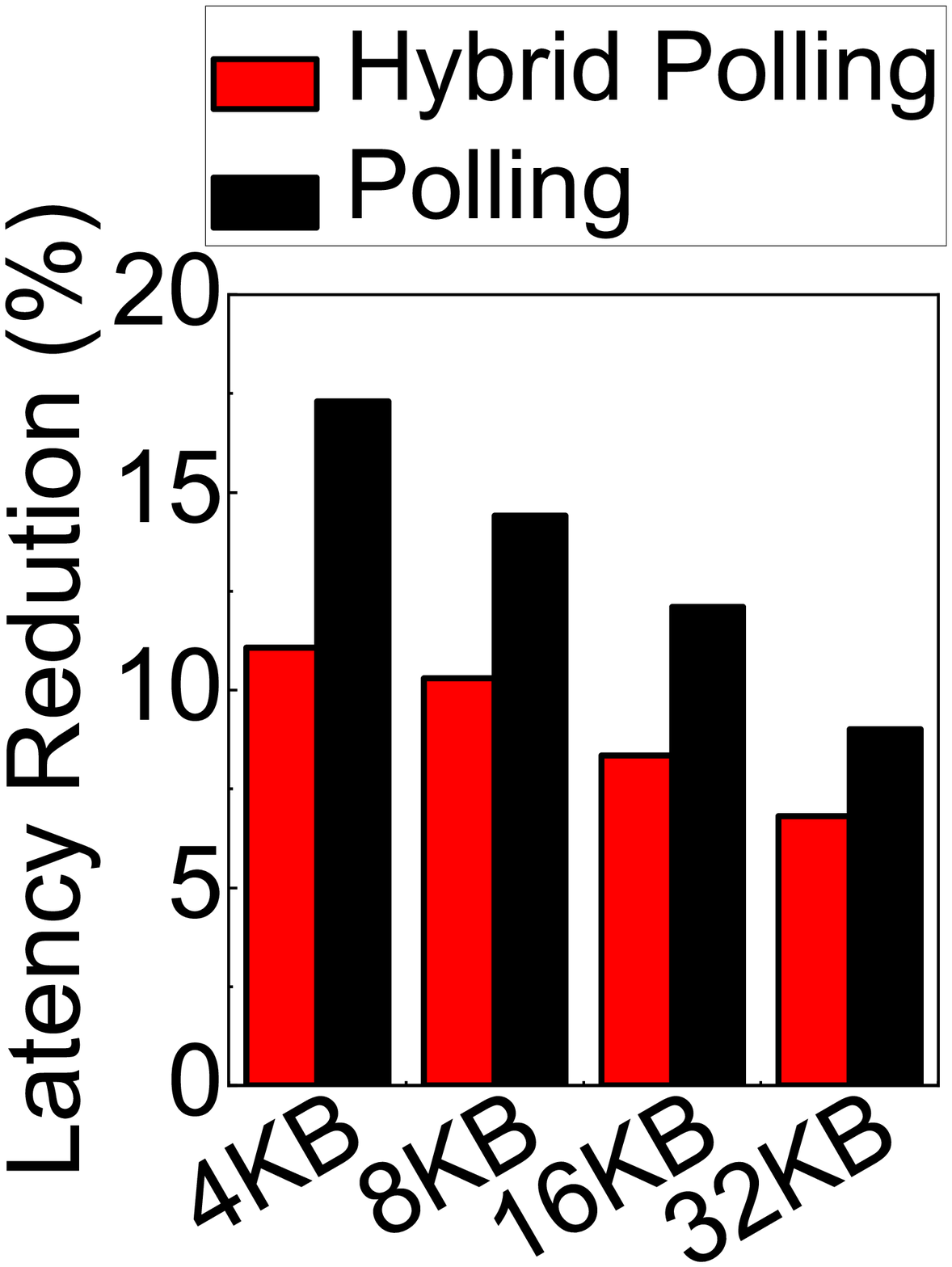}
		\caption{Rnd. writes.\centering}
		\label{fig:rndwr_gain_hybrid}
	\end{subfigure}
	\vspace{-5pt}
	\caption{Latency reduction of hybrid polling.}
	\vspace{-10pt}
	\label{fig:hybrid_gain}
\end{figure}
\begin{figure}
	\centering
	\begin{subfigure}{0.24\linewidth}
		\includegraphics[width=\linewidth]{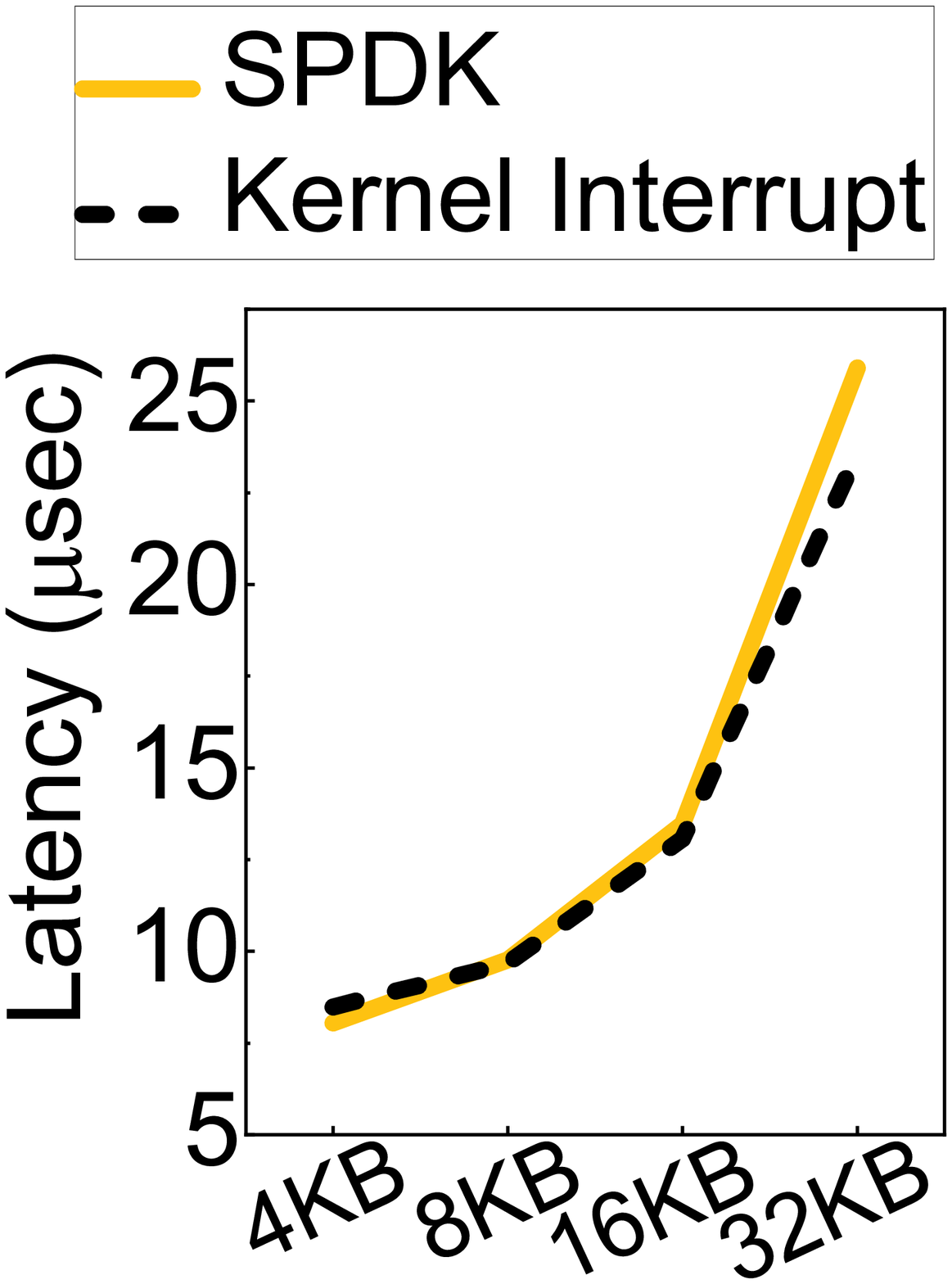}
		\caption{Seq. reads.\centering}
		\label{fig:750_seqrd_spdk}
	\end{subfigure}
	\begin{subfigure}{0.24\linewidth}
		\includegraphics[width=\linewidth]{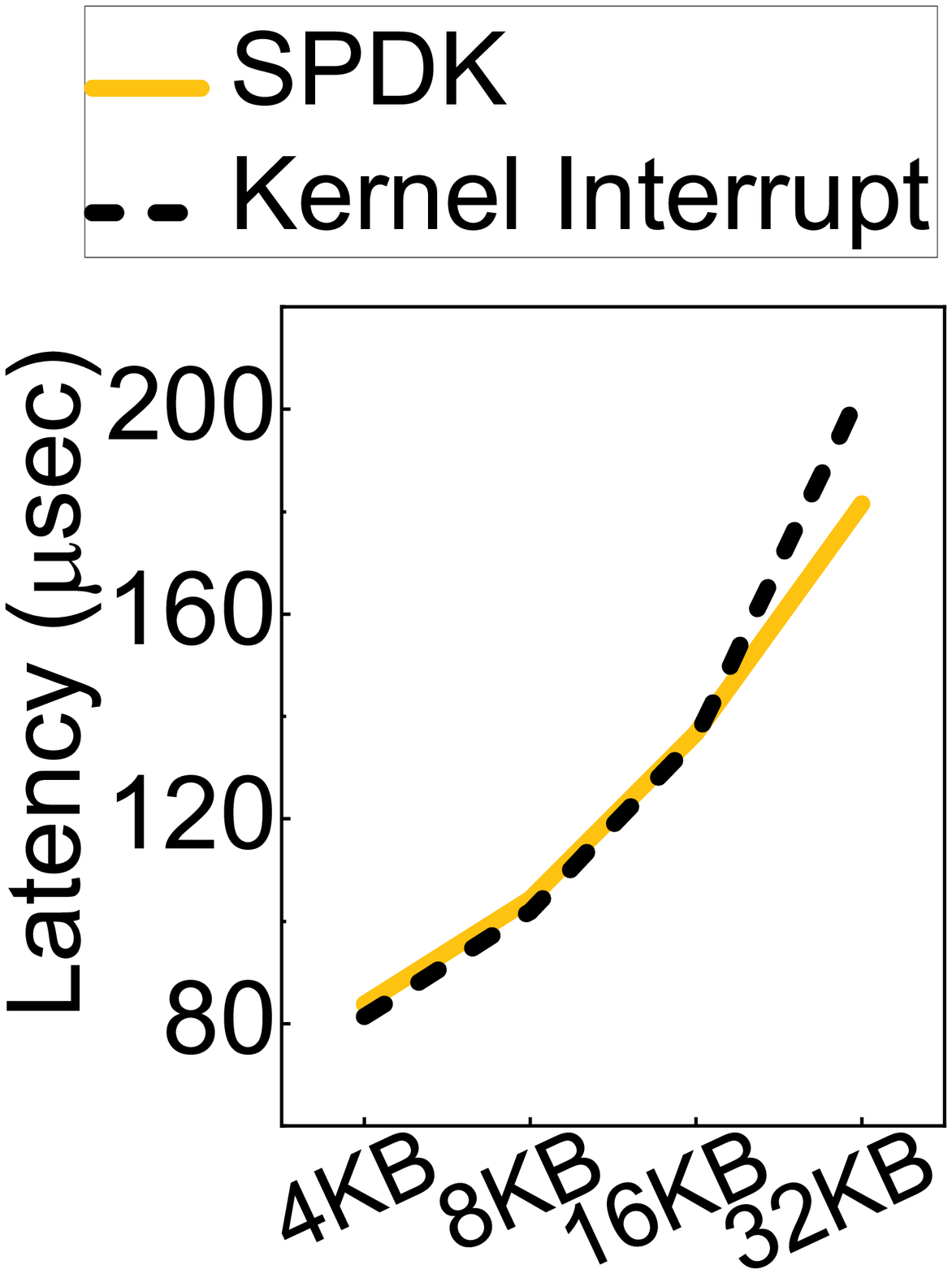}
		\caption{Rnd. reads.\centering}
		\label{fig:750_rndrd_spdk}
	\end{subfigure}
	\begin{subfigure}{0.24\linewidth}
		\includegraphics[width=\linewidth]{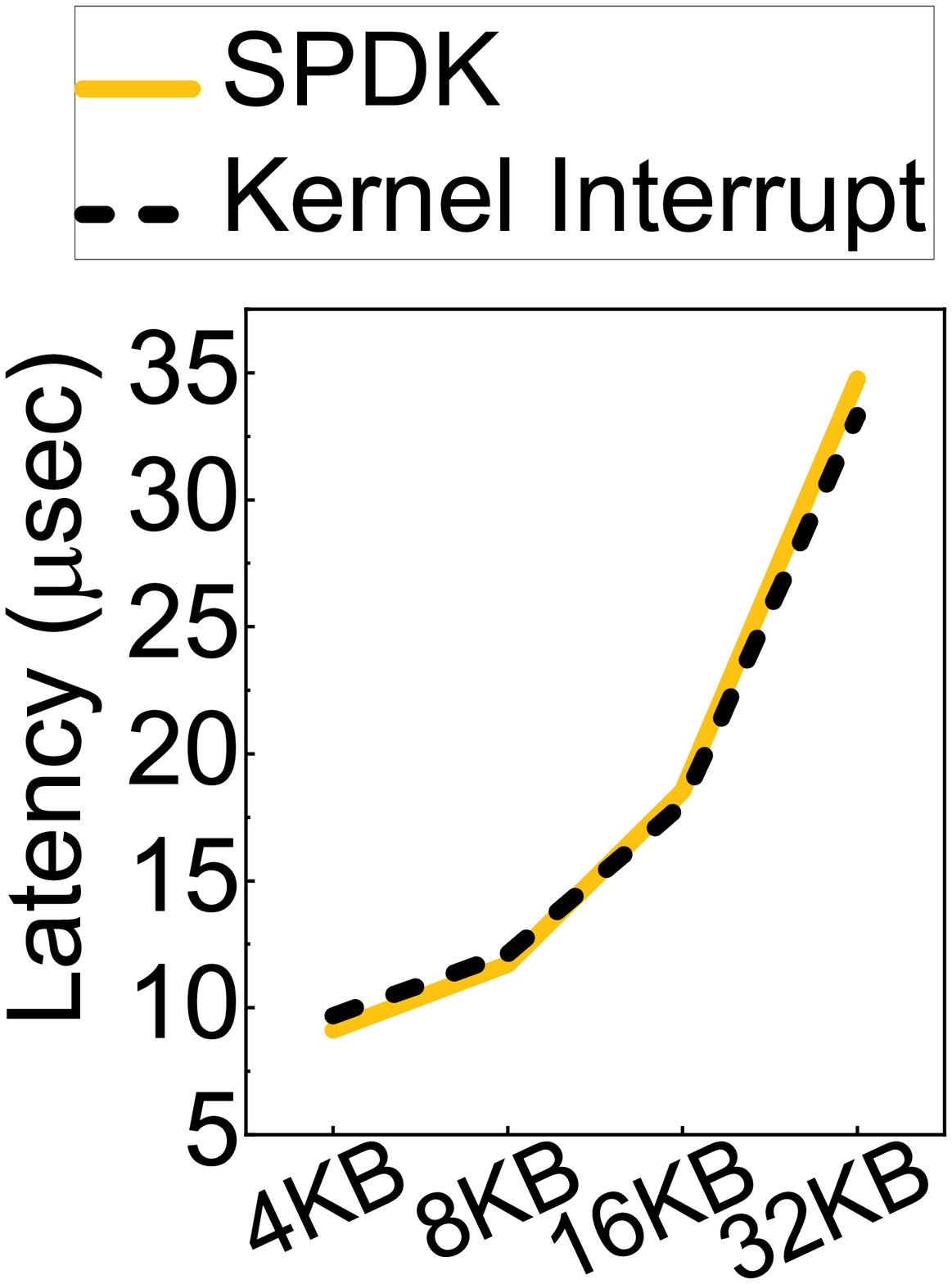}
		\caption{Seq. writes.\centering}
		\label{fig:750_seqwr_spdk}
	\end{subfigure}
	\begin{subfigure}{0.24\linewidth}
		\includegraphics[width=\linewidth]{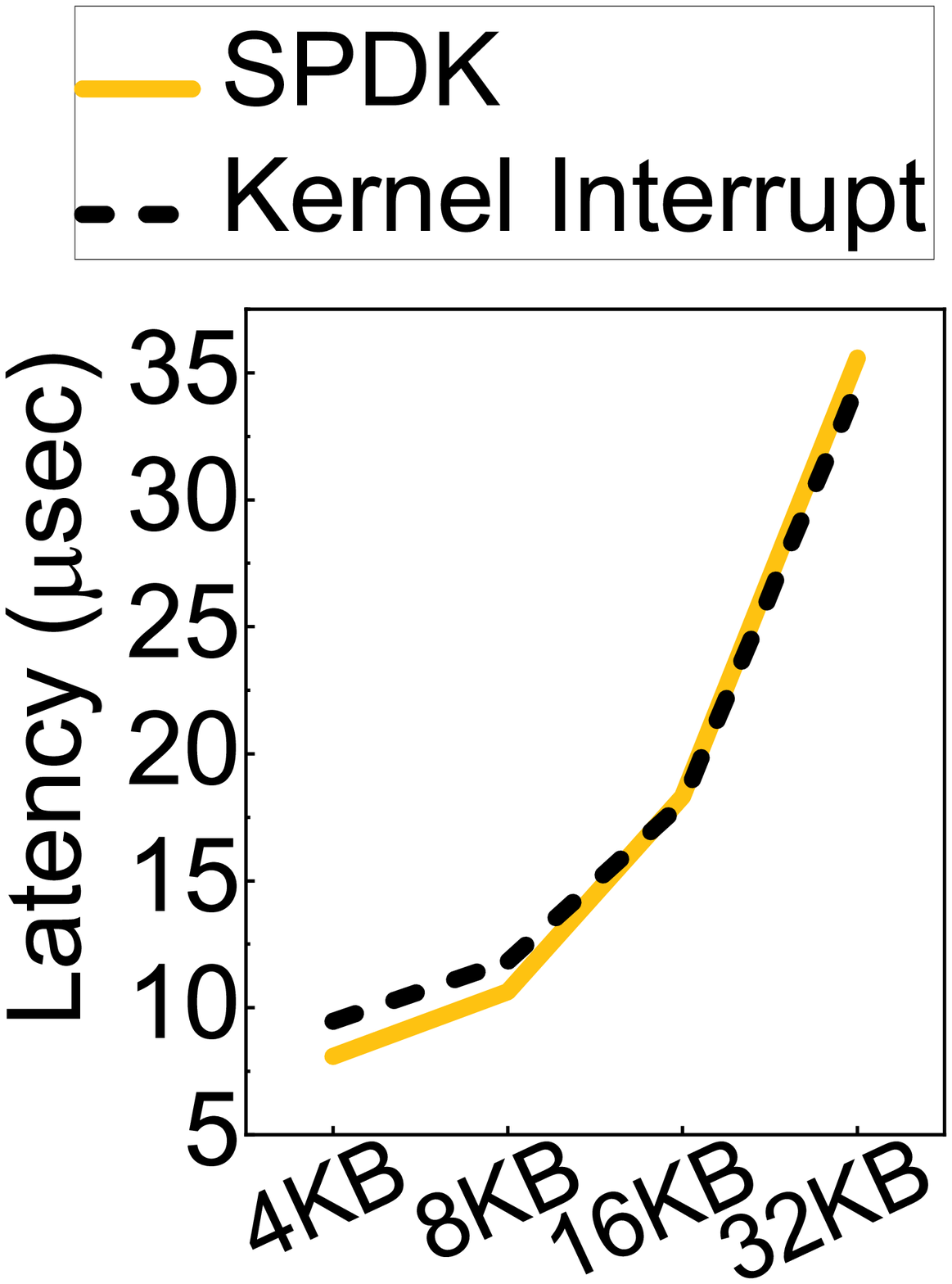}
		\caption{Rnd. writes.\centering}
		\label{fig:750_rndwr_spdk}
	\end{subfigure}
	\vspace{-5pt}
	\caption{Avg. latency of interrupt and SPDK in NVMe SSD.}
	\vspace{-15pt}
	\label{fig:750_spdk}
\end{figure}

\subsection{Will be a hybrid polling better than the polled-mode?}
The hybrid polling is expected to reduce the CPU cycles, wasted by intensively polling CQs. Figure \ref{fig:hybrid_gain} shows how much the latency can be reduced by the polled-mode I/O completion method and hybrid polling, compared to the interrupt-driven I/O completion. In this evaluation, we tested on ULL SSD with the block sizes varying from 4KB to 32KB. One can observe from this figure that compared to the interrupts, the hybrid polling only can reduce the I/O latency 8.2\%, at most, while the polled-mode completion can reduce the I/O latency as high as 33\%, on average. Importantly, the reason why the hybrid polling method exhibits 5\% longer latency, compared to the polled-mode completion, is that the expected time to sleep before jumping to its poll process is highly inaccurate. Even though the hybrid polling calculates the sleep epoch by monitoring the average latency of each request, the device-level latency of SSDs varies due to many different parameters such as internal DRAM caching, different I/O operation types, garbage collection, and execution delays of error correction codes. Thus, the hybrid polling can oversleep or wake up the poll process early, which in turn can increase the latency or reduce benefits of the sleep. Figure \ref{fig:hybrid_cpu} analyzes the CPU utilization that the hybrid polling exhibits. While it reduces the CPU cycles, the hybrid polling still consumes CPU cycles to handle sequential and random I/O requests by 58\% and 56\%, on average, respectively; in other words, the hybrid polling still consumes many CPU cycles, which are 2.2$\times$ more than the conventional interrupt-driven method consumes.


\vspace{-5pt}
\section{Advanced Storage Stack Analysis}
\label{sec:storage}
\vspace{-5pt}
Emerging kernel bypass methods are already used in many computing domains, especially network (DPDK) and storage (SPDK). These methods are expected to reduce latency by removing all kernel module interventions while serving the target requests. 

\begin{figure}
	\centering
	\begin{subfigure}{0.24\linewidth}
		\includegraphics[width=\linewidth]{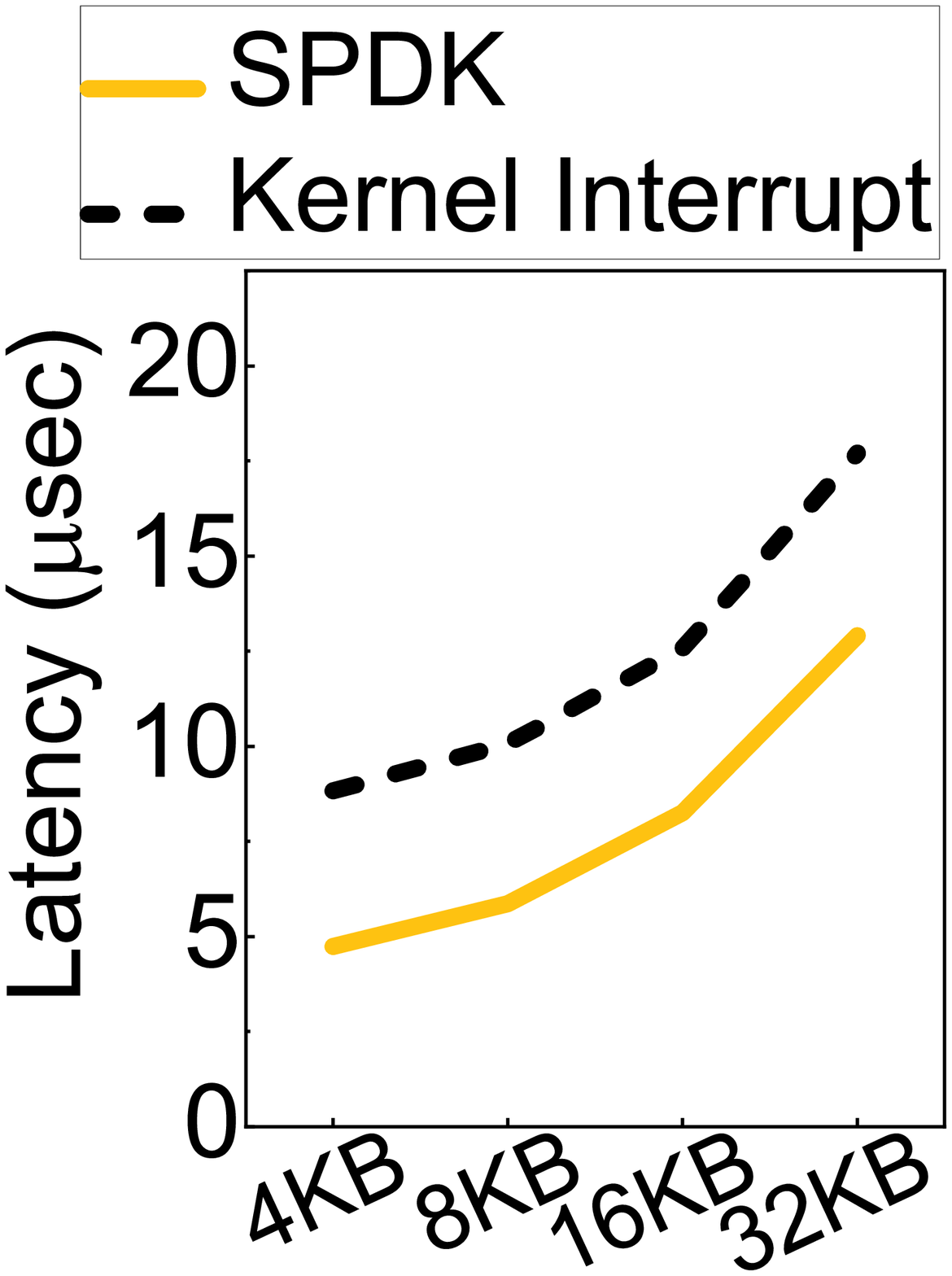}
		\caption{Seq. reads.\centering}
		\label{fig:zssd_seqrd_spdk}
	\end{subfigure}
	\begin{subfigure}{0.24\linewidth}
		\includegraphics[width=\linewidth]{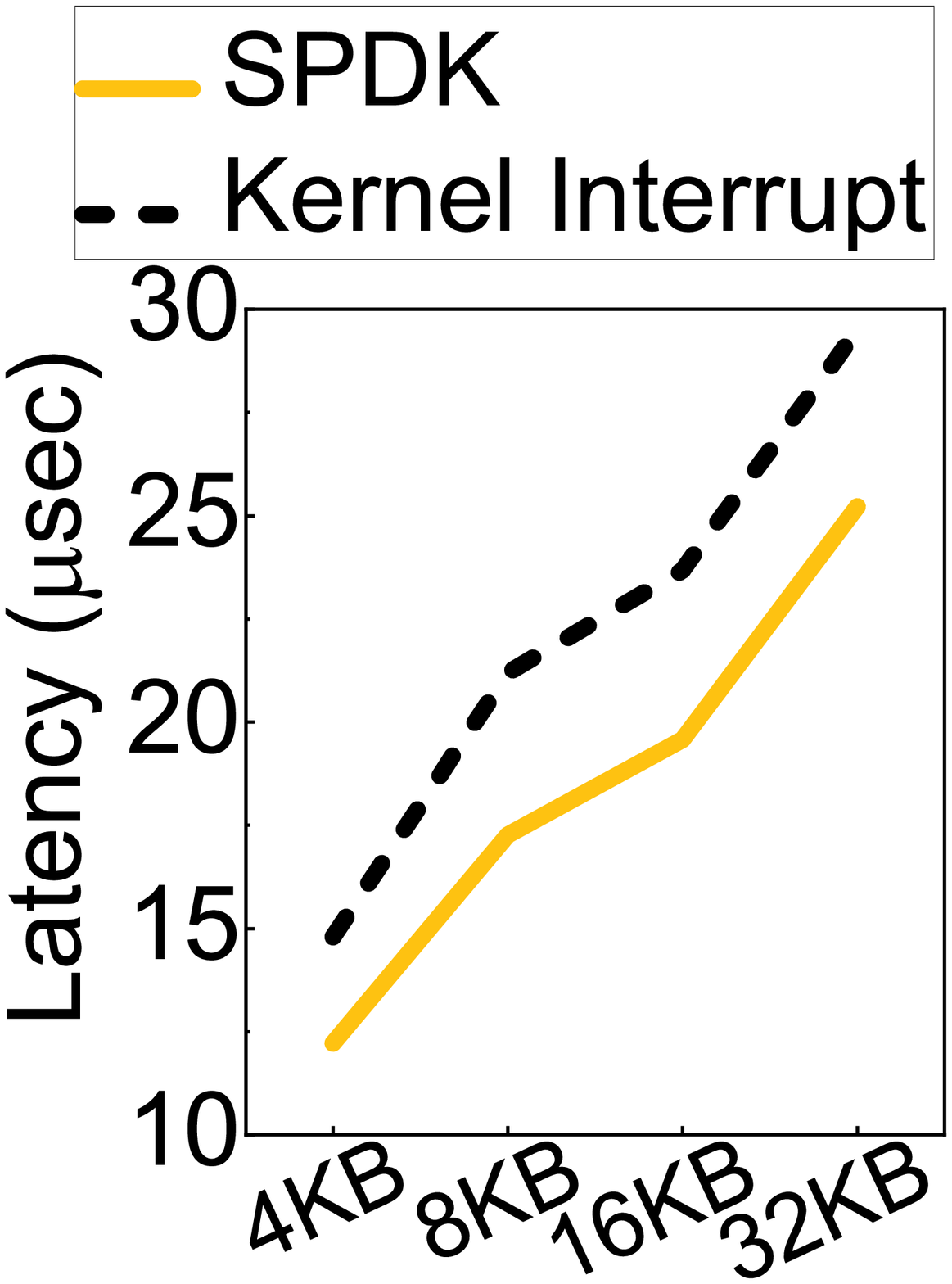}
		\caption{Rnd. reads.\centering}
		\label{fig:zssd_rndrd_spdk}
	\end{subfigure}
	\begin{subfigure}{0.24\linewidth}
		\includegraphics[width=\linewidth]{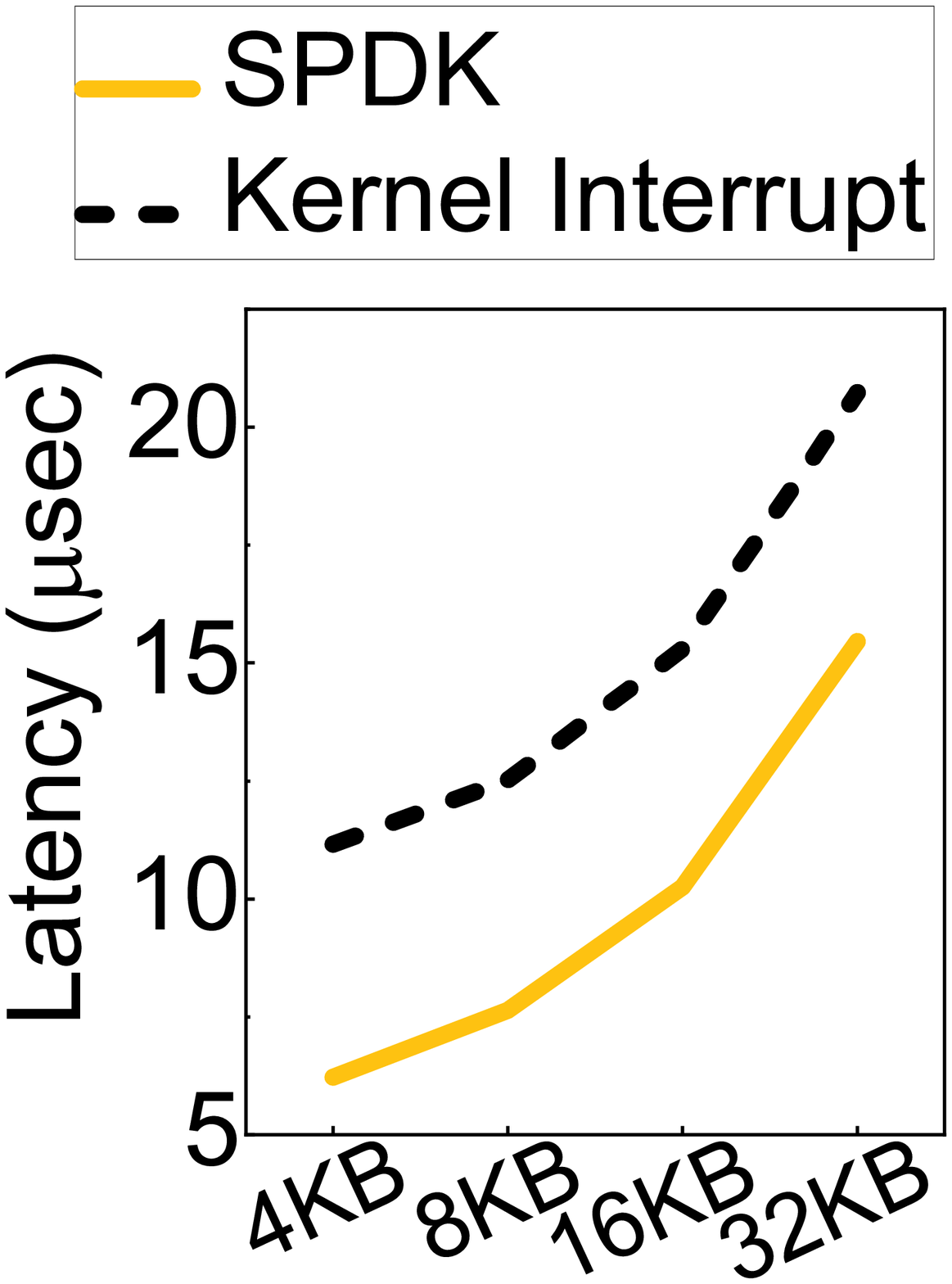}
		\caption{Seq. writes.\centering}
		\label{fig:zssd_seqwr_spdk}
	\end{subfigure}
	\begin{subfigure}{0.24\linewidth}
		\includegraphics[width=\linewidth]{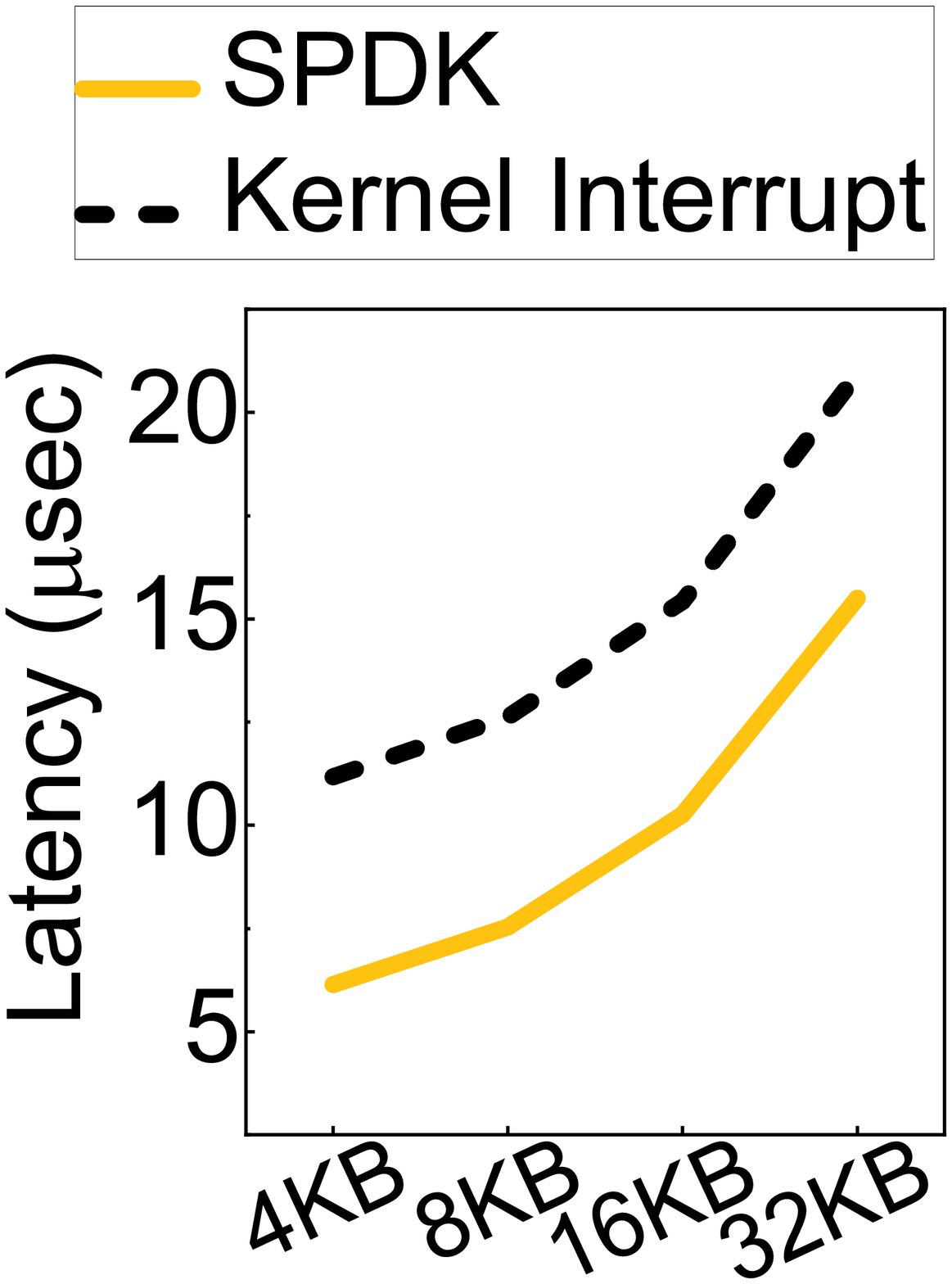}
		\caption{Rnd. writes.\centering}
		\label{fig:zssd_rndwr_spdk}
	\end{subfigure}
	\vspace{-5pt}
	\caption{Avg. latency of interrupt and SPDK in ULL SSD.}
	\vspace{-10pt}
	\label{fig:zssd_spdk}
\end{figure}
\begin{figure}
	\centering
	\begin{subfigure}{0.24\linewidth}
		\includegraphics[width=\linewidth]{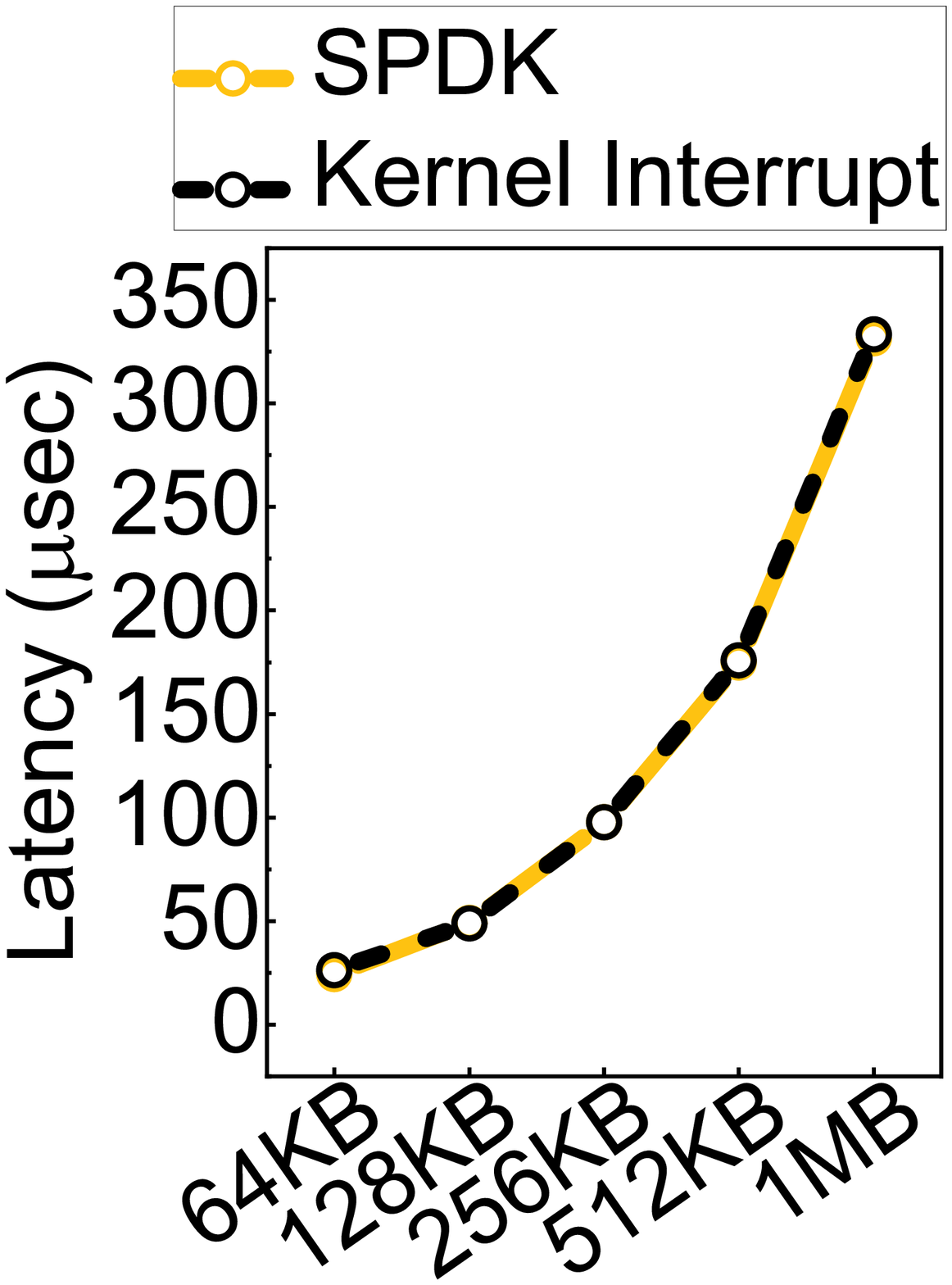}
		\caption{Seq. reads.\centering}
		\label{fig:zssd_spdk_big_seqrd}
	\end{subfigure}
	\begin{subfigure}{0.24\linewidth}
		\includegraphics[width=\linewidth]{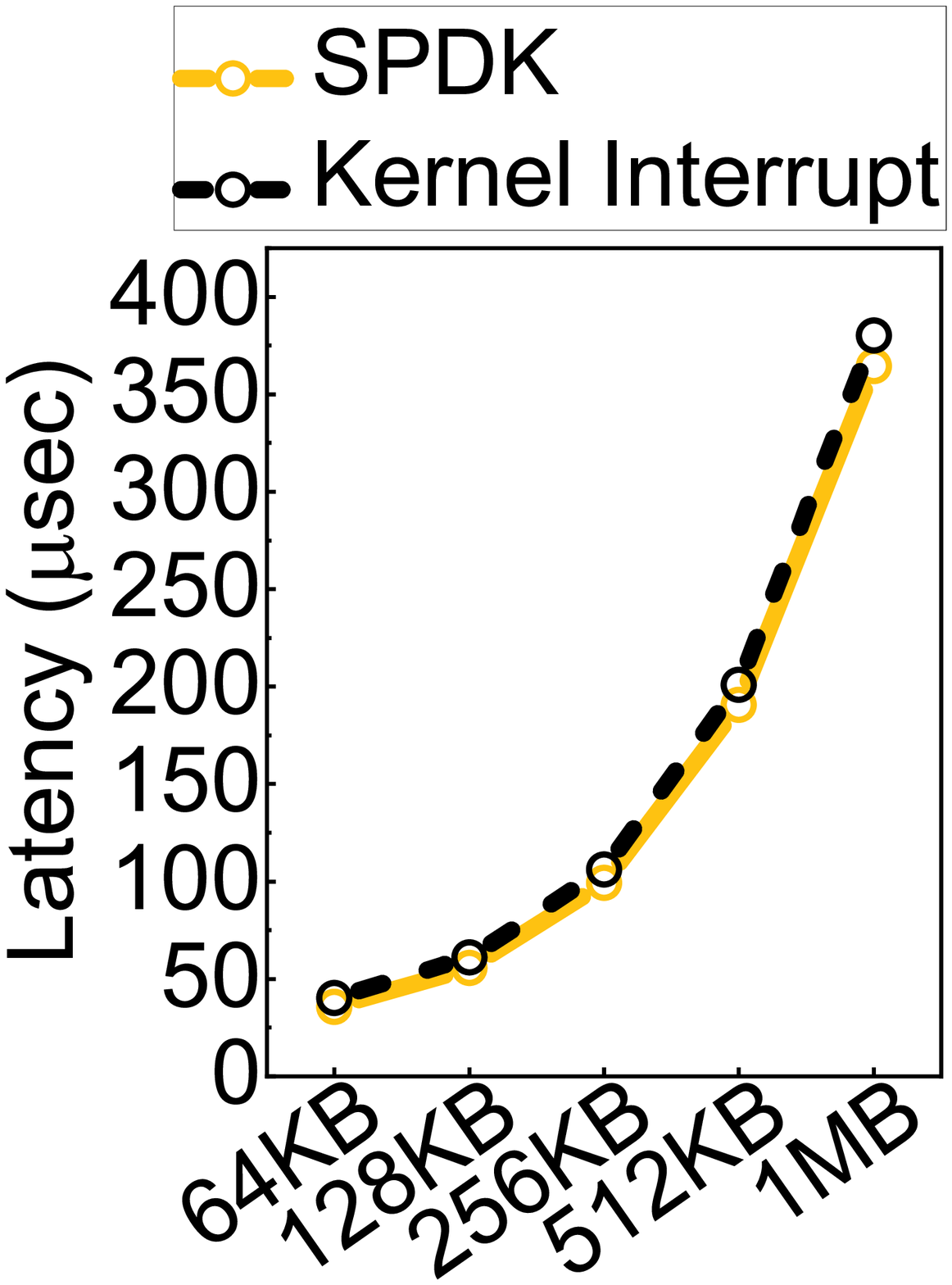}
		\caption{Rnd. reads.\centering}
		\label{fig:zssd_spdk_big_rndrd}
	\end{subfigure}
	\begin{subfigure}{0.24\linewidth}
		\includegraphics[width=\linewidth]{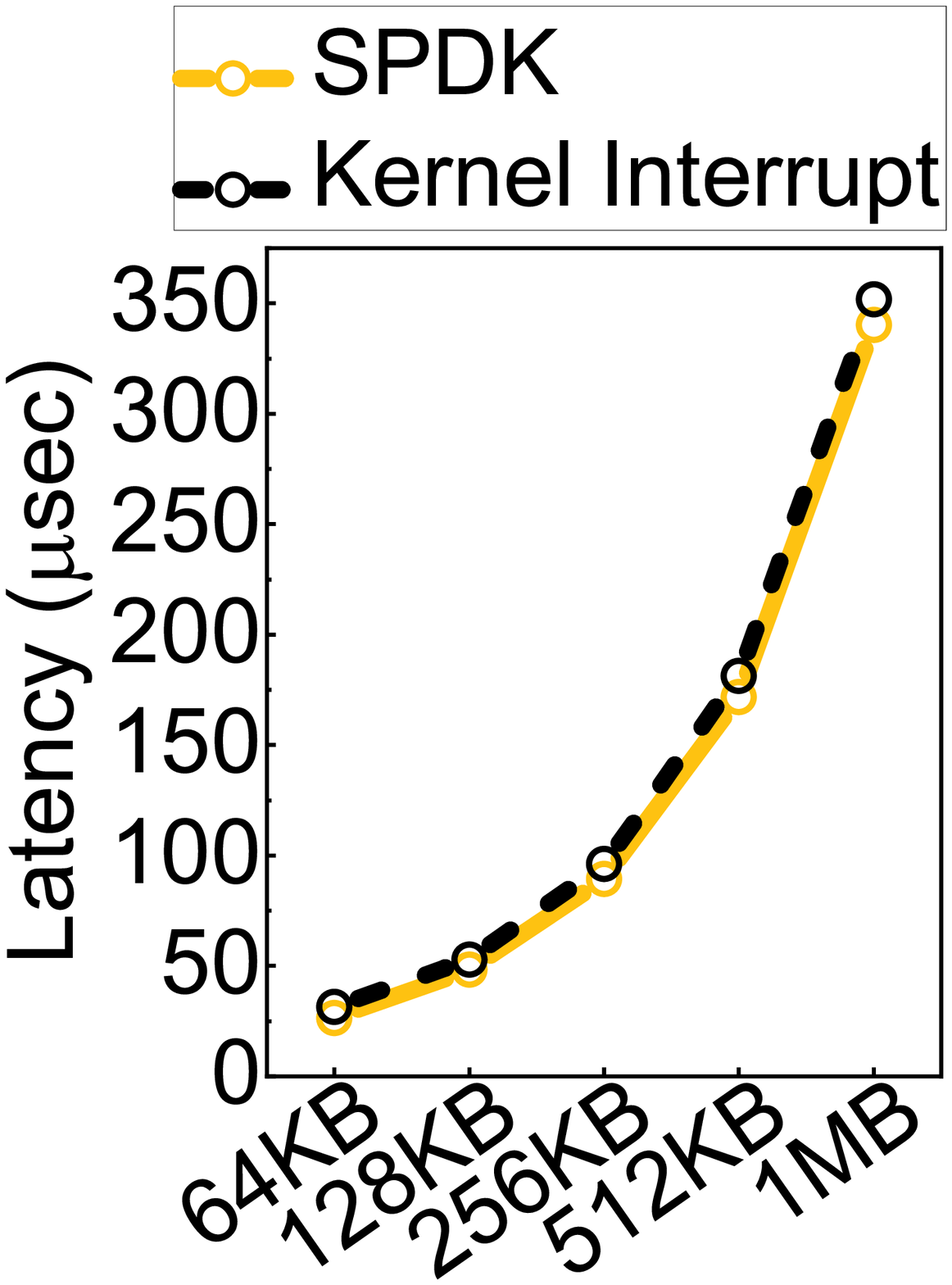}
		\caption{Seq. writes.\centering}
		\label{fig:zssd_spdk_big_seqwr}
	\end{subfigure}
	\begin{subfigure}{0.24\linewidth}
		\includegraphics[width=\linewidth]{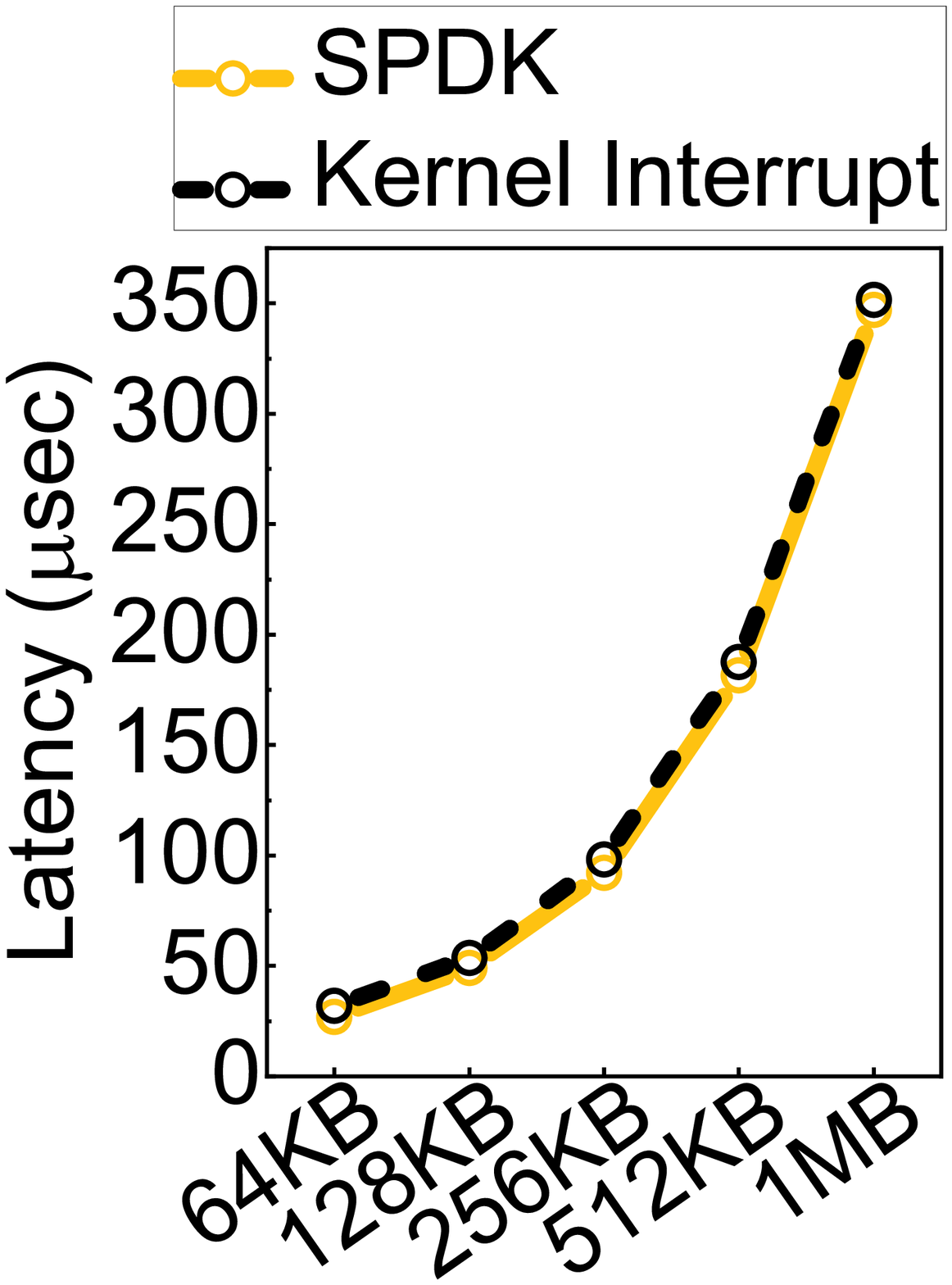}
		\caption{Rnd. writes.\centering}
		\label{fig:zssd_spdk_big_rndwr}
	\end{subfigure}
	\vspace{-5pt}
	\caption{Average latency of interrupt and SPDK with big size requests in ULL SSD.}
	\vspace{-10pt}
	\label{fig:zssd_spdk_big}
\end{figure}

\vspace{-5pt}
\subsection{Can SPDK really eliminate the latency overheads, imposed by the storage stack? What about ULL SSDs?}
\label{750_spdk}
Figure \ref{fig:750_spdk} compares NVMe SSD's latency, exhibited by SPDK and a conventional interrupt-based method that requires going through the entire storage stack. In contrast to the common expectations of SPDK \cite{kim2016nvmedirect, yang2017spdk}, the latency difference between reads and writes of NVMe SSD are only 4.3\% and 11.1\%, on average, respectively, which are almost similar to each other and negligible. We agree that the CPU burst used for a complicated software stack can be a performance bottleneck, but one of the main roles that operating systems (OSs) need to perform is to schedule the CPU burst and I/O burst by nicely overlapping them. While the current NVMe SSDs override PCIe and provide high performance by having the internal DRAM cache, the latency of underlying flash media is not short enough yet to reduce the latency more than that of host's CPU burst.

Figure \ref{fig:zssd_spdk} compares the latency of ULL SSD with the same scenario tested for NVMe SSD. The latency using SPDK can reduce by 25.2\%, 6.3\%, 13.7\% and 13.3\% for sequential reads, random reads, sequential writes, and random writes, on average, respectively. We can learn from this evaluation that current storage system can benefit from removing the complicated software stack from the I/O path only when the latency becomes shorter than that of most conventional flash exhibits.



\vspace{-5pt}
\subsection{Will SPDK be the best option for future low-latency storage? Would it have any side effects?}
While SPDK can reduce the latency on ULL SSDs, we observe several issues that system designers or application users need to pay attention. First, the benefits of SPDK with ULL SSD become negligible when the request size increases. Figure \ref{fig:zssd_spdk_big} shows the latency comparison with SPDK and the conventional system under the execution of I/O requests whose block size is greater than 64KB (64KB$\sim$1MB). One can observe from the figure, the latency trend of SPDK is completely overlapped with that of the conventional system. Based on the results, SPDK is only meaningful with ULL, and even with ULL, users may selectively use SPDK in particular for small-sized I/O requests, as SPDK itself exhibits poor efficiency of host system resource management. 

\begin{figure}
	\centering
	\begin{subfigure}{0.24\linewidth}
		\includegraphics[width=\linewidth]{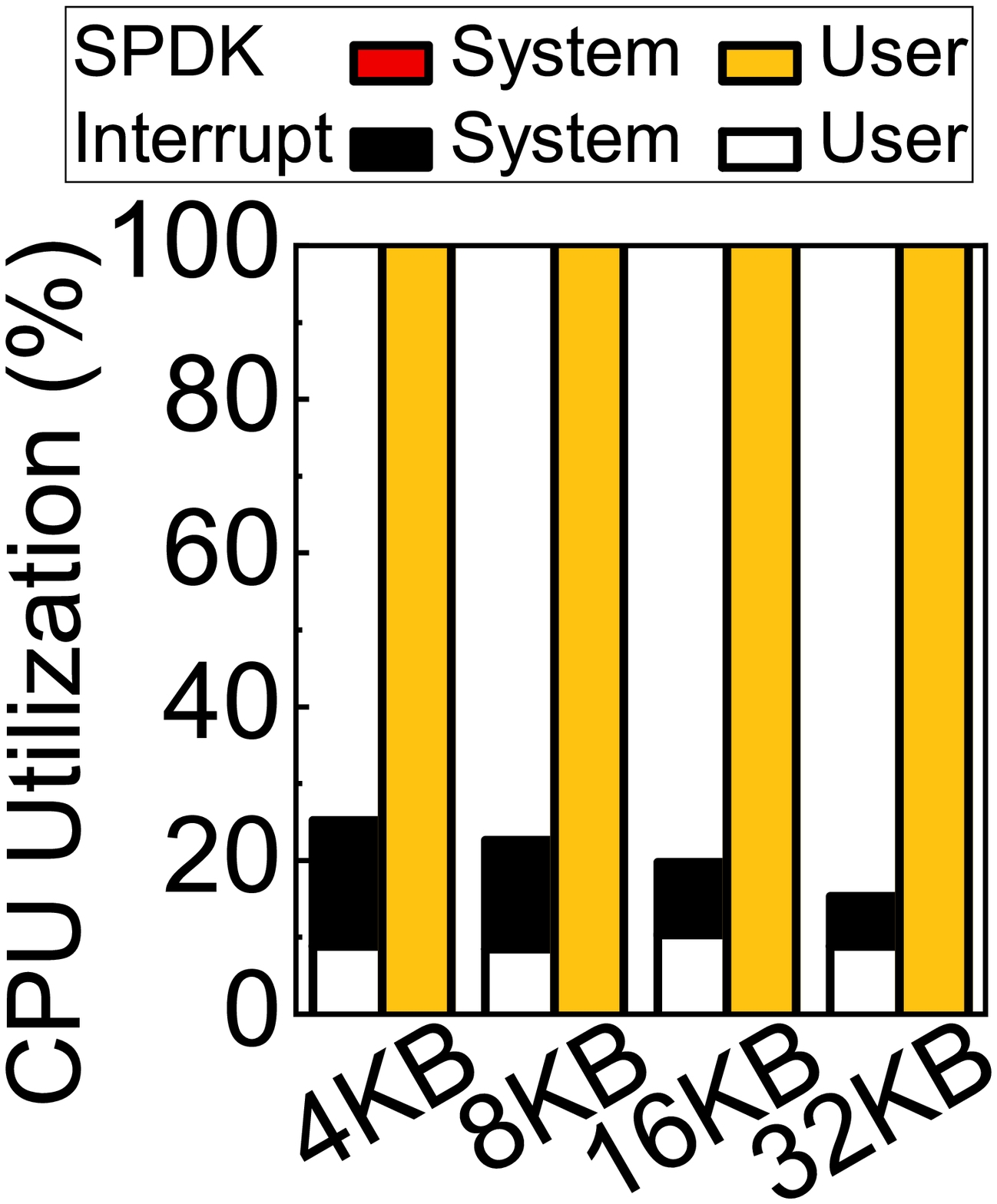}
		\caption{Seq. reads.\centering}
		\label{fig:seqrd_spdk_CPU}
	\end{subfigure}
	\begin{subfigure}{0.24\linewidth}
		\includegraphics[width=\linewidth]{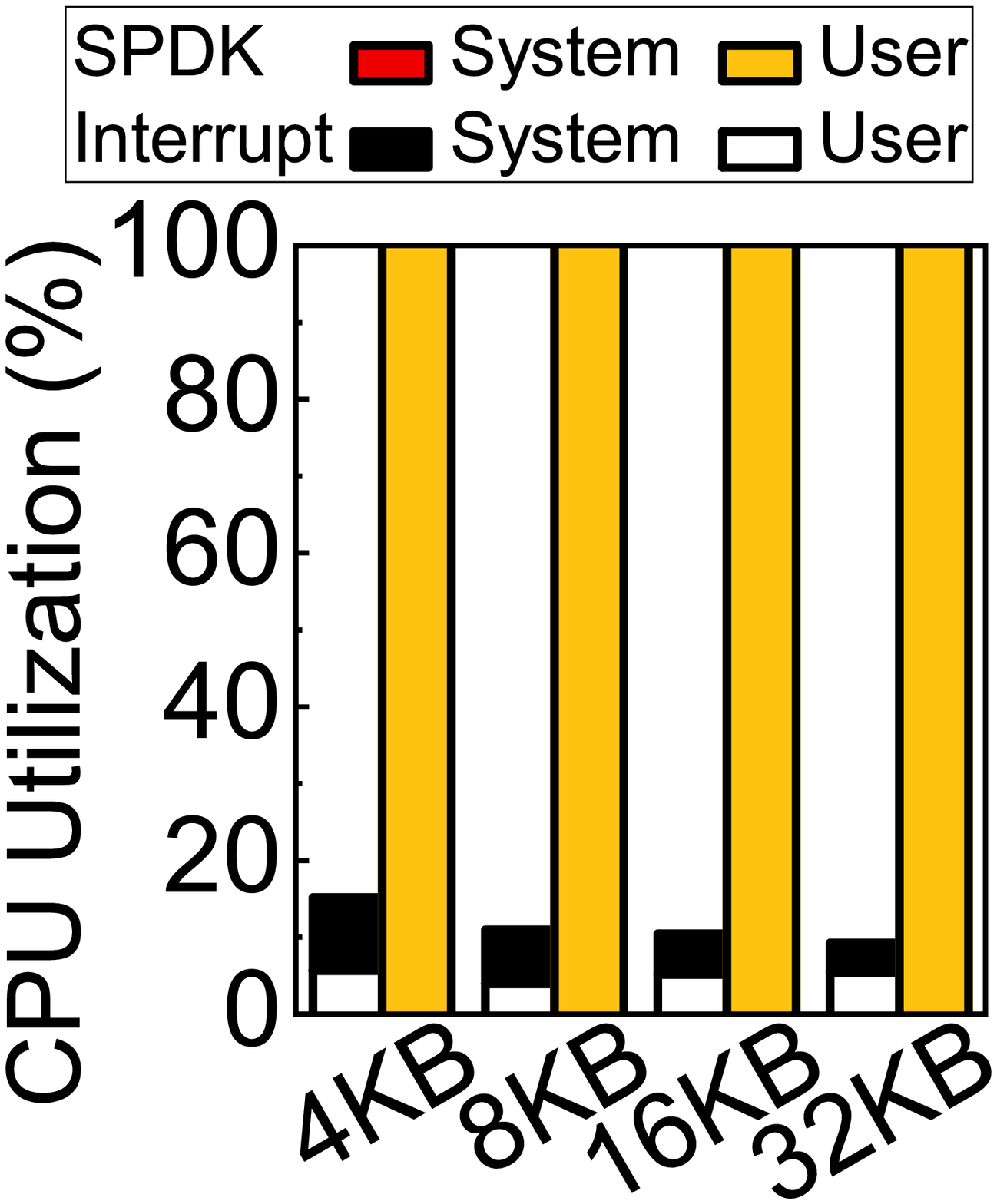}
		\caption{Rnd. reads.\centering}
		\label{fig:rndrd_spdk_CPU}
	\end{subfigure}
	\begin{subfigure}{0.24\linewidth}
		\includegraphics[width=\linewidth]{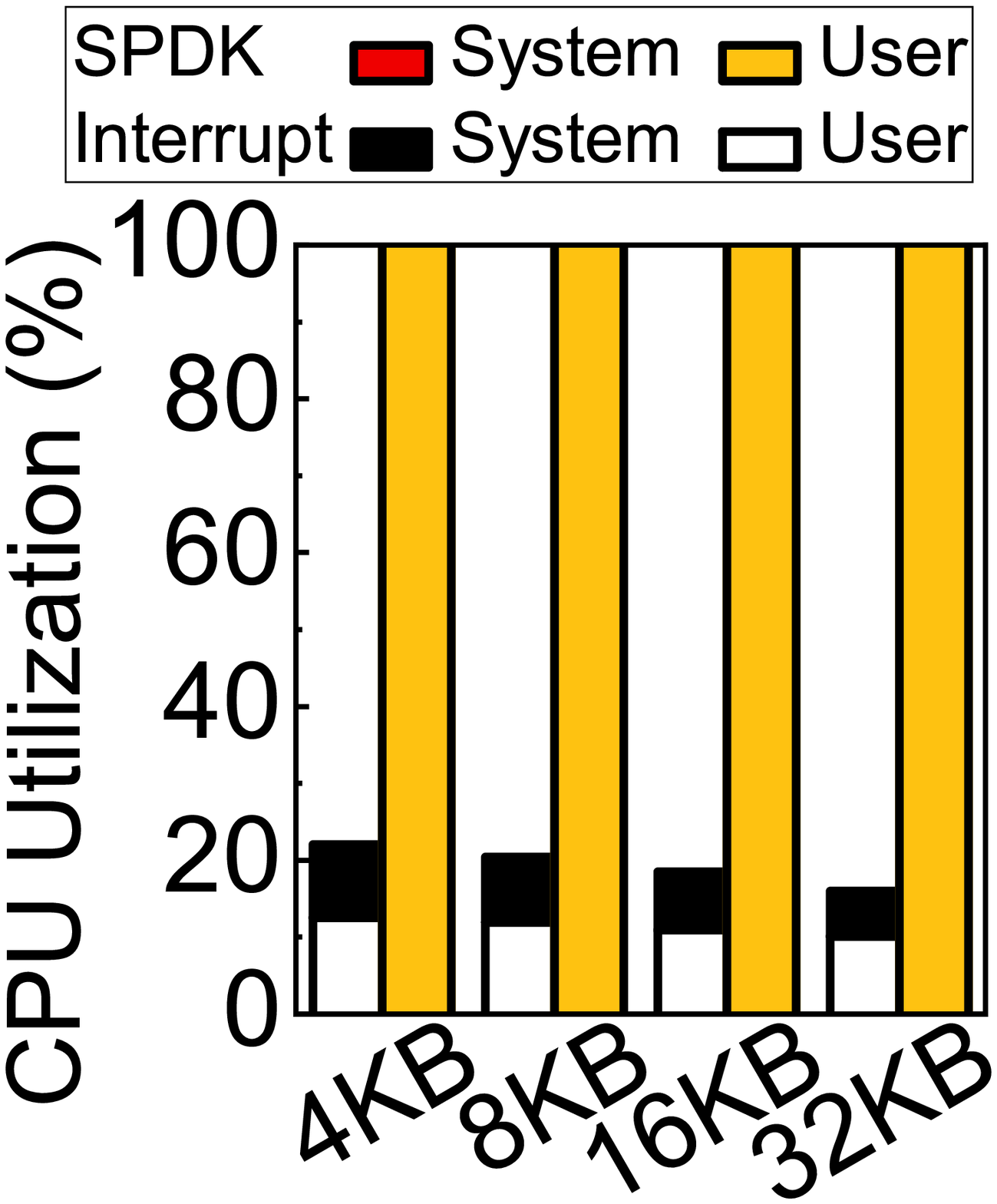}
		\caption{Seq. writes.\centering}
		\label{fig:seqwr_spdk_CPU}
	\end{subfigure}
	\begin{subfigure}{0.24\linewidth}
		\includegraphics[width=\linewidth]{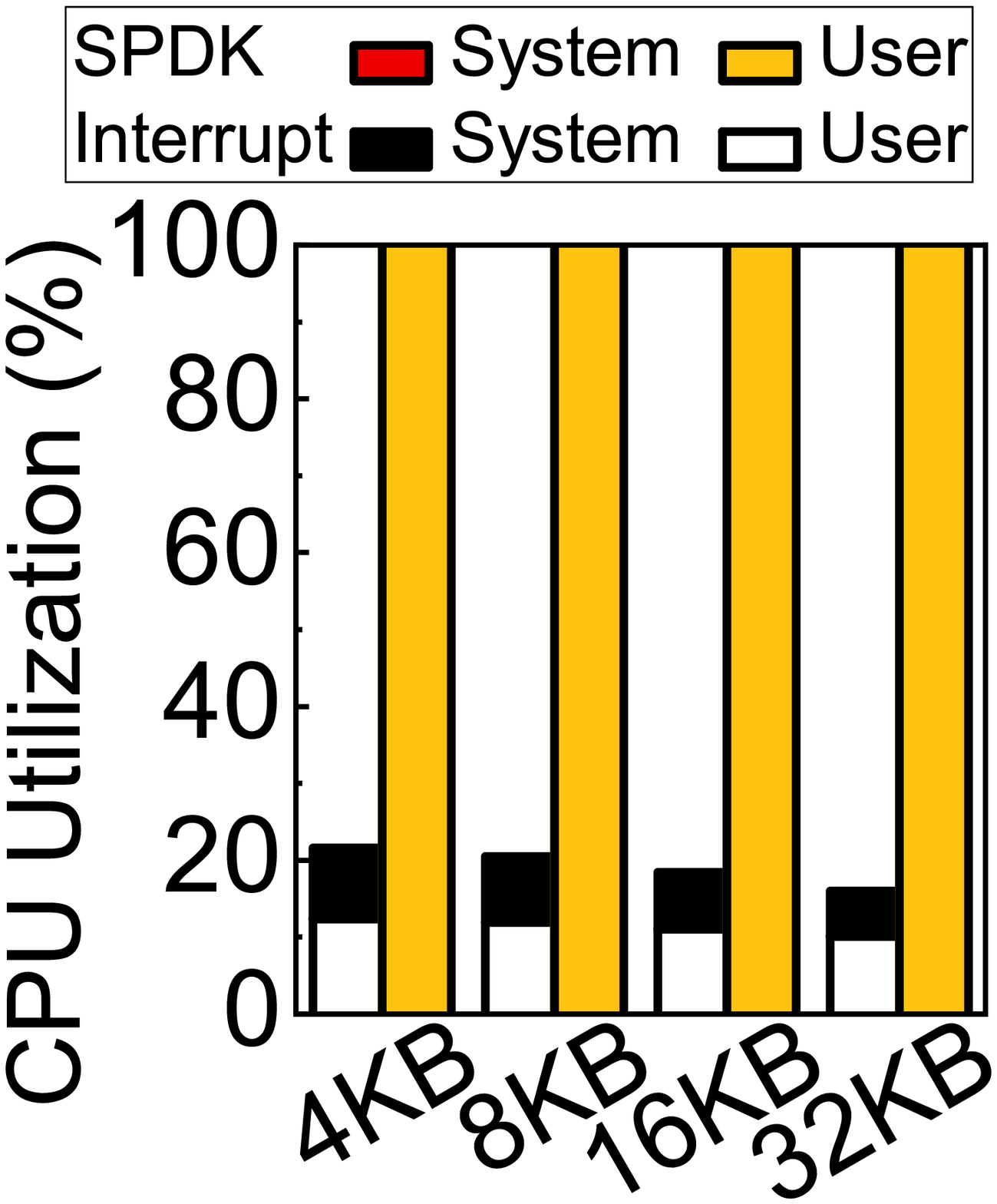}
		\caption{Rnd. writes.\centering}
		\label{fig:rndwr_spdk_CPU}
	\end{subfigure}
	\vspace{-5pt}
	\caption{CPU utilization in SPDK.}
	\vspace{-10pt}
	\label{fig:spdk_CPU}
\end{figure}
\begin{figure}
	\centering
	\begin{subfigure}{0.24\linewidth}
		\includegraphics[width=\linewidth]{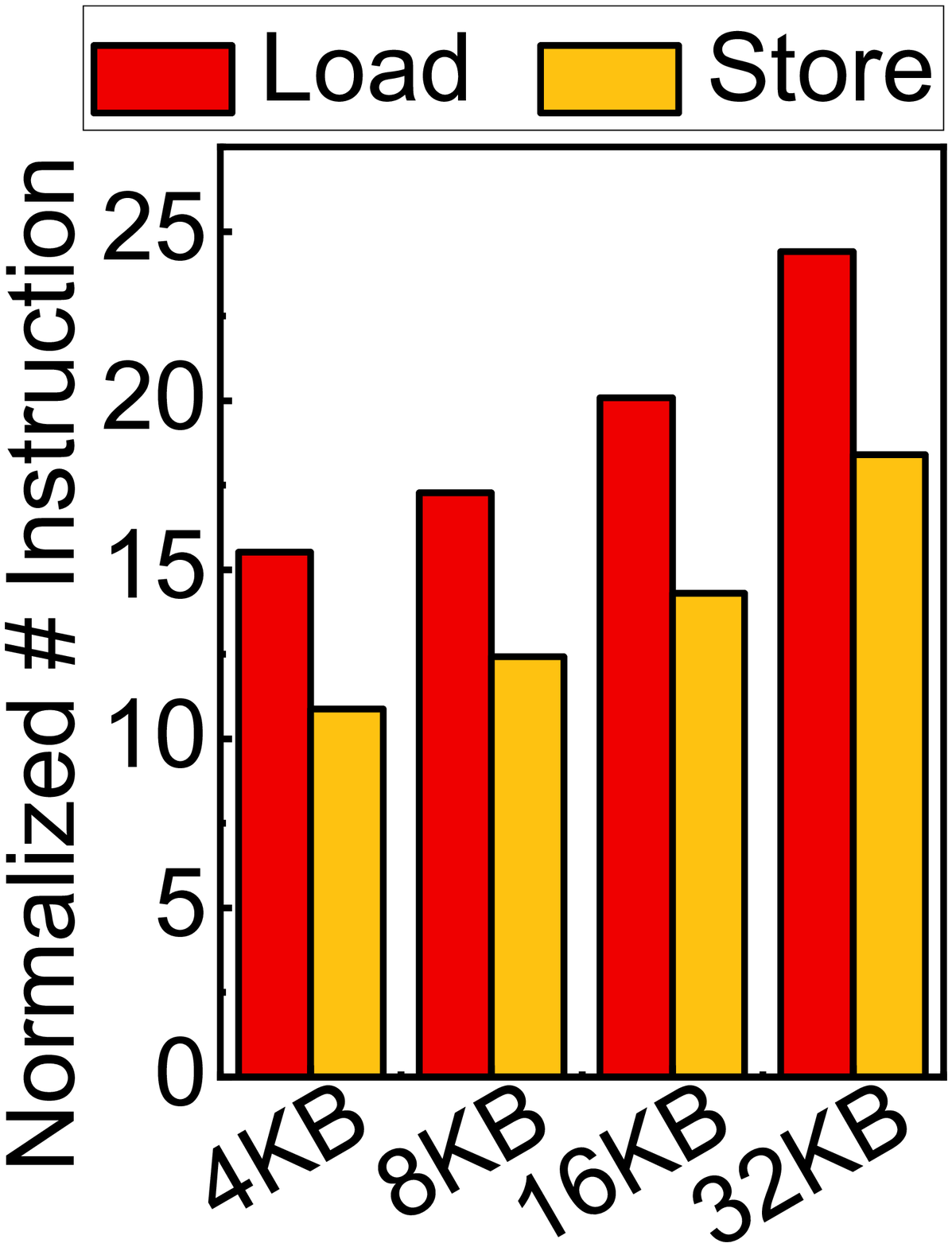}
		\caption{Seq. reads.\centering}
		\label{fig:spdk_seqrd_norm_load}
	\end{subfigure}
	\begin{subfigure}{0.24\linewidth}
		\includegraphics[width=\linewidth]{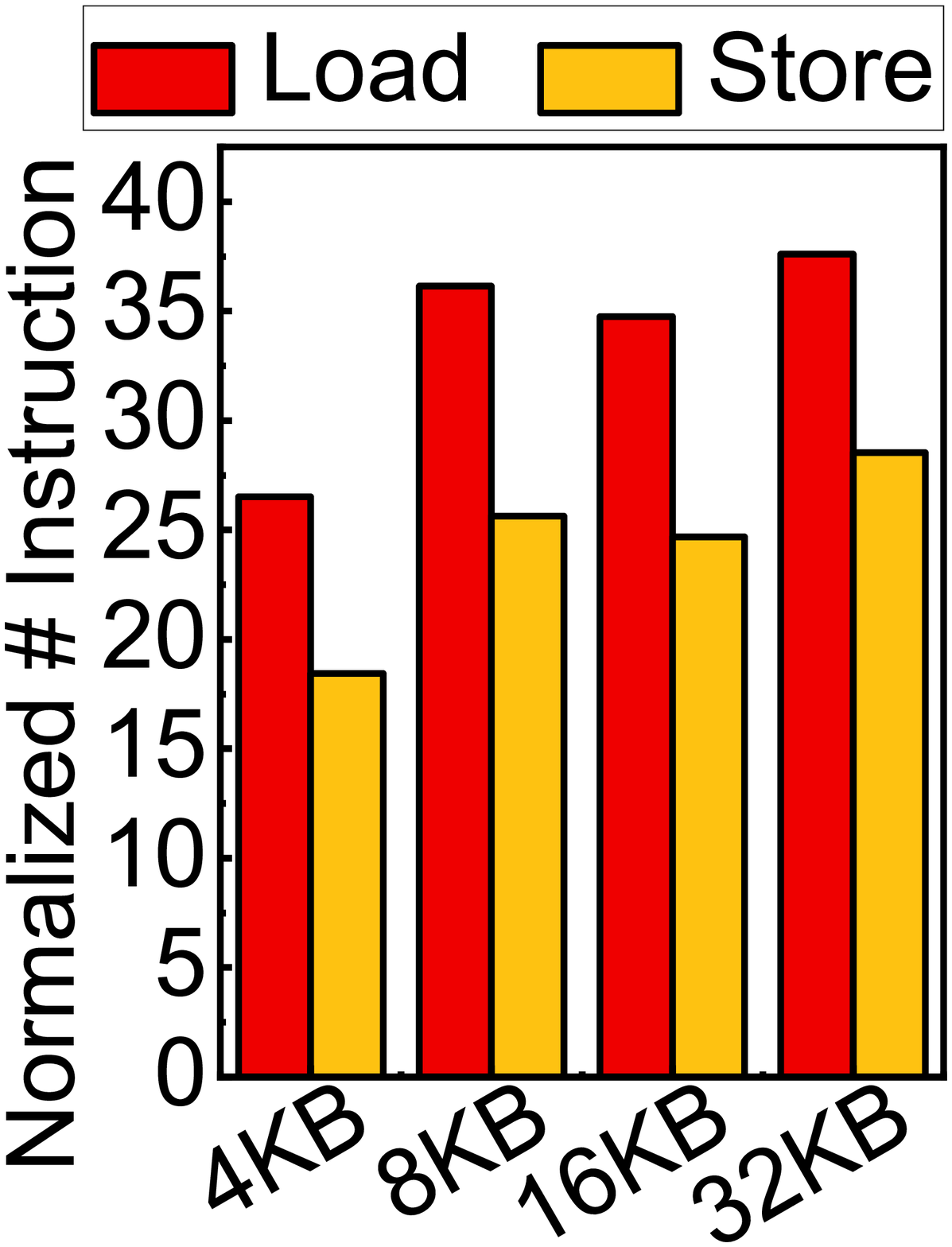}
		\caption{Rnd. reads.\centering}
		\label{fig:spdk_rndrd_norm_load}
	\end{subfigure}
	\begin{subfigure}{0.24\linewidth}
		\includegraphics[width=\linewidth]{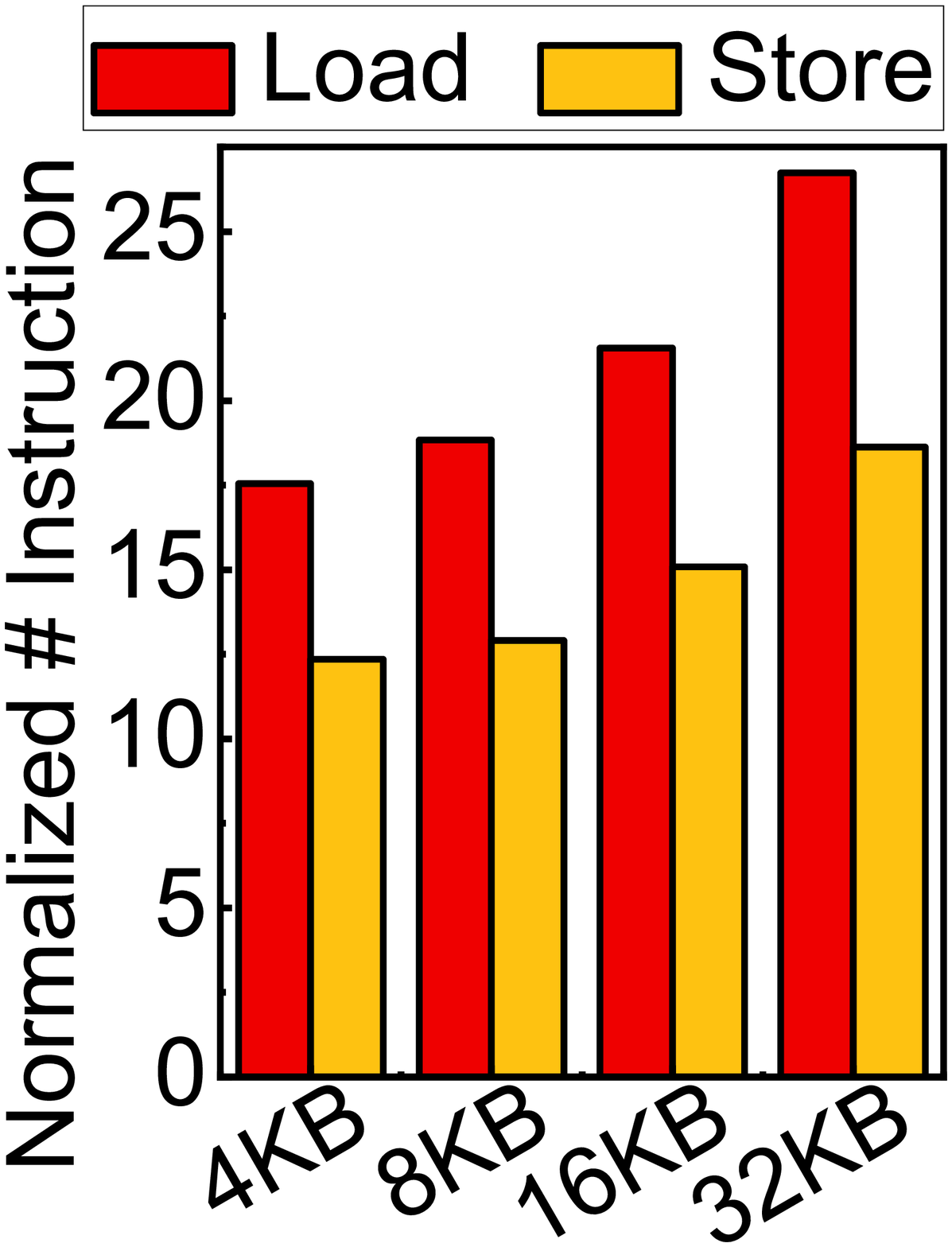}
		\caption{Seq. writes.\centering}
		\label{fig:spdk_seqwr_norm_load}
	\end{subfigure}
	\begin{subfigure}{0.24\linewidth}
		\includegraphics[width=\linewidth]{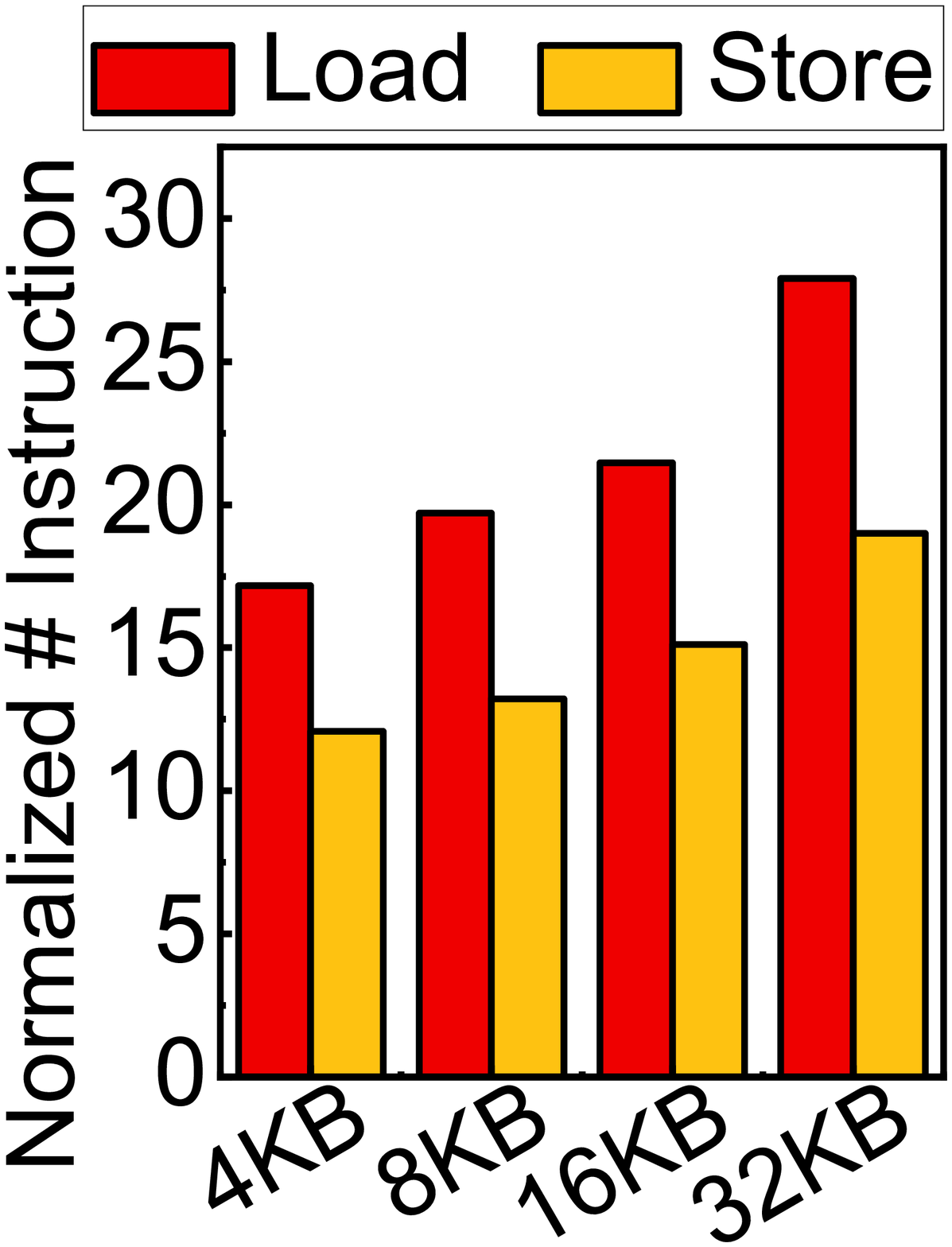}
		\caption{Rnd. writes.\centering}
		\label{fig:spdk_rndwr_norm_load}
	\end{subfigure}
	\vspace{-5pt}
	\caption{Normalized memory instruction count of SPDK.}
	\vspace{-15pt}
	\label{fig:spdk_loadstore}
\end{figure}

Figure \ref{fig:spdk_CPU} studies the CPU utilization of SPDK, which decomposes it into the CPU usages of userland and kernel spaces by comparing SPDK with the conventional system. Even though SPDK removes all kernel module involvements from its datapath, SPDK consumes all the CPU resources while the conventional system only uses, on average, 10\% and 15\% for userland and kernel executions, respectively. These excessive CPU usages are coming from polling process of the user-level NVMe driver. Since SPDK takes the NVMe driver from the kernel and puts it into \texttt{uio}, SPDK cannot handle ISR and therefore it cannot use interrupts as the I/O completion method. Thus, SPDK wastes CPU cycles to poll the huge page, which is mapped to BARs, and it increases power consumption. As multiple user applications need to communicate with \texttt{uio}, 100\% CPU occupancies (took by SPDK) may be insufficient to process data from the userland directly (as there is no room for such applications).

Similar to the overheads, imposed by polling-based I/O completion methods, SPDK also exhibits heavy memory accesses. Figure \ref{fig:spdk_loadstore} shows the number of load/store instructions for SPDK, which are normalized to the ones for the conventional system. SPDK generates load and store instructions more than those of the conventional system by 23$\times$ and 16.2$\times$, on average, respectively. 
Note that one of the reasons why the number of loads/stores with random reads is bigger than that of sequential reads or writes is that random read latency of ULL SSD is relatively slower than other patterns; the longer time it waits for I/O completion, the more polling is executed.

\begin{figure}
	\centering
	\begin{subfigure}{0.49\linewidth}
		\includegraphics[width=\linewidth]{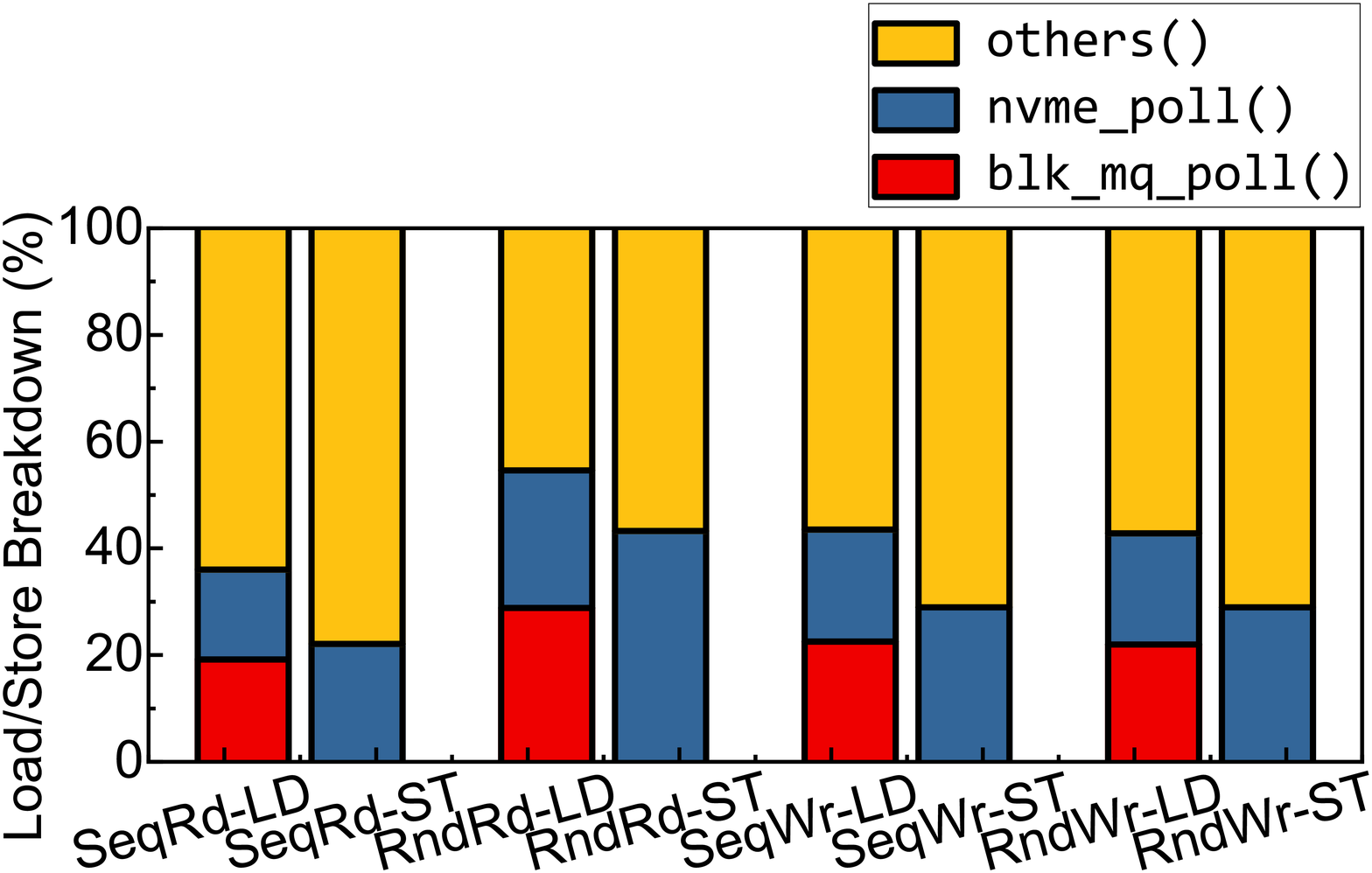}
		\caption{Polling.}
		\label{fig:polling_load}
	\end{subfigure}
	\begin{subfigure}{0.49\linewidth}
		\includegraphics[width=\linewidth]{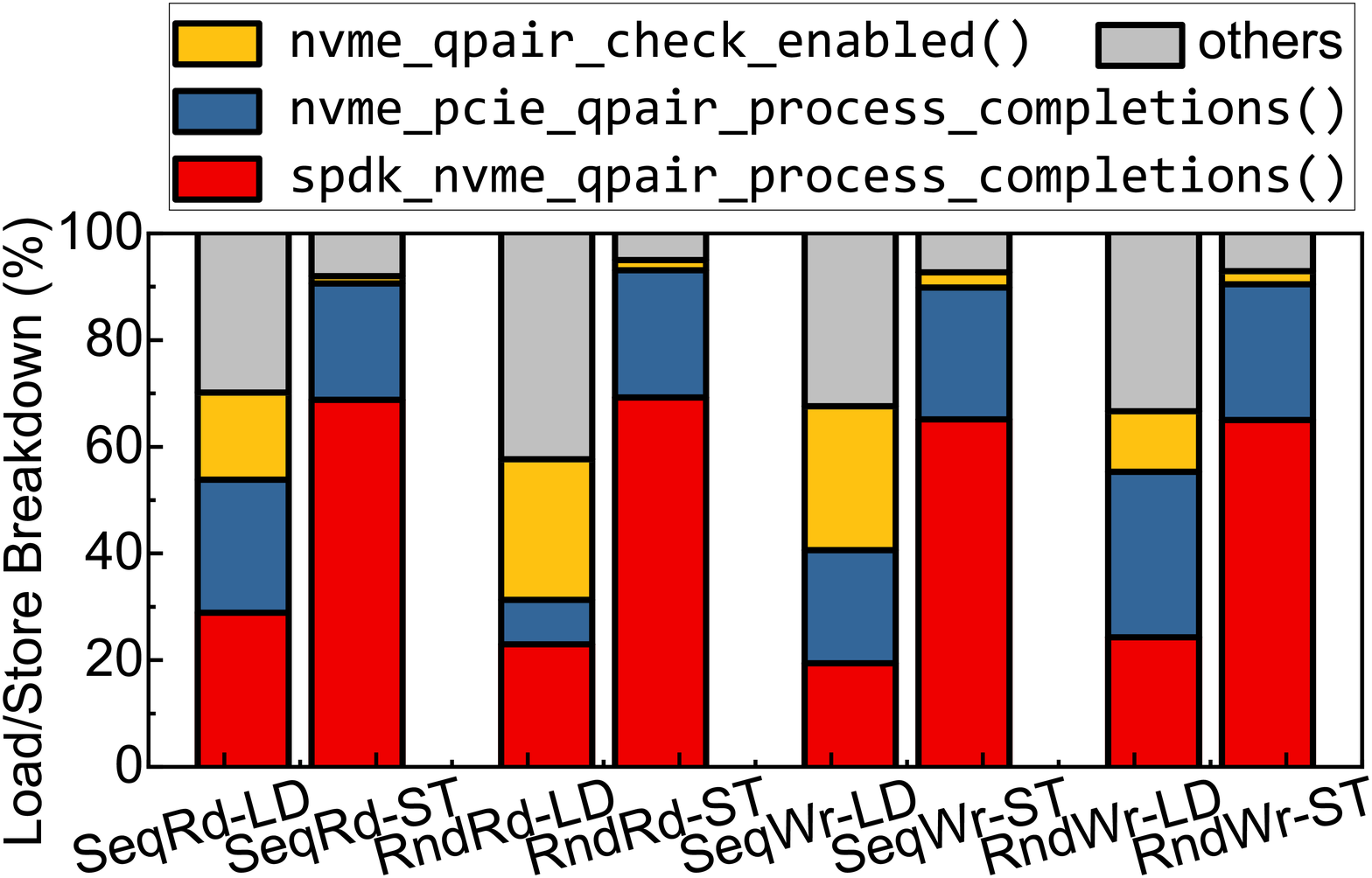}
		\caption{SPDK.}
		\label{fig:spdk_load}
	\end{subfigure}
	\vspace{-5pt}	
	\caption{Memory instruction breakdown.}
	\vspace{-20pt}
	\label{fig:ldst_breakdown}
\end{figure}

Figures \ref{fig:polling_load} and \ref{fig:spdk_load} analyze the contributor functions that generate most load/store instructions for the polled mode and SPDK, respectively, while we perform the previous memory access evaluations. In the conventional system, its polling takes 39\% of the total load/store instructions whereas SPDK's NVMe driver generates 45\% and 91\% of total load and store instructions, on average, respectively. Polling of the conventional system spends most of the execution time at \texttt{blk\_mq\_poll} and \texttt{nvme\_poll}, as discussed in Section \ref{pollcpu}. Surprisingly, even though SPDK also uses the polling as it cannot handle ISR, the number of load and store instructions is greater than that of the kernel side polled mode I/O completion by dozens of times (cf. Figure \ref{fig:poll_loadstore}). Specifically, \texttt{spdk\_nvme\_qpair\_process\_completions()} and \texttt{nvme\_pcie\_qpair\_process\_completions()} generates 37\% and 22\% of total memory instructions (loads/stores), on average, respectively. Note that the role of \texttt{spdk\_nvme\_qpair\_process\_completions()} and \texttt{nvme\_pcie\_qpair\_process\_completions()} are same with those of \texttt{blk\_mq\_poll} and \texttt{nvme\_poll}, respectively. One of the reasons behind two functions of SPDK generate more memory instructions is that SPDK's \texttt{uio} does not need to traverse the software and hardware queues (managed by \texttt{blk-mq}) to check up the CQ entries, and therefore more frequent memory accesses can be introduced. Interestingly, in contrast to kernel-side polling, SPDK uses \texttt{nvme\_qpair\_check\_enabled()}, which is implemented by an inline function. This inline function checks if the queue pairs of CQ and SQ are valid whenever an I/O request is issued. This in turn introduces more loads, which account for 20\% of the total load instructions. \texttt{nvme\_qpair\_check\_enabled()} is used for SPDK to avoid memory accesses during a controller reset, which is triggered under a certain conditions, such as PCIe reset \cite{nvmespec}.
Note that a hybrid polling may be applied to the future storage system in an attempt to reduce CPU and memory access overheads. However, it is in the practical difficulty to be applied since the hybrid polling only supports synchronous operations, while SPDK supports both synchronous and asynchronous operations.

\subsection{How much can SPDK reduce the system-level latency under a real server execution environment?}
\begin{figure}
	\centering
	\begin{subfigure}{0.24\linewidth}
		\includegraphics[width=\linewidth]{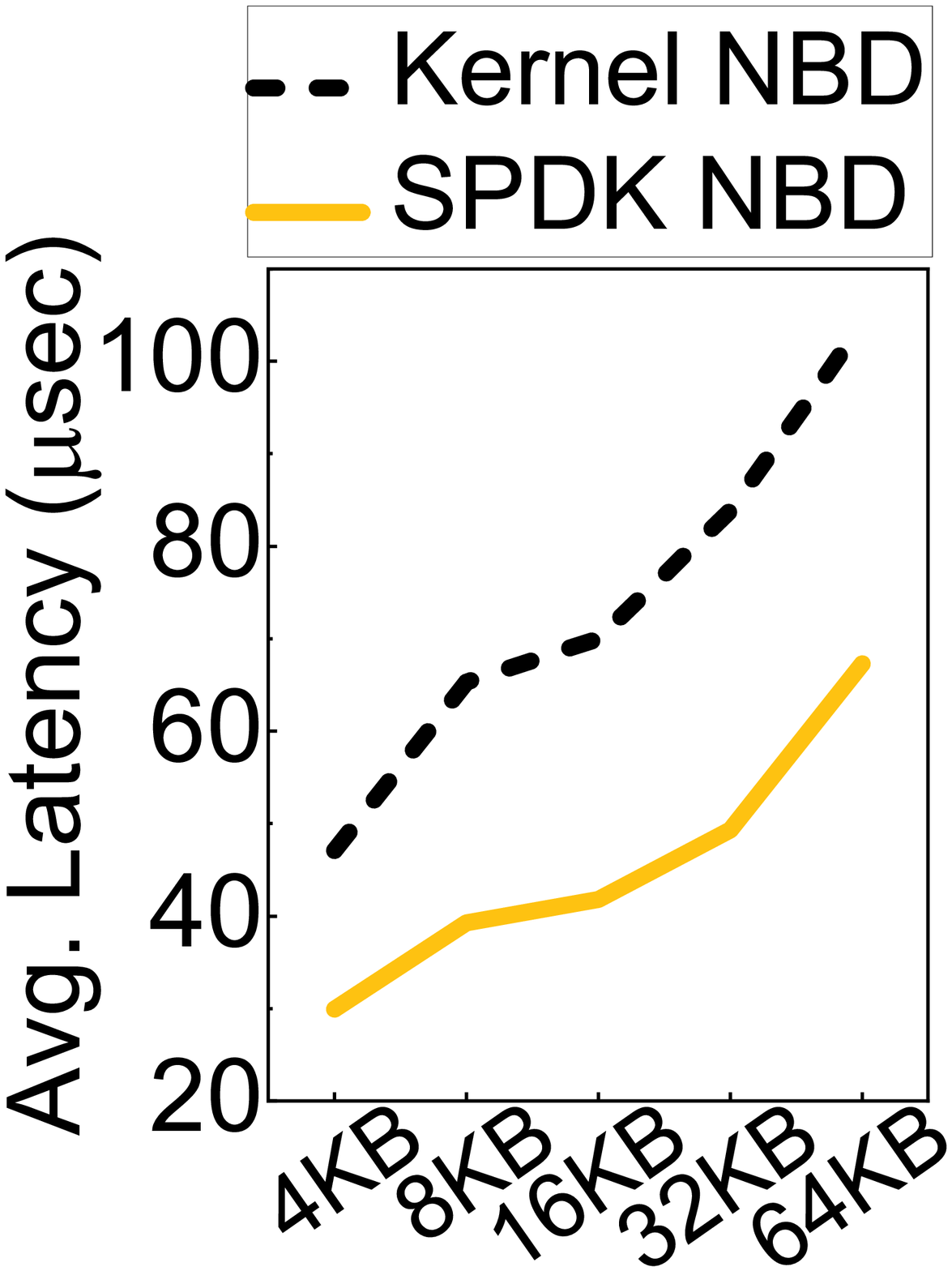}
		\caption{Seq. reads.\centering}
		\label{fig:seqrd_spdk_files}
	\end{subfigure}
	\begin{subfigure}{0.24\linewidth}
		\includegraphics[width=\linewidth]{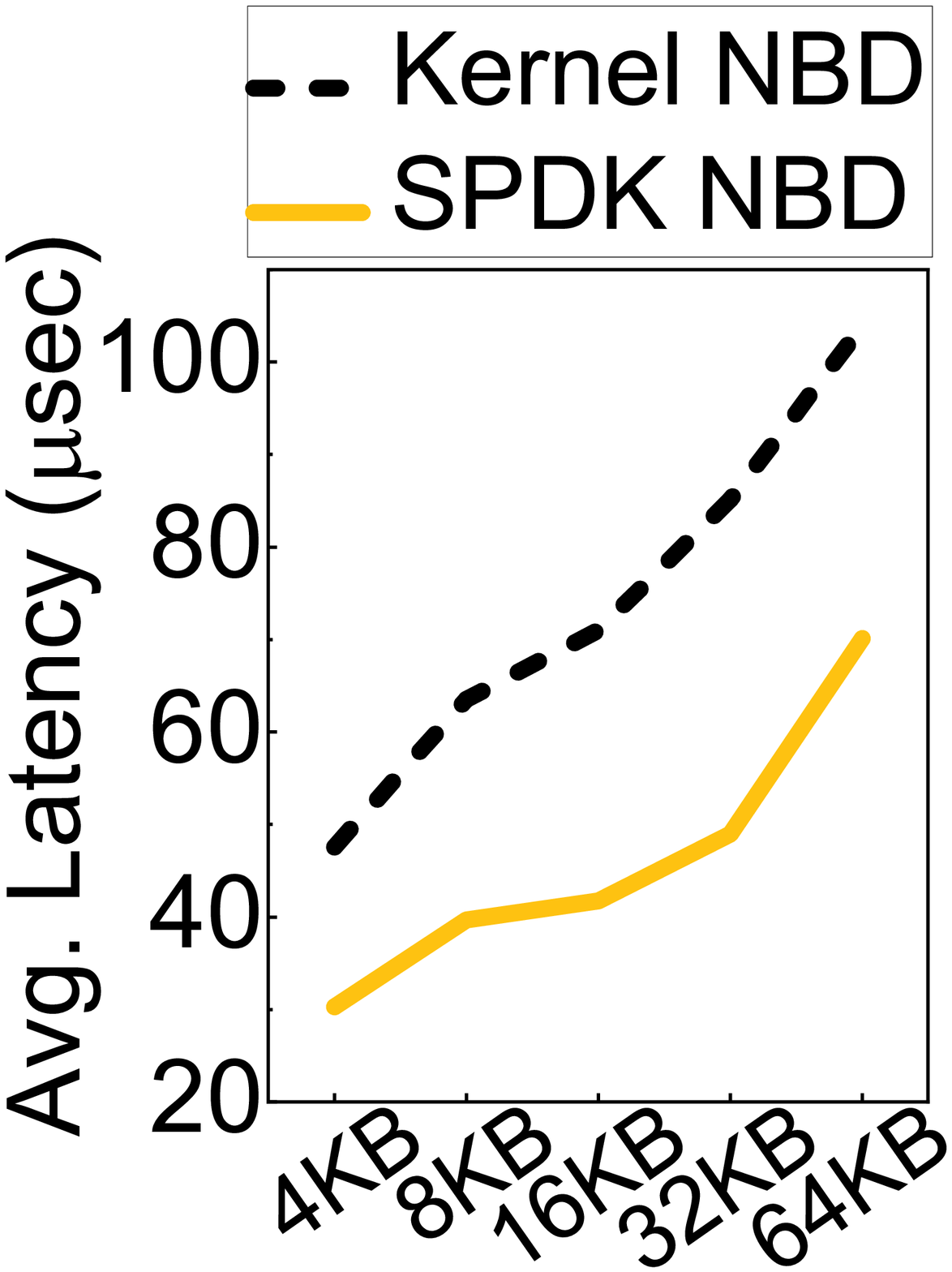}
		\caption{Rnd. reads.\centering}
		\label{fig:rndrd_spdk_files}
	\end{subfigure}
	\begin{subfigure}{0.24\linewidth}
		\includegraphics[width=\linewidth]{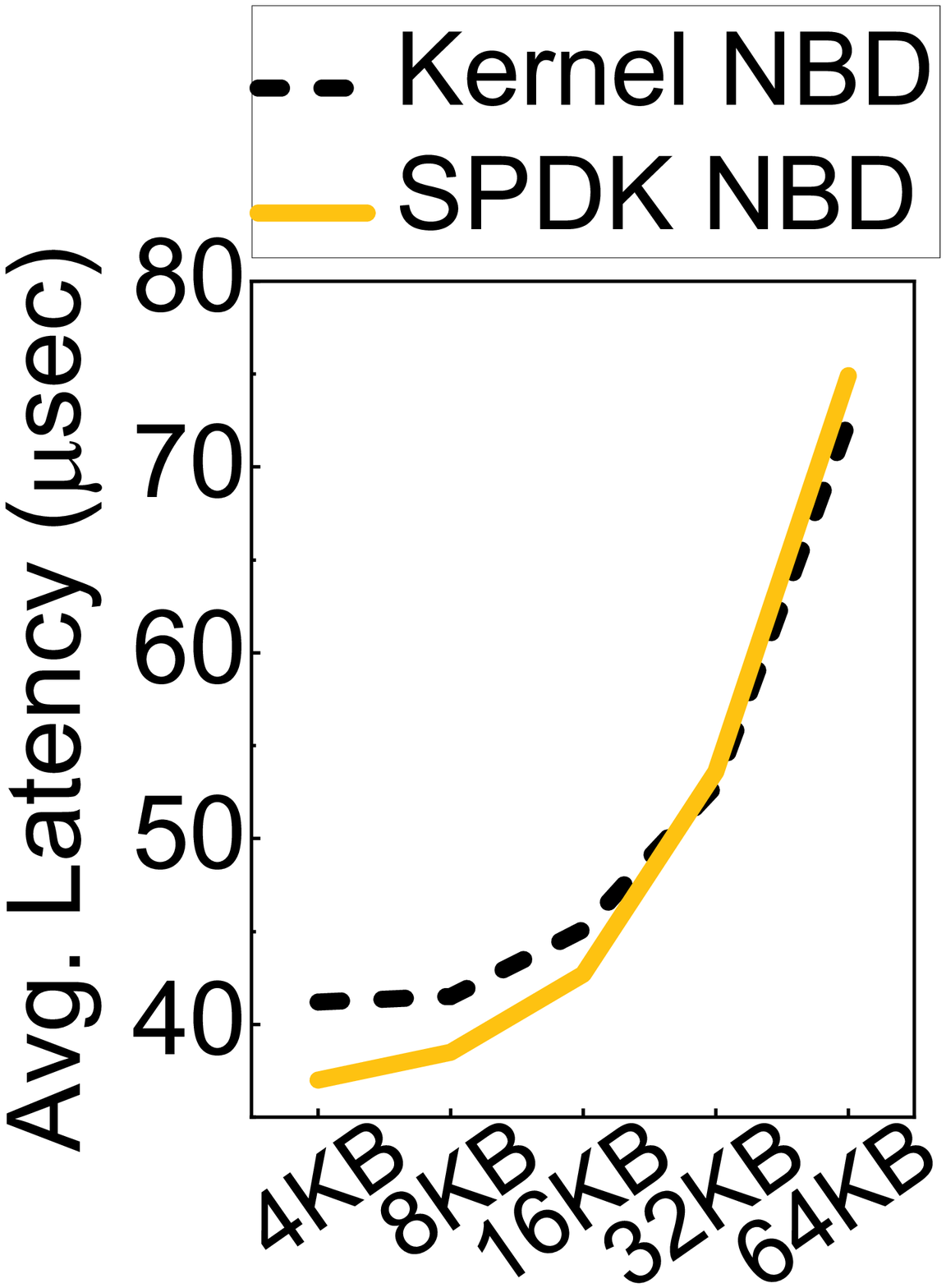}
		\caption{Seq. writes.\centering}
		\label{fig:seqwr_spdk_files}
	\end{subfigure}
	\begin{subfigure}{0.24\linewidth}
		\includegraphics[width=\linewidth]{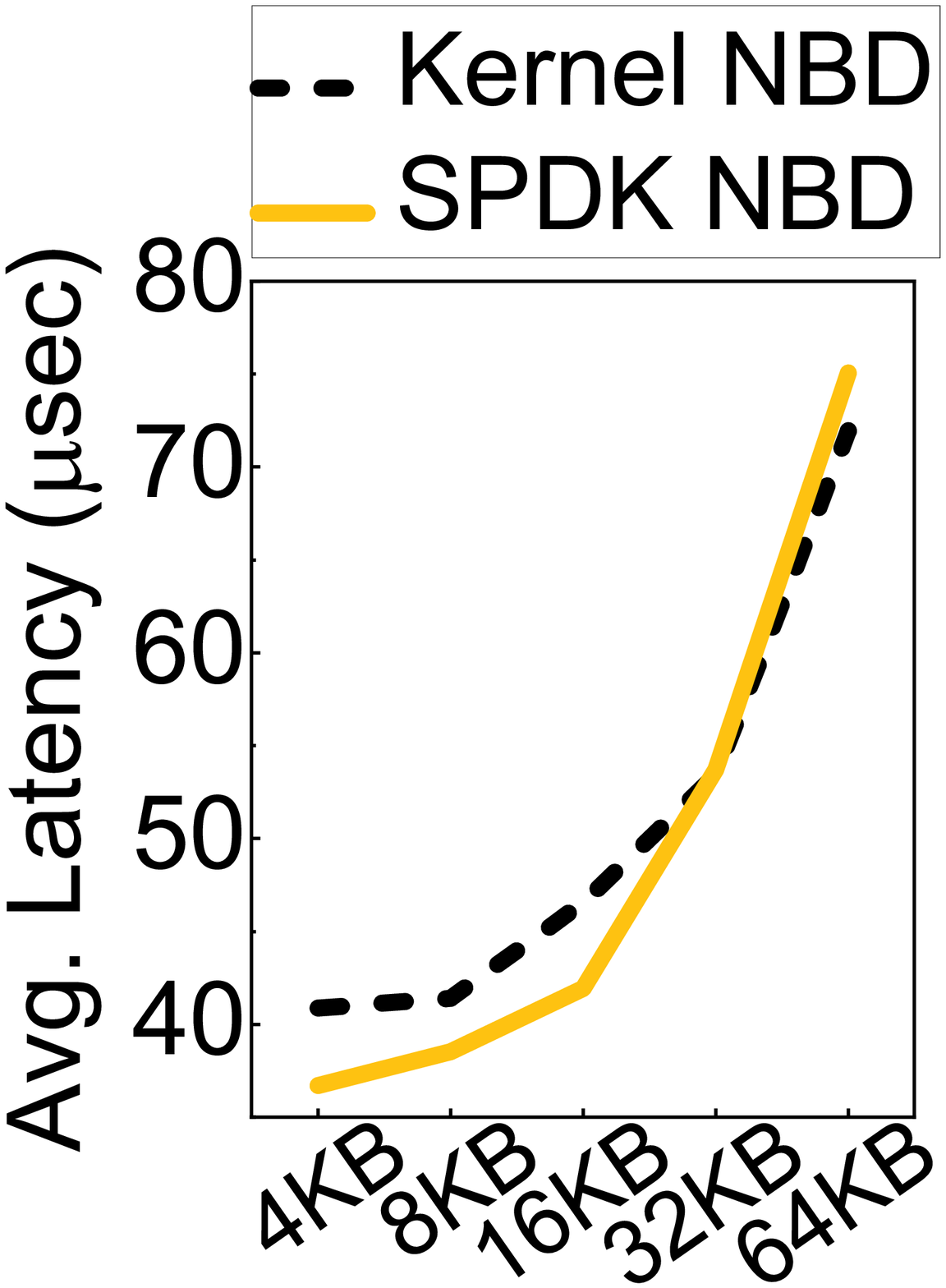}
		\caption{Rnd. writes.\centering}
		\label{fig:rndwr_spdk_files}
	\end{subfigure}
	\vspace{-5pt}
	\caption{Analysis of SPDK for a server-client scenarios.}
	\vspace{-15pt}
	\label{fig:spdk_real_usage}
\end{figure}

Many prior studies \cite{walker2016spdk, yang2018spdk, cao2018polarfs, spdkblobstore, spdkceph, an2017optimizing} use SPDK to directly access the underlying SSD by bypassing all kernel modules in the storage stack. Even though the kernel-bypass is a very promising and attractive approach, it is limited to apply the kernel-bypass to a system that does not have a file system or ACID support. However, in the server-client model, a file system should be employed by the client side node, which cannot be bypassed. We configured a server-client system to evaluate the performance improvement of SPDK when the file system exists at the client-side. Note that, in Section \ref{750_spdk}, we demonstrated that the current high-end NVMe SSD has almost no benefit of using SPDK, and therefore, we only test this server-client model with ULL SSD. To be precise, we configure two network block devices: \emph{Kernel NBD} and \emph{SPDK NBD}. Kernel NBD uses a conventional network block device implemented in Linux kernel, while SPDK NBD is a server-client system that enables ULL SSD as a network block device by employing SPDK and DPDK. In this evaluation, both kernel NBD and SPDK NBD employ ext4 as their file system. 
ACID of I/O services for both kernel NBD and SPDK NBD is always guaranteed since every I/O request heading to the server side SSD passes through the file system. In other words, client side kernel (i.e., storage stack) cannot be bypassed even in case of SPDK NBD; SPDK and DPDK only bypass the server-side kernel modules. 
Figure \ref{fig:spdk_real_usage} compares the latency brought by those two kernel NBD and SPDK NBD when FIO accesses a set of files, connected to the network block devices. In this test, we build 10 million files whose size varies, ranging from 4KB to 64KB. 
For read services, SPDK NBD can reduce the read latency, compared to kernel NBD by 39\% (sequential) and 38\% (random), on average, even though the execution time of client-side file system is included in its I/O path. 
However, for writes, SPDK cannot reduce the latency as much as what we observed in Figure \ref{fig:zssd_spdk}. Specifically, the latency of sequential and random writes exhibited by SPDK are respectively 3.7\% and 4.6\% shorter than those of kernel NBD, on average. 
We believe that this is because the writes need to handle more file system level operations than the reads. The writes should perform creating or modifying multiple metadata (i.e., inodes and bitmaps) and manage system journaling to guarantee the system consistency, while the reads only exhibit minor changes on the metadata (i,e., access time). These file system level operations in writes in turn can unfortunately introduce significant kernel module involvement in accessing ULL, which lose the benefits of SPDK NBD on the writes.  

\vspace{-5pt}
\section{Related Work}
While there is no new flash archive (i.e., ULL SSD) publicly available in the market, it is essential to revamp the current storage stack for such future storage. 
\cite{yang2012poll} pointed out that the existing interrupt-driven method cannot make systems utilize the time that CPUs are released since the latency of NVMe devices is shorter than that of the existing storage. The study proposed to employ a polling-based synchronous I/O service in high performance systems. Similarly, \cite{caulfield2010moneta, liu2009virtualization, swanson2013refactor, polling17, zhang2018flashshare, eisenman2018reducing, yang2017spdk, kim2016nvmedirect} advocate to employ the polled-mode system for fast NVMe devices. To use the polling in an efficient manner, \cite{polling17, zhang2018flashshare, eisenman2018reducing} proposed different types of a hybrid polling, which combines the current interrupt-driven method with the polling by being aware of application characteristics or runtime environment. While these studies are mostly performed by simulation or DRAM emulation, of which the actual datapath exists on the memory controller side rather than block storage controller side. In this work, we performed an in-depth analysis of the entire storage stack, including NVMe driver and \texttt{blk-mq}, and revealed several system-level characteristics and implications, which are not observed by the previous work. 

On the other hand, the kernel-bypass and user-level I/O management are also popularly studied in both academia and industry. Specifically, \cite{kim2016nvmedirect, shin2014path, caulfield2012providing} consider using user-space I/O framework to directly access the underlying NVMe devices and employing a dedicated polling thread to reduce the system-level latency, which are similar to the industry-driven I/O frameworks such as SPDK \cite{walker2016spdk, yang2017spdk}, OpenMPDK \cite{openmpdk2018} and uNVMe \cite{openmpdk2018, unvme19, unvmemicron}. As an application of this advanced storage stack, \cite{yang2017spdk} leverages SPDK to reduce the latency, imposed by multiple virtual machines; this study uses a userland driver to connect hypervisor to the underlying SSD and distributes queue pairs across individual virtual machine instances. As this work pointed out, the current NVMe device is not fast enough to take the full benefits of employing SPDK. However, in the near future, SPDK-applied systems can reduce the overheads imposed by software kernel involvement from their I/O path. To maximize the advantage of SPDK, we also provide several in-depth studies such as file system overheads, which should be reconsidered in a real server-client usage.

\vspace{-5pt}
\section{Conclusion}
\label{sec:conclusion}
\vspace{-5pt}
We analyzed the performance behaviors of ULL SSDs and brought several system-level challenges of the current and future storage stack to take full advantages of ULL devices. Specifically, we observed that ULL SSDs can offer their maximum bandwidth with only a few queue entries, which contradicts with the design direction of the current rich NVMe queue. While it is beneficial to employ a polling-based I/O completion routine for ULL SSDs, system-level overheads delivered by polling, such as high CPU cycles and frequent memory accesses, incur frequent CPU stalls and increase overall system power consumption.

\section*{Acknowledgment}
This is a full version of a 5-page workshop paper \cite{koh2018exploring}. In this work, we completely revised all the previous evaluation results from scratch by updating Linux kernel, Z-SSD firmware and hardware testbeds with the latest software/hardware version. The authors thank Samsung for the engineering sample donations and technical support. This research is mainly supported by NRF 2016R1C1B2015312. This work is also supported in part by NRF2015M3C4A7065645, DOE DE-AC02-05CH 11231, NRF2017R1A4A1015498, and Samsung grant (G01190271). The authors also thank Omesh Tickoo, Prof. J. W. Lee (SNU) and, J. Hwang (Samsung) for shepherding this paper and cooperative collaborations.



{\footnotesize \bibliographystyle{acm}
\bibliography{ref}}

\end{document}